\documentclass[manuscript]{acmart}

\AtBeginDocument{%
  }

\usepackage{subcaption}

\begin{document}

\newcommand{\minititle}[1]{\noindent
\textbf{#1}}

\title[Generative AI and Perceptual Harms]{Generative AI and Perceptual Harms: Who’s Suspected of using LLMs?}

\author{Kowe Kadoma}
\email{kk696@cornell.edu}
\affiliation{%
  \institution{Cornell University}
  \city{Ithaca}
  \state{NY}
  \country{USA}
}

\author{Danaé Metaxa}
\email{metaxa@seas.upenn.edu}
\affiliation{%
  \institution{University of Pennsylvania}
  \city{Philadelphia}
  \state{Pennsylvania}
  \country{USA}
  }

\author{Mor Naaman}
\email{mor.naaman@cornell.edu}
\affiliation{%
  \institution{Cornell Tech}
  \city{New York}
  \state{New York}
  \country{USA}
}

\renewcommand{\shortauthors}{Kadoma et al.}

\begin{abstract}
Large language models (LLMs) are increasingly integrated into a variety of writing tasks.
While these tools can help people by generating ideas or producing higher quality work, like many other AI tools, they may risk causing a variety of harms, potentially disproportionately burdening historically marginalized groups. 
In this work, we introduce and evaluate \textit{perceptual harms}, a term for the harms caused to users when others perceive or suspect them of using AI. 
We examined perceptual harms in three online experiments, each of which entailed participants evaluating write-ups from mock freelance writers. 
We asked participants to state whether they suspected the freelancers of using AI, to rank the quality of their writing, and to evaluate whether they should be hired.
We found some support for perceptual harms against certain demographic groups.  
At the same time, perceptions of AI use negatively impacted writing evaluations and hiring outcomes across the board.
\end{abstract}



\received{}
\received[revised]{}
\received[accepted]{}

\maketitle

\section{Introduction}
The public release of generative artificial intelligence (AI) technologies like ChatGPT, Copilot, Claude, 
Stable Diffusion, and Midjourney has led to widespread adoption across various sectors. 
People are using generative AI for a variety of tasks, from inspiring ideas~\cite{singh2023elephant,gero2022sparks,yuan2022wordcraft} and revising text~\cite{cui2022correct} to creating images~\cite{epstein2023art,zhou2024artcreativity}
and popular music~\cite{chu2022empirical,dervakos2021heuristics,frid2020music}. 
Recent scholarship in a variety of domains has shown several benefits to using generative AI tools. For example, in the workplace, generative AI can increase workers' productivity and reduce the skill gap between employees~\cite{Zhang-2023-productivity,haslberger2023no}.
In education, it has been argued that generative AI can support engaging, personalized learning~\cite{baidoo2023education,chan2023students}.

Despite these possible benefits, generative AI technologies also have the potential to cause harm.
Researchers have identified several types of harms in generative AI systems, many of which disproportionately burden historically marginalized groups~\cite{weidinger-etal-2022-risks,shelby2023}.
Many of the identified harms stem from biases that arise during the model's development when decisions about training data have downstream impacts on the model's outputs~\cite{mehrabi2021survey,jiang2020identifying}.
Furthermore, several of these identified harms directly impact people who are interfacing with AI systems.
For example, users may encounter stereotypical text in language models~\cite{abid2021muslimbias} or stereotypical images in text-to-image models~\cite{gautam2024melting,ghosh2023person,sun2024smile} when prompting the model, often due to biases in the training data.
However, there are some harms that are not caused by model outputs.

In this work, we define a new category of potential AI harms: perceptual harms.
Perceptual harms, we argue, occur when the appearance (or perception) of AI use---regardless of whether AI was actually used---results in differential treatment between social groups.
Because perceptual harms are not caused by the model's outputs, they can be categorized as a potential societal harm~\cite{shelby2023}.

Perceptual harms are likely to occur as AI-generated content continues to proliferate in various domains, given the fact that AI disclosure statements are not common or practical and that people cannot reliably identify AI-generated content~\cite{draxler2024ghostwriter,jakesch2023human}.
Previous research in AI-mediated communication~\cite{hancock2020ai} has demonstrated, in multiple settings, that the suspicion that a communication partner used AI to create content results in reduced evaluations of that partner~\cite{jakesch2019ai,mieczkowski2021ai}. 
Furthermore, when people use AI, they are seen as less intentional~\cite{hong2018biasart}.
All these results from prior work can be seen as examples of perceptual harms, but these papers did not reflect on any differential treatment among groups.
In this context, we offer a framework to understand the various dimensions of perceptual harms and how to study the disparate impact the perception of AI use may have on different groups. 

We measure perceptual harms across three dimensions of increasing consequence for the person being evaluated: suspicion of AI use, evaluation of content quality, and potential outcomes (e.g., hiring). 
These dimensions were chosen based on the literature highlighting people's inability to distinguish between AI-generated and human-generated content (inducing suspicion), which can influence how people assess the quality of said content~\cite{kobis2021artificial,jakesch2023human}.
The suspicion and reduced quality evaluations potentially have an impact on outcomes for the people whose content is being evaluated. 
Furthermore, it is possible that perceptual harms may function independently across these three dimensions.  
For example, men might be more readily suspected of using AI than women, but women's content could still be judged more negatively than men's when AI use is suspected. 
Investigating perceptual harms across these three dimensions, this paper asks the following questions:
\begin{itemize}
  \item[\textbf{RQ1}] Are different groups of people suspected of using AI at different rates?
  \item[\textbf{RQ2}] How does the suspicion impact the evaluation of the work across different groups? 
  \item[\textbf{RQ3}] How does the suspicion impact the outcomes for people of different groups?
\end{itemize}

We present three experiments, each of which addresses all of the research questions, with each experiment focusing on a different identity category: 
Experiment~1 focuses on gender, Experiment~2 on race, and Experiment~3 on nationality.
We chose these three identity categories because they are legally protected categories along the lines of which individuals in the U.S. context experience marginalization and thus are likely to display perceptual harms. 

Specifically, we investigate perceptual harms in the context of online freelance marketplaces, which both provide a plausible environment and context for an online experiment of perceptual harms in LLMs and serve as an important context of study. 
Online freelance marketplaces are popular platforms where customers can hire workers for a variety of tasks~\cite{hannak2017marketplaces}.
These marketplaces have a significant economic impact by providing additional income for many workers, with many relying on them as their primary source of income~\cite{hannak2017marketplaces}.
While LLMs could help freelancers complete tasks more efficiently (and potentially increase earnings), these technologies have also increased competition within the marketplace as traditional customers are opting to use the LLM as opposed to human services~\cite{liu2023generate,hui2023short}.
In an already competitive marketplace, perceptual harms could add to the disenfranchisement of specific groups, as evident, for example, by prior work showing that women and Black workers face discrimination in online platforms~\cite{hannak2017marketplaces}.

This paper addresses the research questions above using a series of carefully designed online experiments to test the differences in outcomes for various demographic groups in settings that are as realistic as possible.
In the experiments, participants evaluated the profiles of freelance professionals for a purported hiring decision.
We used a within-subjects experimental design, with participants exposed to the (fictional) profiles from all demographic groups (e.g., women and men in Experiment~1) in a randomized order.
After reading the written content from each freelancer, we asked participants whether they thought the freelancer had used AI-assistance in their writing sample and also asked them to evaluate the quality, content, and structure of the writing.
We also asked participants about the likelihood of hiring the freelancer for a task similar to the writing sample. 

The results are mixed. 
There is some evidence that different social groups may be suspected of using AI more than others.
We found that socially dominant groups (men in Experiment~1) and non-dominant groups (foreign nationals in Experiment~3) {may be impacted differentially on the basis of AI suspicion.
However, we do not see evidence of differential evaluations or differential outcomes between groups.
As expected, we find evidence that people associate AI writing with lower quality and that lower-quality work negatively impacts future job opportunities.
Aside from these empirical results, this work also offers a framework for considering and evaluating perceptual harms in Generative AI and LLM applications and outlines a new way to think about the adverse impacts of these technologies.

\section{Background and Hypotheses}
We situate this study in two main areas of related literature: harms in sociotechnical systems, specifically in the context of AI, and AI-mediated communication. Building on both these areas of prior work, we present and explain the hypotheses driving our study. 

\subsection{Harms in Sociotechnical Systems}
Sociotechnical systems---systems that are a combination of technical and social components---have been extensively studied for harms and biases, especially in the realm of machine learning and artificial intelligence~\cite{selbst2019fairness,shelby2023}. 
Many potential harms caused by AI systems are seen as byproducts of the models' construction and, in particular, the training data used~\cite{mehrabi2021survey,shelby2023,weidinger-etal-2022-risks,jiang2020identifying}. 
When socially constructed beliefs and unjust hierarchies in the data are reflected as model outputs, it is a particular type of harm known as representational harms~\cite{shelby2023,weidinger-etal-2022-risks}. 
Representational harms may occur in language models when stereotypical text or demeaning language is generated~\cite{shelby2023,weidinger-etal-2022-risks}. 
Oftentimes, this harmful language is directed at historically marginalized groups like women~\cite{lucy-bamman-2021-gender,kotek2023genderbias} and racial and ethnic minorities~\cite{abid2021muslimbias,nadeem-etal-2021-stereoset}.
With text-to-image models, representational harms may occur when models produce stereotypical images or fail to recognize particular identities~\cite{gautam2024melting,ghosh2023person,sun2024smile}.

Aside from the technical components of these systems, social aspects that impact users' experiences can also cause harms.
For example, users of marginalized backgrounds may require an increased effort for the tool to work as well for them, a type of harm known as quality-of-service harms~\cite{shelby2023}.
When these tools do not perform as well for users of marginalized backgrounds, it can cause deep feelings of frustration, self-consciousness, and shame, which users from non-marginalized backgrounds may not experience~\cite{wenzel2023voice,mengesha2021inclusive}.
Additionally, users of marginalized backgrounds may also not feel included by the tool, which can undermine their sense of agency~\cite{kadoma2024writer}.
Prior work has even found that visual cues in UIs can signal belongingness (or lack thereof) to users of different gender groups~\cite{metaxa2018gender}.

In this paper, we propose a new category of potential harms, one that is situated neither in these technologies themselves nor in user interactions with them. Unlike model and user interaction-related biases, we propose that public opinion about these tools, combined with existing social stereotypes about groups or individuals, can lead to negative judgments and perceptual harms when someone is perceived or assumed to have used an AI system (whether or not they actually did so).

\subsection{AI-Mediated Communication} 
AI-mediated communication (AI-MC)~\cite{hancock2020ai} represents a growing body of work examining the effects of introducing AI into human-to-human exchanges. 
Researchers have provided robust evidence for some of the adverse outcomes when AI modifies, augments, or generates messages.
One notable consequence of AI involvement in communication is the \textit{Replicant Effect}, or rather the decrease in trust between people when someone suspects AI was involved~\cite{jakesch2019ai}. 
The Replicant Effect has been shown in several settings such as chats~\cite{hohenstein2023artificial} and online dating~\cite{wu2020dating}, and even in email, where recipients' trust decreased when they were told AI was involved in the writing process~\cite{liu2022console}.
Most notably, for the context of our experiment, prior work has shown that the perception a self-presentation profile was written by AI can negatively impact the perceived trustworthiness of the profile writer's~\cite{jakesch2019ai}.

Additional research in AI-MC shows that most people cannot reliably differentiate between human-written text and AI-generated text~\cite{jakesch2023human,ippolito2019automatic,clark-etal-2021-thats,dou-etal-2022-scarecrow}. 
Oftentimes, people rely on false heuristics such as verbose language and grammatical errors~\cite{jakesch2023human,fu2024aimc} to try to differentiate between human-written text and AI-generated text.
While we know that AI involvement leads to decreased trust and that people can't differentiate between AI-generated text and human text, we do not know if these effects have differential impacts on people of different social groups.

\subsection{Hypotheses}
While most harms impact historically marginalized groups, in the context of AI use, we hypothesize that the dominant group will be most often suspected. 
White men are overrepresented in the technology sector~\cite{crawford2016artificial,landivar2013disparities,house2016preparing} and have a more positive view toward technology~\cite{kim2023one}. 
Thus, while counterintuitive, we hypothesize that the perception of these trends in society will cause participants to more readily think White men are likely to use AI over groups like White women or Black men. 
At the same time, we hypothesize that a national identity suggesting non-US and presumed English-as-second-language speakers will be suspected of using AI more than the baseline comparison group (white, presumed native English speakers).
While nationality is a political construct, it is deeply intertwined with culture and language~\cite{xueliang1988language}. 
Although there is no official language in the United States, English remains the dominant medium of communication~\cite{usalanguage}.
Prior work has shown that non-native English speakers are less likely to produce English without mistakes and could, therefore, be seen as needing more help in the form of AI assistance~\cite{hwang2023chatgpt,giglio2023use,buschek2021nonnative}.
In this case, we believe the evaluators' assumptions about nationality and language will overpower the effect we expect in Experiment~1 and~2. 

Our first hypotheses is thus:
\begin{itemize}
   \item H1: In race and gender contexts, freelancers from the dominant group will be more suspected of using AI; however, contributors presented as non-English nationalities will be more suspected of using AI.
\end{itemize}

In the quality evaluation measurements, while all evaluated people could be affected by AI suspicion, we hypothesize that people from the non-dominant social groups will receive lower evaluations.
Prior work has shown that while controlling for ability, age, and experience, there are still significant differences in performance evaluations between people of different genders and races. 
For example, in healthcare, female physicians likely unduly received lower teaching evaluations than their male counterparts~\cite{morgan2016student}.
Regarding race, it has been shown, for example, that in the US labor force, across a variety of occupations, Black workers received lower evaluations than White workers (more than likely due to biases)~\cite{waldman1991race}.

\begin{itemize}
   \item H2: Controlling for suspicion, freelancers from the non-dominant group will receive lower quality evaluation scores. 
\end{itemize}

Our hypothesis for the outcomes measurement is similar to the evaluation measurement---controlling for whether they were suspected of using AI, people from the non-dominant social groups will be less likely to be hired.
Correspondence audits in sociology have shown that while controlling for education and experience, there are still significant differences in hiring likelihood between people of different genders and races.
For example, it has been shown that women with children received fewer callbacks, whereas men with children were not disadvantaged during the hiring process~\cite{correll2007mothers}.
Additionally, for race, Bertrand and Mullainathan showed that Black job seekers are significantly less likely to be hired than White job seekers~\cite{bertrand2004emily}. In both of these examples, women and Black job seekers had worse outcomes, likely due to biases in hiring.

\begin{itemize}
   \item H3: Controlling for suspicion, freelancers from the the non-dominant group will be less likely to be recommended for hiring.   
\end{itemize}

\section{Methods}
\label{sec:methods}
We addressed our research questions using a series of within-subjects online experiments, aiming to mimic a realistic scenario where crowd workers are asked to help review the profiles of freelance marketing contributors, including whether these freelancers used generative AI to create their content. We first describe how we created the profiles (Section~\ref{sec:profile}) and the content associated with the freelancers  (Section~\ref{sec:content}), before describing our measurements (Section~\ref{sec:measures}) and experimental procedure (Section~\ref{sec:procedure}) in more detail.
We then describe participant recruitment in Section~\ref{sec:participant}.
The research design was approved by Cornell University IRB and preregistered on OSF\footnote{\url{https://osf.io/sa4ce/?view_only=bdfcec7b419041cd9fa695a2153ef275}}\textsuperscript{,}\footnote{\url{https://osf.io/ugnak/?view_only=363c5f8efe3b4a2985b8571db126d894}} where the experiment data is also available. See Section~\ref{sec:dev} in the appendix for a preregistration deviation report.

\subsection{Profile Creation} 
\label{sec:profile}
We manipulated the demographic information of the ``freelancers'' by changing the freelancers' stated names and the photos in their profiles shown to participants. 
Our presentation parallels the design of popular freelancing sites like Fiverr or TaskRabbit, which do not explicitly state workers' gender, race, or nationality but feature names and photos on each profile.

For Experiment~1 (gender) and Experiment~2 (race), we created the displayed names from a list of common first and last names complied by Gaddis~\cite{gaddis2017black}.
Gaddis used the list to test the racial perception of names used in online correspondence audits while controlling for socioeconomic status (SES)~\cite{gaddis2017black}.
Gaddis selected 160 first names from the New York State birth record data categorized by gender, race, and SES.
Gaddis also provides a set of 20 common surnames from the 2000 US Census Bureau that are associated with a skewed racial composition, i.e., significantly likely to be White or Black (e.g., Of people with the surname ``Nielson'', 95.6\% of them are White).

For Experiment~1, we randomly selected two masculine-associated first names (i.e., Brett and Graham) and feminine-associated first names (i.e., Emily and Meredith) that are of the same race (White) and SES (high) from the initial list of first names provided by Gaddis~\cite{gaddis2017black}. 
We also selected four surnames that have a high occurrence for Whites in the US Census, like Walsh or Meyer.
We randomly paired the set of first names with the set of common surnames to create the full names of our freelancers as either White men or White women (e.g., ``Meredith Walsh'').

For Experiment~2, where we focused on race, we reused the White masculine-associated names from Experiment~1 and created a set of Black names.
We used a procedure similar to Experiment~1.
From Gaddis, we randomly selected two Black masculine-associated first names that are of a high SES and paired the first names with two surnames that have a high occurrence for Blacks in the US Census.

\begin{figure*}[t]
    \centering
    \begin{subfigure}[h]{0.28\textwidth}
        \centering
        \includegraphics[width=\textwidth,height=5cm,keepaspectratio]{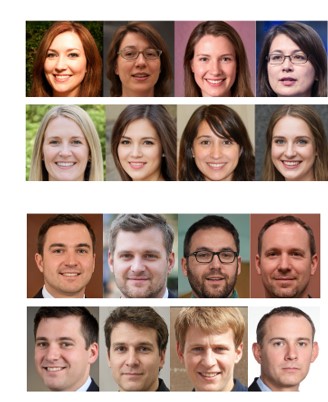}
        \caption{Experiment~1 (gender)} 
        \label{fig:1_profile}
    \end{subfigure} 
    \begin{subfigure}[h]{0.28\textwidth}
        \centering
        \includegraphics[width=\textwidth,height=5cm,keepaspectratio]{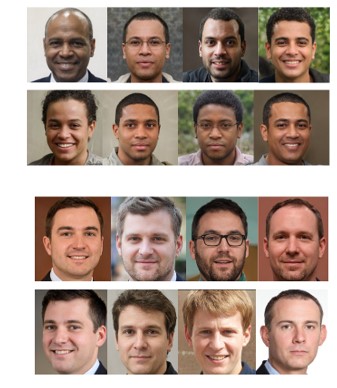}
        \caption{Experiment~2 (race)}
        \label{fig:2_profile}
    \end{subfigure}
    \begin{subfigure}[h]{0.27\textwidth}
        \centering
    \includegraphics[width=\textwidth,height=5cm,keepaspectratio]{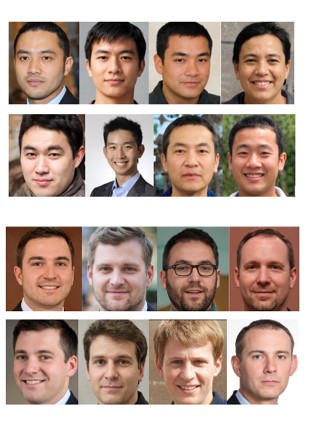}
        \caption{Experiment~3 (nationality)}
        \label{fig:3_profile}
    \end{subfigure}
    \caption{The different sets of profile photos used in each experiment, with two demographic groups in each.}
    \label{fig:profile_photos}
\end{figure*}

In Experiment~3 (nationality), we used a list of names compiled by Hogan~\cite{hogan2011racial}.
While the list does not have nationality metadata, the names were chosen to reflect the demographics of Toronto, Canada---a large, multi-ethnic city with many foreign-born, non-native English speakers.
In our experiment, we randomly chose two distinctly East and Southeast Asian masculine-associated names from this list and used them along with the White masculine-associated names used in Experiment~1 and~2.
The final set of names used in each experiment is shown in Table~\ref{table:names}.

\begin{table}[h]
\caption{\textbf{Profile Names}. Each column contains the profile names used in each experiment.}
\begin{tabular}{|c|c|c|}
\hline
Experiment 1 & Experiment 2 & Experiment 3 \\ \hline
\begin{tabular}[c]{@{}c@{}}Meredith Walsh\\ Emily Becker\\ Brett Larsen\\ Graham Meyer\end{tabular} & \begin{tabular}[c]{@{}c@{}}Andre Booker\\ Darius Washington\\ Brett Larsen\\ Graham Meyer\end{tabular} & \begin{tabular}[c]{@{}c@{}}Jun Liu\\ Fai Zhang\\ Brett Larsen\\ Graham Meyer\end{tabular} \\ \hline
\end{tabular}
\label{table:names}
\end{table}

To create the visual representations of our freelancers, we used the AI-based image generator ThisPersonDoesNotExist~\cite{Sashaborm_2021}.
The image generator created photorealistic images of fictional people based on the specified gender, age range, and ethnicity.
We created a distinct set of eight images in a professional headshot style for each demographic group in our studies.
The photos featured (fake) individuals in their mid-to late-30s, which is the age of most workers on freelance platforms~\cite{Segal_2023}. 
To create the profiles shown in our experiment interface, we randomly matched, within each demographic group, one of the two names to one of the eight photos.
Figure~\ref{fig:profile_photos} shows our final set of photos for each experiment.  

The interface screen shown to participants was modeled after the profile pages of workers on Guru, a popular freelance site.
Figure~\ref{fig:interface} shows an example profile as it was shown in the experiment interface, where each page features the name, photo, and location in the top banner, with the content below.

\begin{figure*}[h]
\includegraphics[width=\linewidth,height=6cm,keepaspectratio]{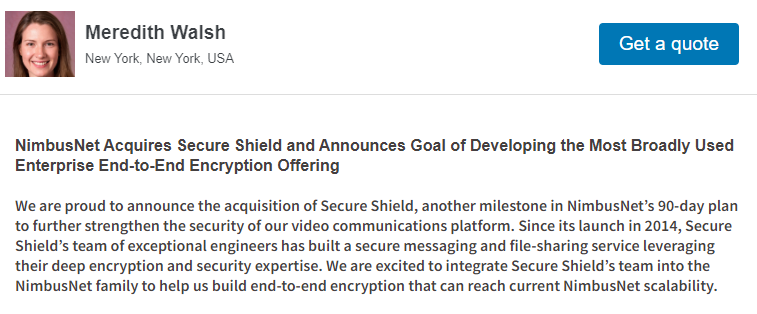}
\caption{\textbf{Screenshot of the Profile Interface.} The top panel contains the profile photo, name, and location of the (fake) freelancing marketing professional. The sample press release, supposedly written by this freelancer, is below.
}
\label{fig:interface}
\end{figure*}

\subsection{Content Creation}
\label{sec:content}
We asked participants to evaluate our freelancers based on the provided writing sample within the marketing and digital advertising domain. We selected this domain for our writing sample since marketing and digital advertising services are highly requested on freelance marketplaces~\cite{Segal_2023,upworkdata}. 
Within marketing, we chose press releases for our evaluation content, rather than other writing tasks, because they are intended to be read by a general audience. Additionally, press releases have elements of creativity in self-promotion while also being informative~\cite{catenaccio2008press}, which ensures content variety.

We created a set of sixteen press releases for our freelancer profiles.
To expand the generalizability of the experiment, we created four types of press releases---product launch, event announcement, acquisition or partnership, and new hire.
We started with press release templates from the public relations websites Prowly and PR Lab so that our writing samples would mirror the visual and written structure of varied, real-life press releases.  
We then simplified the structure of the existing templates by removing datelines, subject lines, logos, and company contact information.
The resulting set of templates for each press release type can be found in Table~\ref{table:templates}. 
After creating our templates, we identified existing press releases
on Prowly, PR Lab, and Business Wire websites and modified them to match our template designs by replacing all people, products, and companies with fictional counterparts. 
All of our press releases were between 300-400 words long.

\begin{table*}[h]
\caption{Press release templates used in the experiments and the structure of each press release type.}
\small
\centering
\begin{tabular}{|p{4cm}|p{4cm}|p{4cm}|p{4cm}|} 
\hline
\textbf{Product Launch} & \textbf{Event} & \textbf{Acquisition or Partnership} & \textbf{New Hire} \\ \hline
-Title \newline
-Product overview \newline
-Uniqueness of the product &  
-Title \newline
-Event challenges \newline
-Future projection  
&  
-Title \newline
-Company overview \newline
-Message from CEO  
&  
-Title \newline
-Overview of previous leadership \newline
-Announcement of new leadership  
\\ \hline
\end{tabular}
\label{table:templates}
\end{table*}

The content in all of the press releases was human-written; however, to ensure we arouse participants' suspicion of AI use, we manually modified half of the press releases across all types to sound more AI-like. 
Prior work has shown that people cannot distinguish between human-written content and AI-generated content, often relying on false heuristics to identify AI-generated content~\cite{jakesch2023human}. 
We directly used these heuristics, namely, verbose language and rare or long words~\cite{fu2024aimc,jakesch2023human}, to modify the chosen press releases. 
We opted to partially modify the human-written press releases (as opposed to modifying the whole press release), ensuring content that is faithful to the original while still mimicking AI-like characteristics. That kind of process is also similar to many human-AI co-writing situations where people write the majority of the text themselves and selectively incorporate AI suggestions~\cite{bhat2023cogproc}. 

Specifically, to arouse AI suspicion, we modified one sentence in the beginning, middle, and end of half of the press releases to include verbose language or vivid descriptions.
We also used a thesaurus to replace common words with less common synonyms. 
For example, in a product launch press release for a mobile game, we modified the sentence "Meet friendly villagers along the way and help them rebuild their homes and workshops" to "Encounter affable villagers and assist them in reconstructing their homes and workshops."
We call these profiles \textit{AI-inducing} profiles and evaluate below whether they had a different effect than the control set of profiles (controlling for any changes in quality evaluation that result from these edits).

\subsection{Measures}
\label{sec:measures}
We designed the experiment to examine the perceptual harms caused by generative AI. 
Our direct measures included AI suspicion, variables related to the content and its quality, and a measure of potential job (hiring) outcome.

\minititle{AI Suspicion.} Prior work on trustworthiness and AI-mediated communication uses the term `AI Score' to describe whether an evaluator thought the content had been AI-generated~\cite{jakesch2019ai}. 
In our work, we asked participants to review the profiles of online freelancers and state whether `freelancers may have used generative AI'.
By adding the statement about the freelancer's potential AI use, we primed participants to be suspicious of the extent to which the freelancer created the content.
We therefore refer to `AI Score' as `AI suspicion' as we believe this term captures our intended measure.
The response options for this measure range from \textit{definitely human-written} (1) to \textit{definitely AI-generated} (5) on a 5-point Likert. 
We also captured participants' explanations for their AI suspicion scores through a free response. 

\minititle{Quality Evaluations.} 
We used three different measures for quality evaluations. First, 
participants were asked to provide a rating for the \textit{overall quality} of the written content.
Response options ranged from~1, indicating a poor piece of writing, to~5, indicating an excellent piece of writing.
We did not provide examples for what constitutes each score, allowing participants to interpret these ratings based on their own judgment.

We drew from education literature, specifically writing evaluations for English as a Second Language (ESL) learners, to develop additional and more concrete writing evaluation measures.
These content-based measures capture the strength of the ideas or information conveyed in the message~\cite{rothschild1990self,sager1973sager}.
To develop our content measures, we omitted items from Rothschild~\cite{rothschild1990self} that focused on details in the writing~\cite{sager1973sager,rothschild1990self}.
Instead, we focus on first impressions and idea clarity with the following statements: \textit{the ideas and details expressed in the press release create an impression on the reader} and \textit{all ideas are clear and fully developed}.
The response options ranged from \textit{strongly disagree} (1) to \textit{strongly agree} (5).
We simplify our analysis by performing a row-wise average across the responses to the two statements to create the \textit{content index} measure.

While the content index captures \textit{what} ideas are conveyed in the writing, the structure measures capture \textit{how} the idea is expressed~\cite{rothschild1990self}. 
We modified the statements from Rothschild~\cite{rothschild1990self} to omit unnecessary details about sentence mechanics (e.g., explaining a run-on sentence or a fragment).
We measure structure with the following statements: \textit{each sentence is complete}, \textit{sentences are joined in the most effective/meaningful way}, and \textit{there are almost no grammatical errors}. 
The response options were the same as those for the content questions.
Similarly, we create a \textit{structure index} by performing a row-wise average across the three statements to simplify our analysis.

\minititle{Hiring Likelihood.} We wanted to understand the potential outcomes due to perceptual harms, namely, if the perceived use of generative AI would impact hiring decisions. 
We asked participants how likely they would hire the freelancer to complete a similar task on a scale from~0 (very unlikely) to 100 (very likely).

\subsection{Experiment Procedure}
\label{sec:procedure}
In each of our experiments, participants saw a demographically balanced set of four (2 demographic groups x 2 writing styles) (fictional) freelance profiles in a randomized order.
After reading the content on each profile, participants were asked to evaluate if the writing sample was generated by AI (RQ1---AI suspicion), as well as evaluate the writing's overall quality and its quality in terms of content and structure (RQ2---content evaluation).
Participants were then asked about the likelihood they would hire the freelancer for a similar task (RQ3---outcomes).
After the experiment, we asked participants about their demographic background and familiarity \& attitudes toward artificial intelligence. 

\subsection{Participant Recruitment}
\label{sec:participant}
We recruited our participants using a US-representative sample of English-speaking adults on the crowdsourcing service Prolific.
To determine the number of participants in each experiment, we calculated the sample size for a medium effect size using a linear mixed model with 80\% power based on a pilot experiment.
Each experiment had its own unique set of participants. 
We recruited 350 participants for Experiment~1 and Experiment~2 and 300 participants for Experiment~3.
In all of our experiments, to ensure high-quality responses for our analysis, we excluded participants who failed attention checks or correctly guessed the purpose of the experiment.
We had usable data from 334 participants in Experiment~1, the same number in Experiment~2, and 272 participants in Experiment~3.
Across all three experiments, approximately 25\% of participants were between 55-64 years old, 64\% identified as White, 37\% received a bachelor's degree, and 40\% reported being somewhat familiar with coding. 
More than 90\% of participants said they were familiar with generative AI technologies like ChatGPT or Gemini, and roughly 80\% of participants reported using generative AI tools.
More details on participant's demographics can be found in Section~\ref{sec:detailed_demographics} of the Appendix.

\section{Results}
Our results provide some evidence of perceptual harms.
Our three experiments confirmed that people of different groups are suspected of AI use at varying rates (H1), but we did not find conclusive evidence that perceptual harms impacted the quality evaluations of the writing content (H2) or resulted in different hiring outcomes for people of different groups (H3). 
The experiments do, however, provide strong evidence that people associate content suspected as AI with lower quality regardless of the author's gender, race, or nationality, which in turn negatively impacts job opportunities.

\subsection{Experiment~1: Gender}
Experiment~1 examined how perceptual harms may affect different gender groups. 
We found that men were suspected of using AI more than women, and this difference between genders was exacerbated when the writing style was AI-inducing. However, we did not observe gender to have a significant impact on participants' evaluations of writing quality or their hiring recommendations.

\begin{figure}[t]
    \centering
    \begin{subfigure}[h]{0.5\textwidth}
        \centering
        \includegraphics[width=\textwidth,height=4.8cm,keepaspectratio]
        {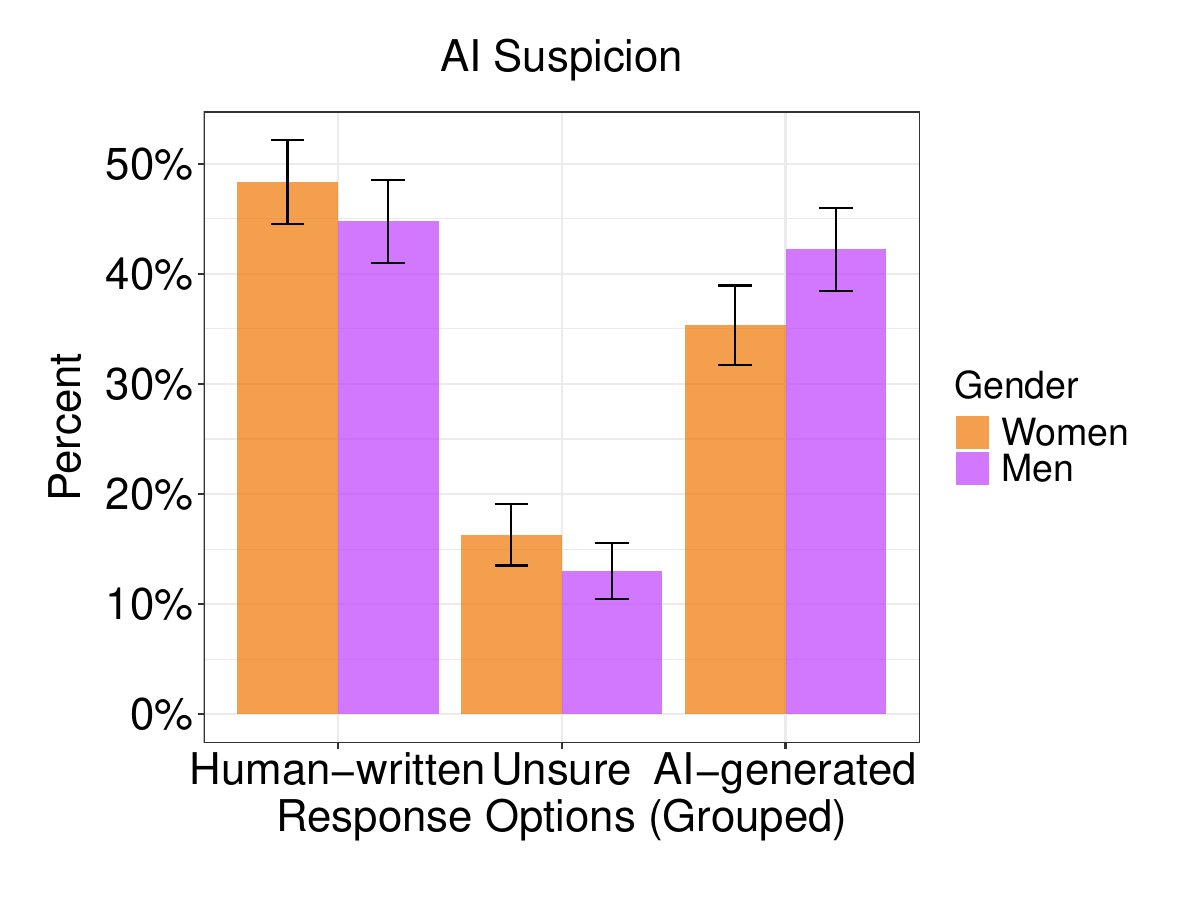}
        \caption{AI Suspicion: overall frequency of responses by gender}
        \label{fig:gender_ai}
    \end{subfigure} 
    \begin{subfigure}[h]{0.49\textwidth}
        \centering
        \includegraphics[width=\textwidth,height=4.8cm,keepaspectratio]
        {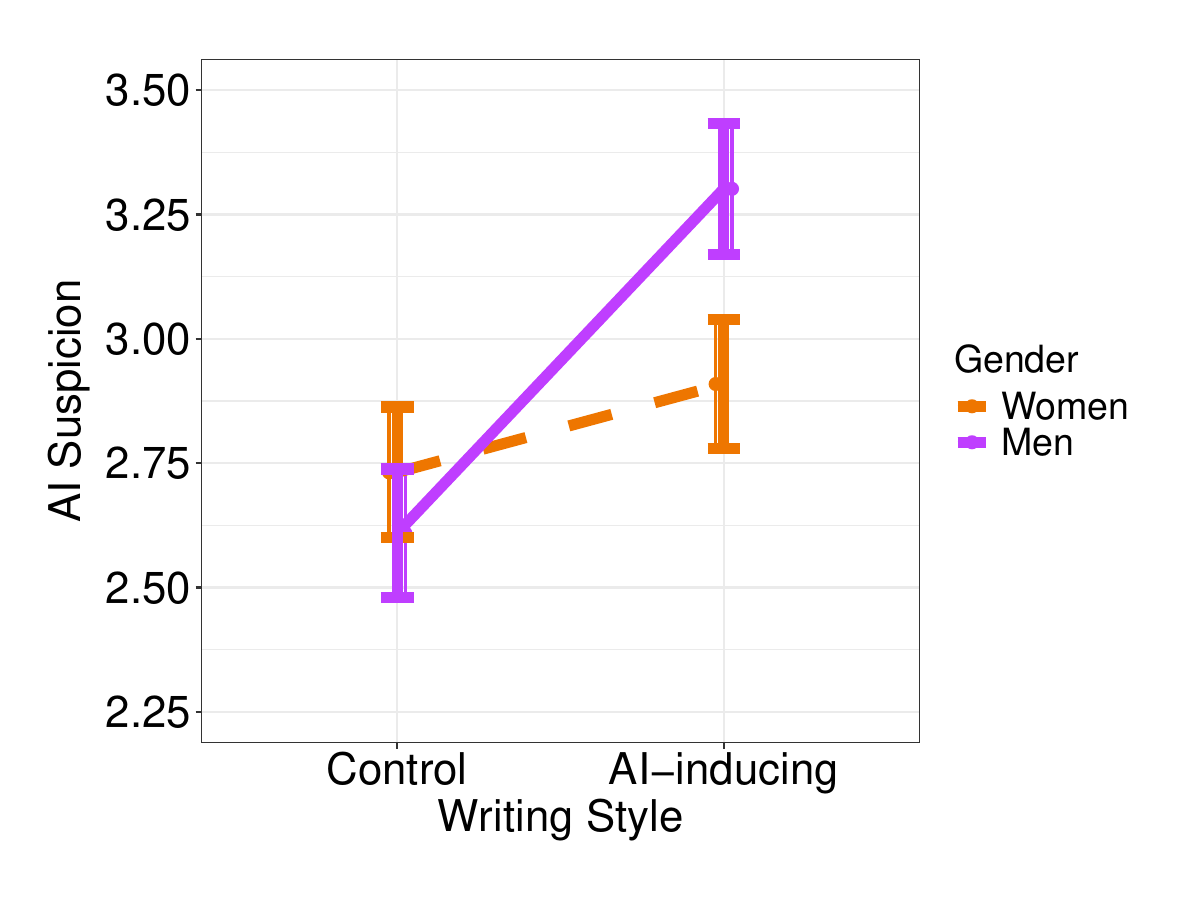}
        \caption{AI Suspicion: interaction between writing style and gender}
        \label{fig:gender_type_ai}
    \end{subfigure}
    \caption{Participants' AI Suspicion evaluation by the presented gender of the evaluated freelancer.}
    \label{fig:gender_ai_overall}
\end{figure}

\begin{figure*}[t]
    \centering
    \begin{subfigure}{0.28\textwidth}
        \centering
        \includegraphics[width=\textwidth,height=5cm,keepaspectratio]{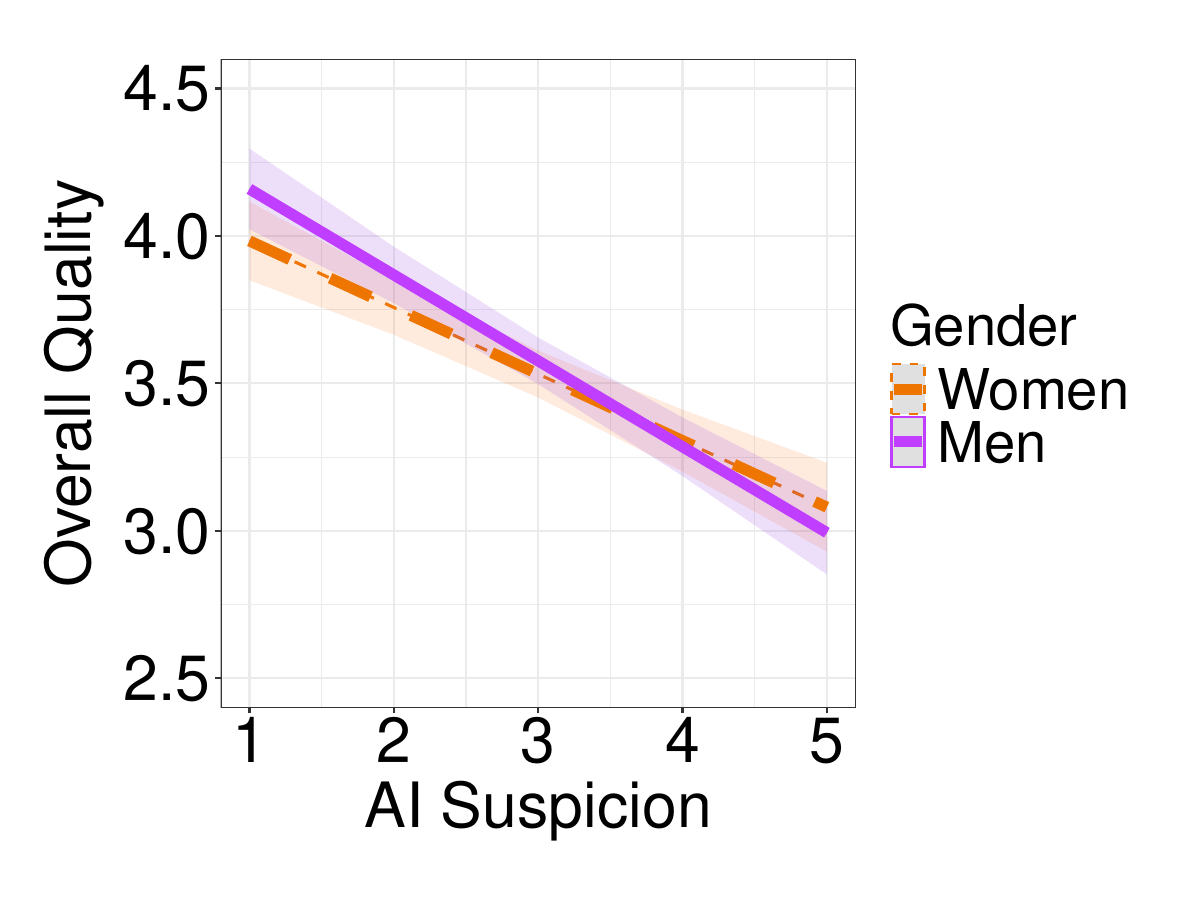}
        \caption{All Writing Styles} 
        \label{fig:gender_all_quality}
    \end{subfigure} 
    \begin{subfigure}{0.28\textwidth}
        \centering
        \includegraphics[width=\textwidth,height=5cm,keepaspectratio]{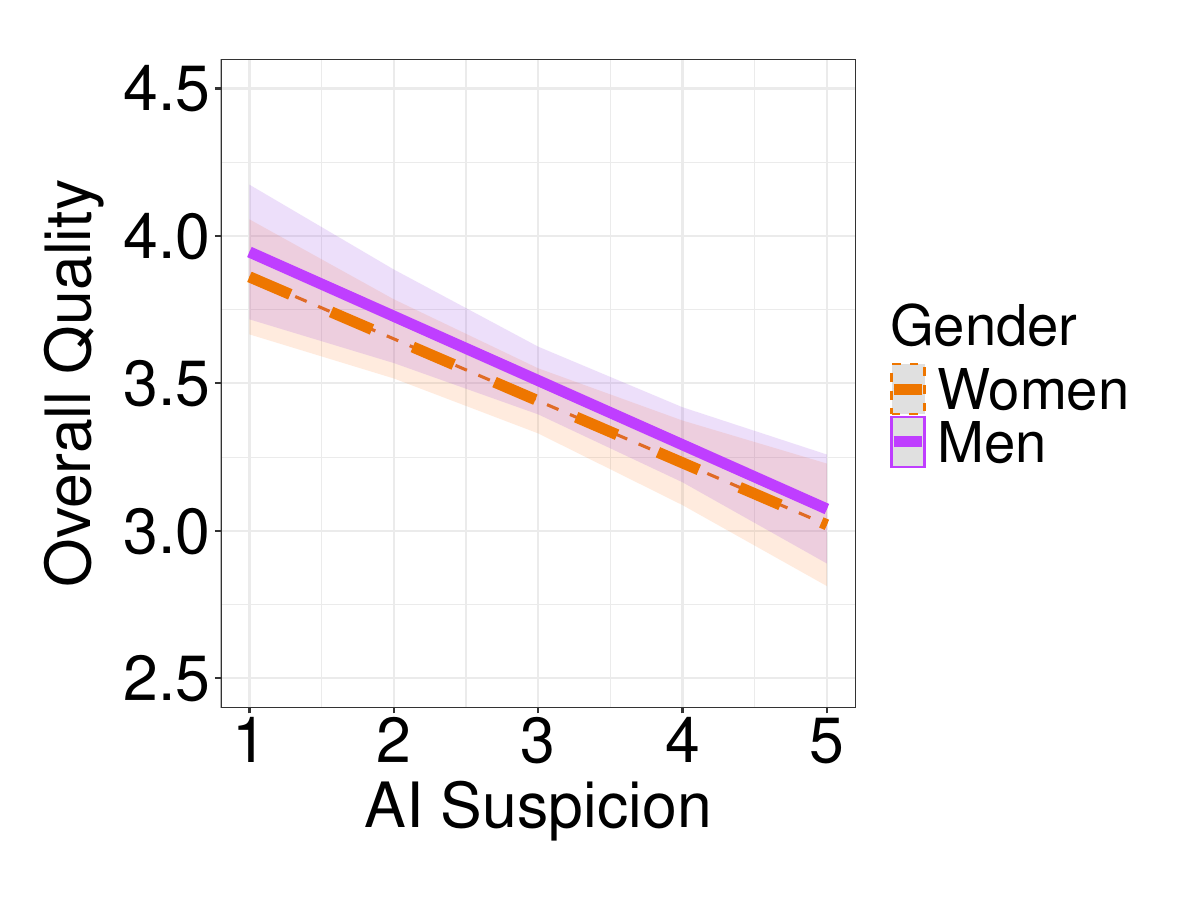}
        \caption{AI-inducing}
        \label{fig:gender_ai_quality}
    \end{subfigure}
    \begin{subfigure}{0.27\textwidth}
        \centering
    \includegraphics[width=\textwidth,height=5cm,keepaspectratio]{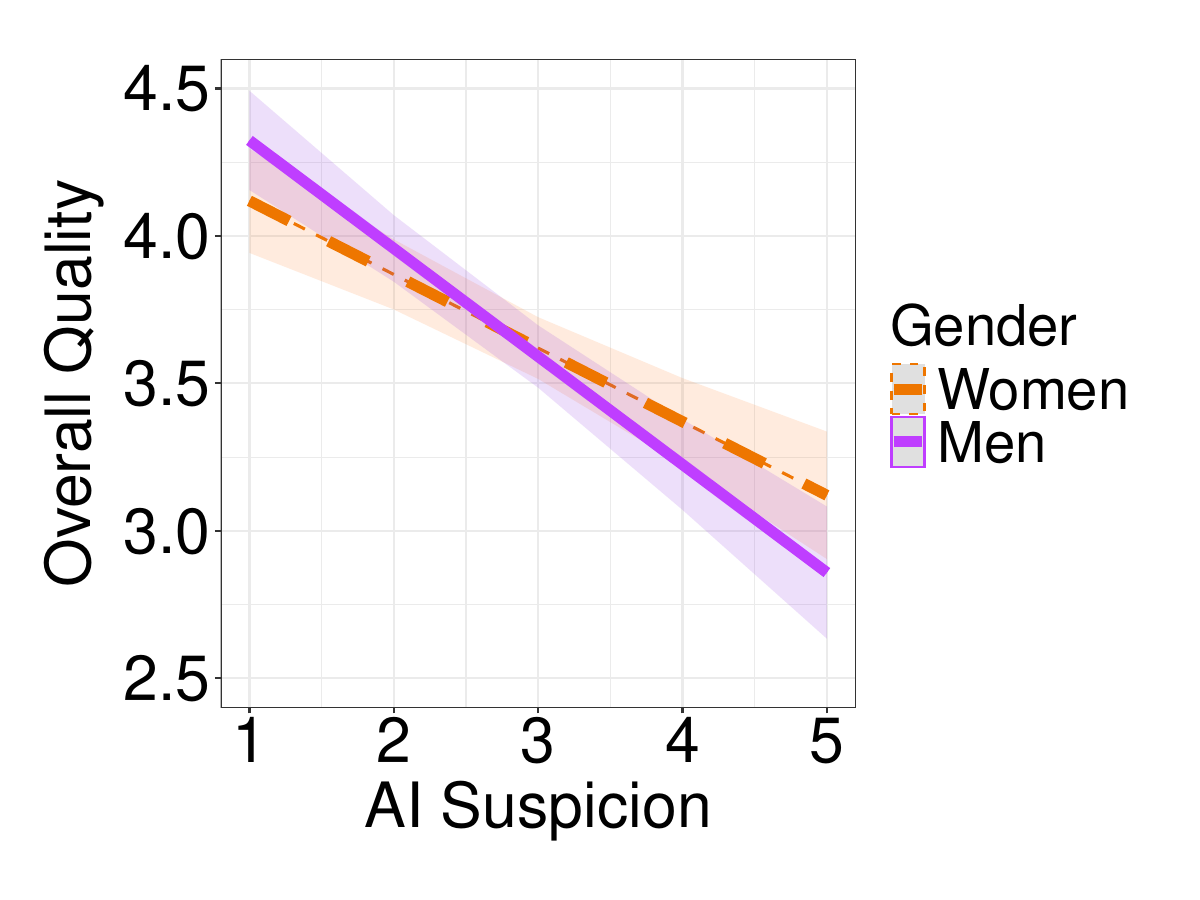}
        \caption{Control}
        \label{fig:gender_control_quality}
    \end{subfigure}
    \caption{The negative effect of AI Suspicion on Quality Evaluations, by gender.}
    \label{fig:gender_interaction_ai_quality}
\end{figure*}

\minititle{AI Suspicion.}
\label{sec:ai_sus}
Figure~\ref{fig:gender_ai} shows the frequency (y-axis) of different survey responses to the AI Suspicion question (x-axis) for each gender. 
Although participants' responses were recorded on a 5-point scale, we grouped the ``definitely'' and ``probably'' responses on each side (Section~\ref{sec:measures}) to create a 3-point scale reflecting the participants' leaning in their rating: human-written, unsure, or AI-generated. 
For example, when the evaluated freelancer was presented as a man, 42.2\% of participants thought the writing was AI-generated (as indicated by the rightmost purple column in the figure).
In contrast, when the freelancer was presented as a woman (in the rightmost orange column), just 35.3\% of participants thought the writing was AI-generated.
To statistically evaluate this difference, we used a linear mixed model with gender as a fixed effect; we included participants as a random effect to account for the repeated measures design and individual variability. 
The model confirmed that men freelancers were statistically significantly more likely to be suspected of using AI compared to women ($p=0.037$).

We next examined how inducing AI suspicion via the language used in the writing samples affected AI suspicion.
Figure~\ref{fig:gender_type_ai} illustrates the interaction of gender (men and women) and writing style (AI-inducing or control).
On the left of the figure, when the writing sample was written in a normal style (control), women (dashed, orange) and men (solid, purple) freelancers were suspected of using AI at about the same rate.
However, when we manipulated the writing style to sound AI-stylized (AI-inducing), men were more suspected of using AI than women.
This effect was confirmed by a linear mixed model analysis with AI suspicion as the dependent variable and the fixed effects of gender, writing style (AI-inducing or control), and their interaction with participants as a random effect.
Our results show that gender had a significant main effect ($p<0.001$) on AI suspicion, and there was also a significant interaction effect between gender and writing style ($p<0.001$), as reflected in the figure. The effect of writing style alone on AI suspicion also trended towards significant (p=$0.058$).

\minititle{Quality Evaluations.} Figure~\ref{fig:gender_interaction_ai_quality} shows the relationship between AI suspicion (x-axis) and overall quality ratings (y-axis) when participants evaluated freelancers presented as men (solid, purple) or women (dashed, orange). 
To examine the effect of the writing style on quality, we present three versions of the data: the full data from both conditions (Fig~\ref{fig:gender_interaction_ai_quality}a, on the left), as well as the results from the two conditions separately, the AI-inducing style treatment (Fig~\ref{fig:gender_interaction_ai_quality}b, in the middle) and the control (Fig~\ref{fig:gender_interaction_ai_quality}c, on the right).
Across all the conditions, the figure shows a strong inverse relationship: higher AI suspicion is associated with lower quality ratings. 
The figure also shows that there is no clear difference between quality evaluations for men versus women when controlling for suspicion. 

We used a linear mixed model to statistically analyze the overall quality data of all writing styles (represented in Fig~\ref{fig:gender_interaction_ai_quality}a). 
In the model, we include AI suspicion, gender, and writing style as fixed effects. 
We also include the interaction effects of AI suspicion and gender, writing style and gender, writing style and AI suspicion, and the three-way interaction between gender, AI suspicion, and writing style as fixed effects.
Lastly, we add participants as a random effect. 
The model confirms the visual trends in the figure.
AI suspicion had a significant main effect on quality ($p<0.001$); however, gender did not have a significant main effect ($p = 0.522$, \textit{n.s.}). 
While writing style was not significant ($p = 0.060$, \textit{n.s.}), the trend indicates that a larger sample could expose quality differences where the AI-inducing writing style is seen as lower in quality (controlling for suspicion). 
There were no significant interactions in the model.

Similar to the analysis for overall quality, we also consider the effect of AI suspicion on the dependent variables of content index and structure index (Section~\ref{sec:measures}). 
We use a linear mixed model for each measure, using the same independent variables as the overall quality model reported above. 
The results are largely the same:
for the \textit{content} dependent variable, the AI suspicion negatively impacted content evaluation ($p<0.001$), and both genders were evaluated similarly ($p=0.264$, \textit{n.s.}). 
Lastly, the AI-inducing writing style was seen as lower quality content compared to the control ($p=0.005$).
The results for the \textit{structure} variable are similar.

\begin{figure}[h]
\includegraphics[height=4.8cm]{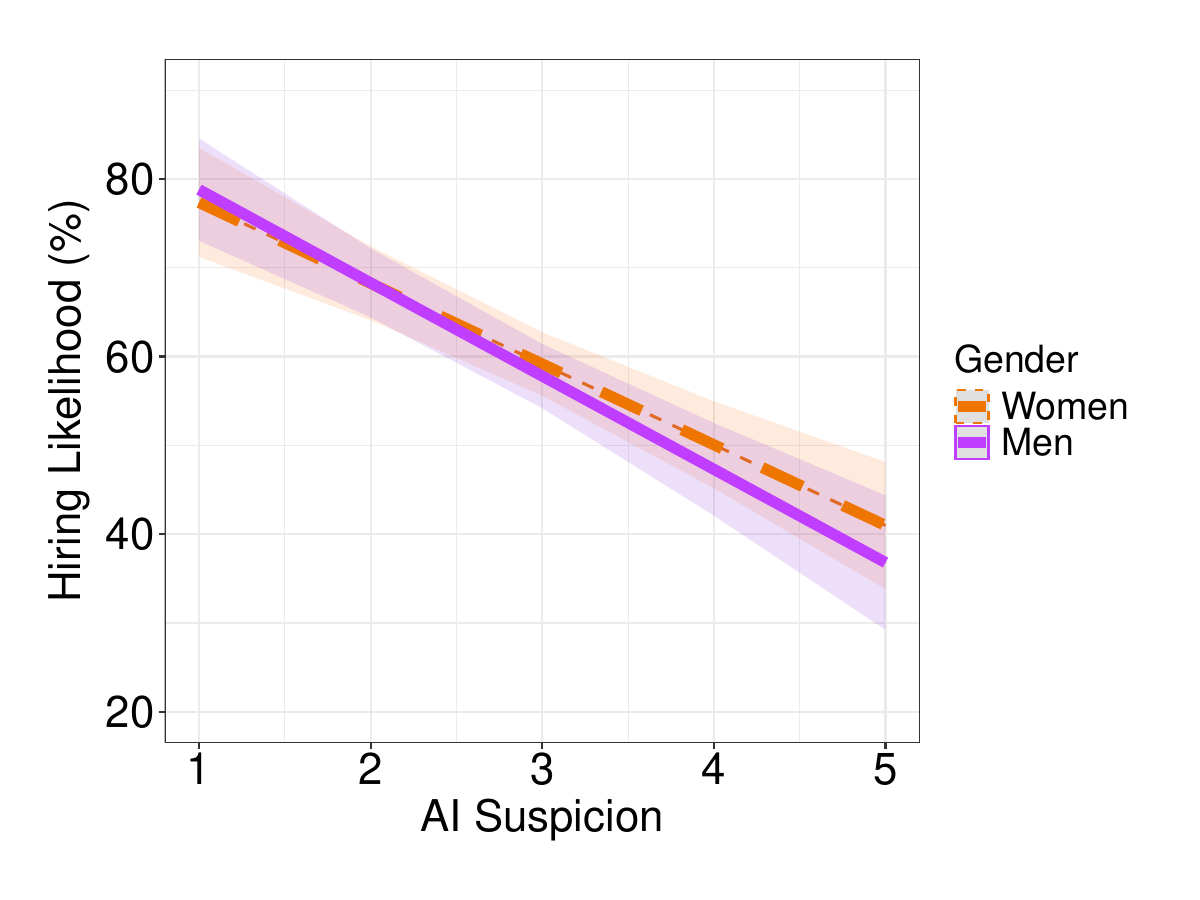}
\caption{The negative effect of AI Suspicion on Hiring Likelihood, by gender.}
\label{fig:gender_hiring_ai}
\end{figure}

\minititle{Hiring Likelihood.} Finally, we show a strong inverse relationship between AI suspicion and participants' hiring recommendations in Figure~\ref{fig:gender_hiring_ai}. 
The figure shows that when AI use was not suspected, participants rated their likelihood to hire around 70-80 on a 100-point scale. However, when AI use was heavily suspected, hiring likelihood dropped by half, to around 40/100, with men and women similarly impacted.

While the hiring likelihood trend looks somewhat different between the genders, a linear mixed model analysis did not show significant differences.  
We used a linear mixed model with AI suspicion, gender, writing style, all two-way interactions, and the three-way interaction as fixed effects to predict the hiring likelihood score\footnote{Quality was highly correlated with the hiring measure and was not included in the model; we were interested to see if the variables that predict these two measures are different.}.
The model confirmed the relationship, with AI suspicion having a significant effect on hiring ($p<0.001$).
The writing style effect again was trending towards significance, suggesting that there may be differences between the AI-inducing writing style and the control ($p = 0.071$, \textit{n.s.}). 
Gender ($p = 0.411$, \textit{n.s.}) did not have a significant main effect, and there were no significant interaction effects.

\subsection{Experiment~2: Race}
Experiment~2 examines the impact of perceptual harms on freelancers from two different racial groups, Black and White. 
We did not find racial differences in participants' suspicions of AI use, quality evaluation, or their likelihood to hire people from these different groups.

\begin{figure}[t]
    \centering
    \begin{subfigure}[h]{0.45\textwidth}
        \centering
        \includegraphics[width=\textwidth,height=4.8cm,keepaspectratio]
        {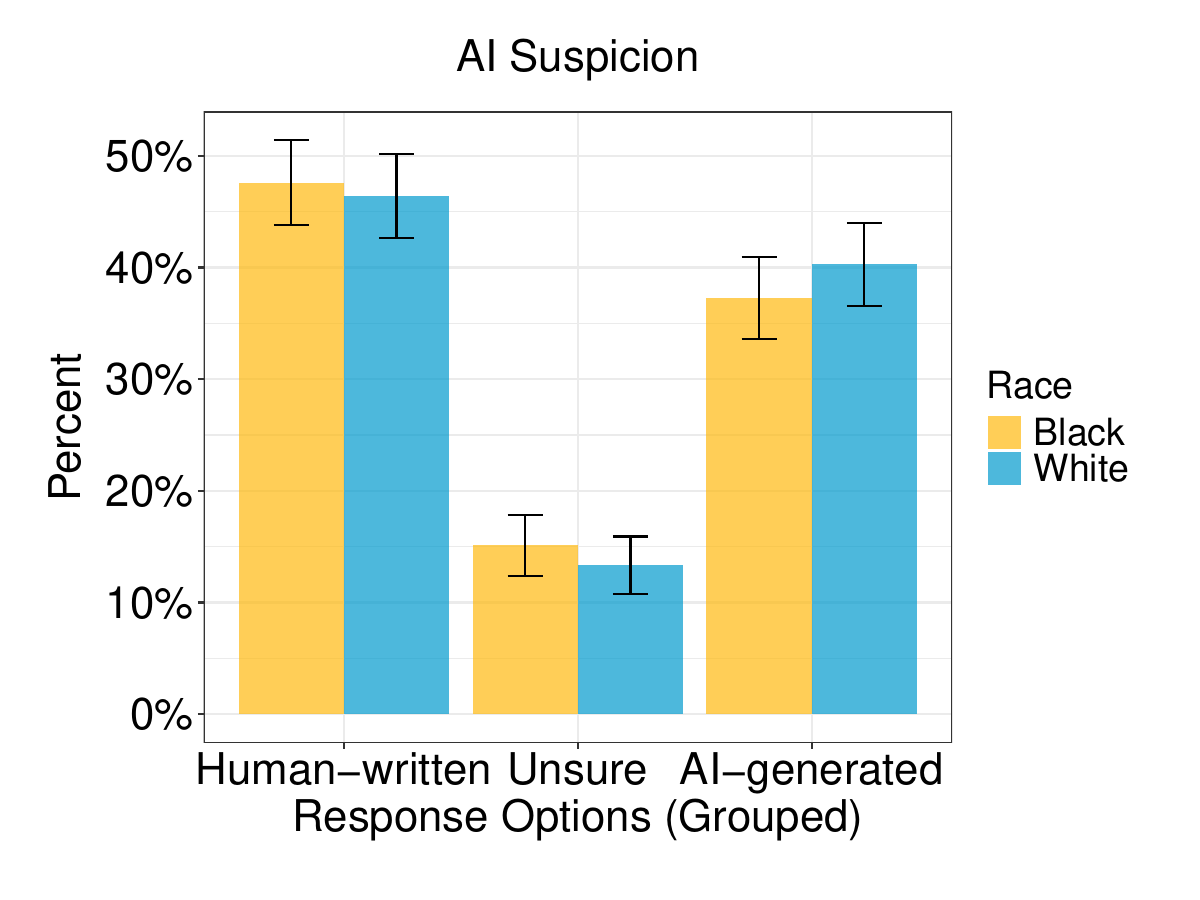}
        \caption{AI Suspicion: overall frequency of responses by race}
        \label{fig:race_ai}
    \end{subfigure} 
    \begin{subfigure}[h]{0.45\textwidth}
        \centering
        \includegraphics[width=\textwidth,height=4.8cm,keepaspectratio]
        {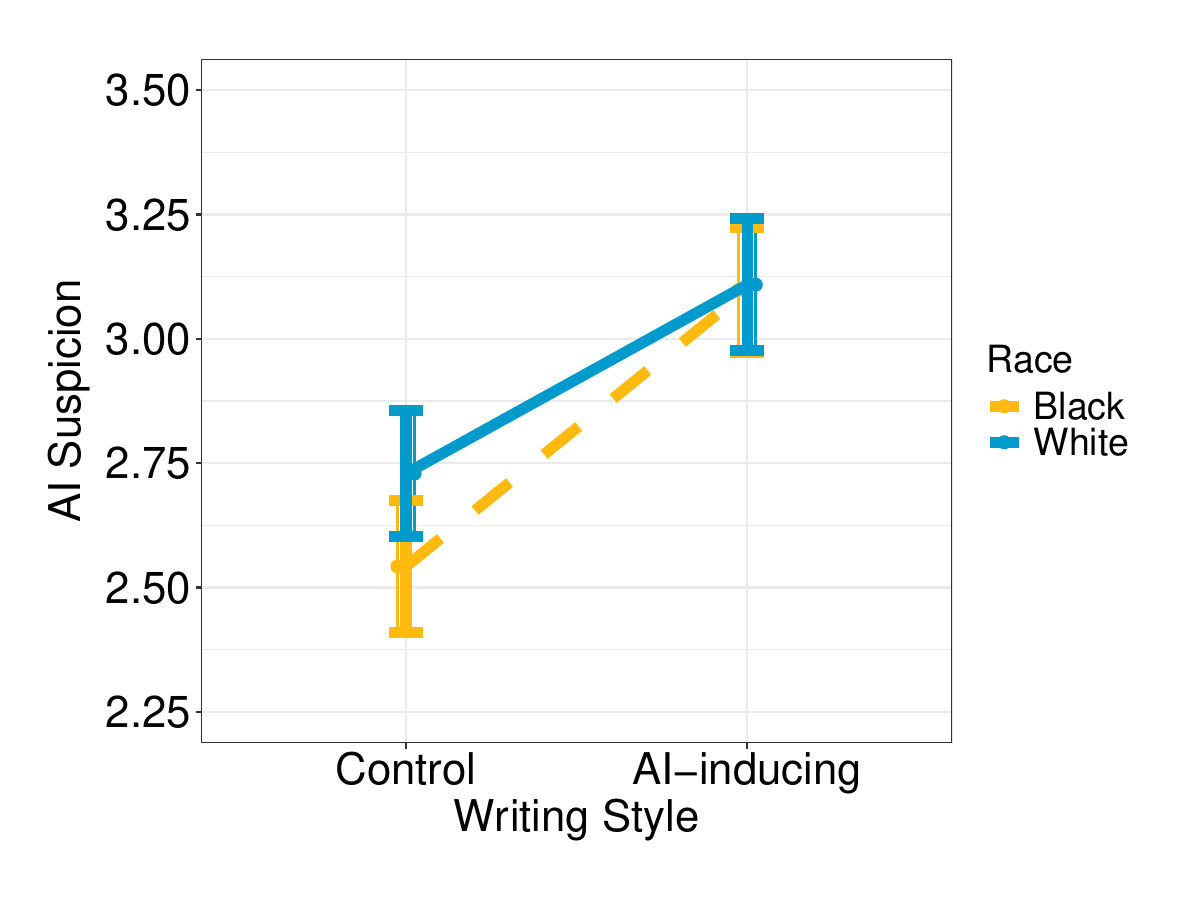}
        \caption{AI Suspicion: interaction between writing style and race}
        \label{fig:race_type_ai}
    \end{subfigure}
    \caption{\textbf{Participant's AI Suspicion}. Participants’ AI Suspicion evaluation by the presented race of the evaluated freelancer.}
    \label{fig:race_ai_overall}
\end{figure}

\minititle{AI Suspicion.}
We did not find racial differences in AI suspicion.
Figure~\ref{fig:race_ai} shows that freelancers presented as White were only suspected of using AI 40.2\% of the time (rightmost, blue column), and when the freelancers were presented as Black, they were suspected 37.2\% of the time (rightmost, gold).
We analyzed the data in a similar manner to Experiment~1, with race as a fixed effect and participant as a random effect, which confirmed race did not have a significant effect on AI suspicion ($p=0.405$, \textit{n.s.}). 
We also examined the relationship between writing style and AI suspicion between the racial groups, as seen in Figure~\ref{fig:race_type_ai}. The figure reflects that there may be slightly less suspicion towards Black freelancers (dashed, gold) in the control condition. Regardless, AI suspicion increases substantially for both groups in the AI-inducing writing style condition. 
A linear mixed model predicting AI suspicion from race, writing style, and their interaction (fixed effects) as well as participants (random effect) confirmed that only writing style had a significant effect on AI suspicion ($p<0.001$).

\begin{figure*}[t]
    \centering
    \begin{subfigure}[h]{0.28\textwidth}
        \centering
        \includegraphics[width=\textwidth,height=5cm,keepaspectratio]{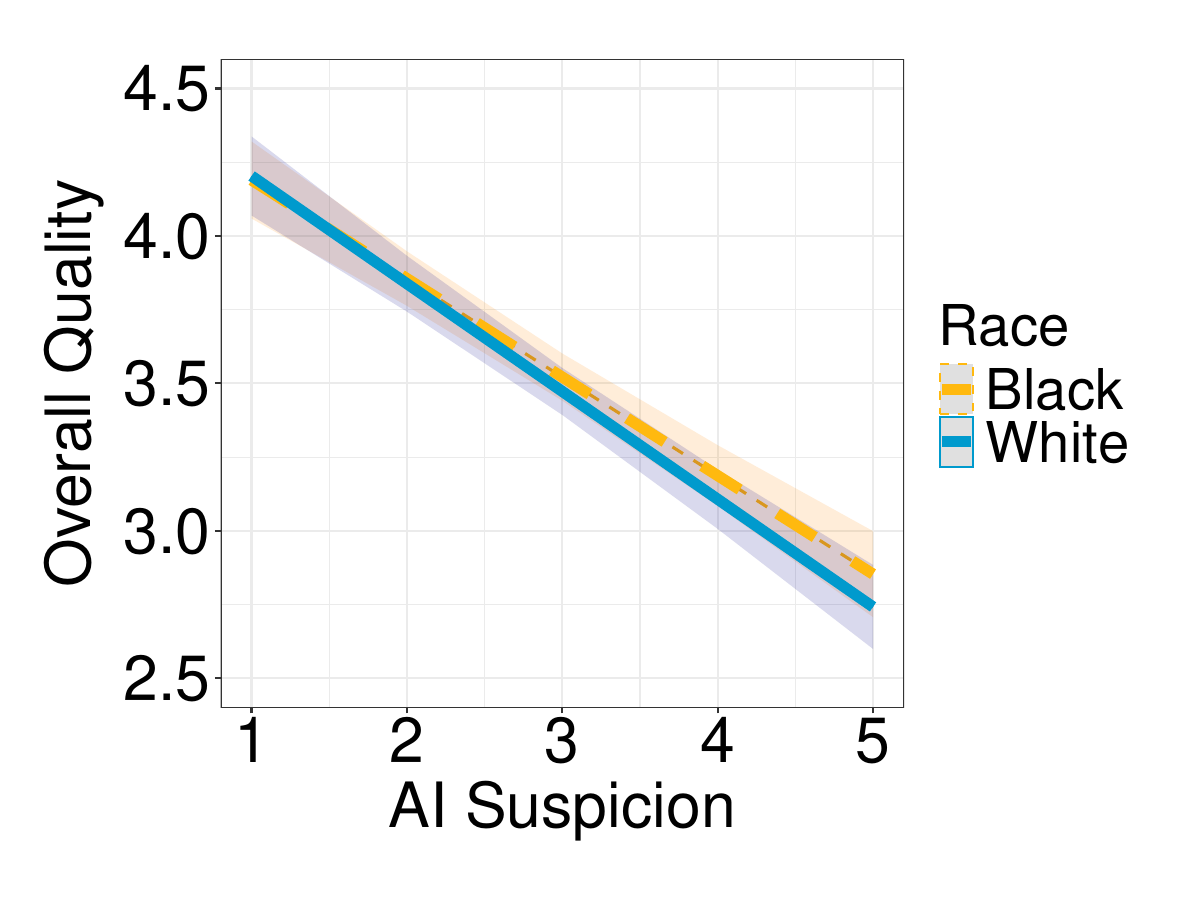}
        \caption{All Writing Styles} 
        \label{fig:race_all_quality}
    \end{subfigure} 
    \begin{subfigure}[h]{0.28\textwidth}
        \centering
        \includegraphics[width=\textwidth,height=5cm,keepaspectratio]{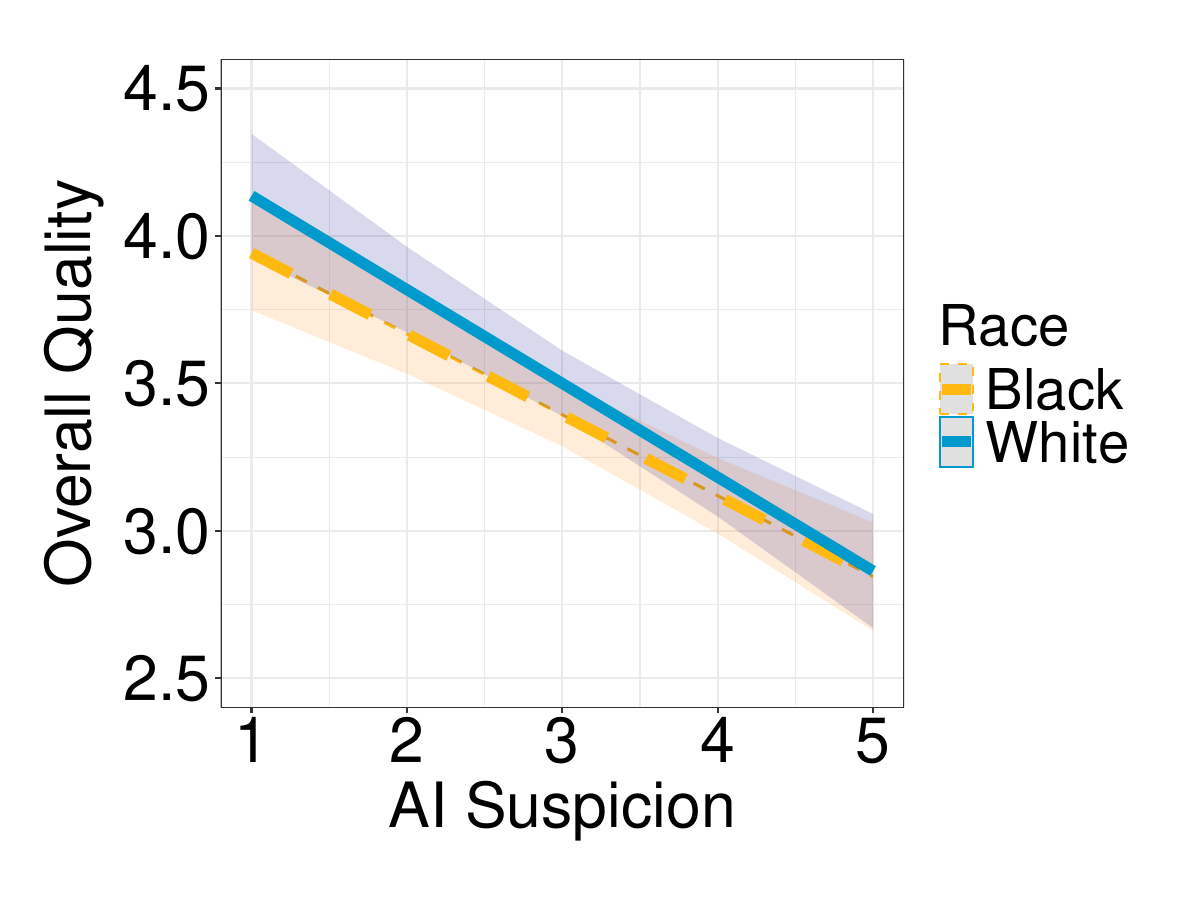}
        \caption{AI-inducing}
        \label{fig:race_ai_quality}
    \end{subfigure}
    \begin{subfigure}[h]{0.27\textwidth}
        \centering
    \includegraphics[width=\textwidth,height=5cm,keepaspectratio]{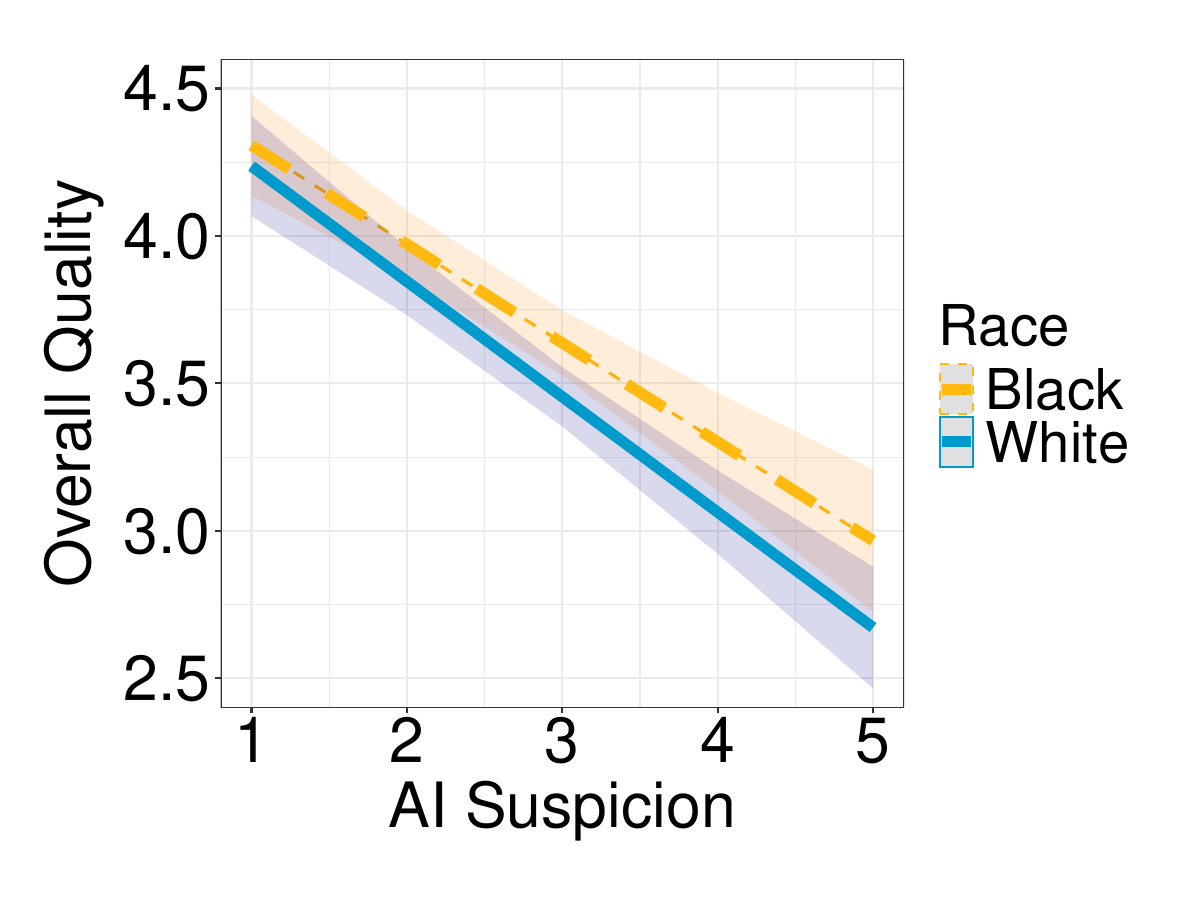}
        \caption{Control}
        \label{fig:race_control_quality}
    \end{subfigure}
    \caption{The negative effect of AI Suspicion on Quality Evaluations, by race.}
    \label{fig:race_interaction_ai_quality}
\end{figure*}

\minititle{Quality Evaluations.}
Next, we move on to the quality evaluations and hiring judgments, as we did in Experiment~1.
Figure~\ref{fig:race_interaction_ai_quality} illustrates a strong inverse relationship between AI suspicion and overall quality (quality evaluations drop when AI suspicion grows) without notable differences by race. 
We confirm this result with a linear mixed model that predicts overall quality for all writing styles (Fig~\ref{fig:race_interaction_ai_quality}a) from AI suspicion, race, writing style, and all interaction effects with participants as a random effect. 
We observe that AI suspicion had a significant main effect ($p<0.001$), and there were no significant racial differences ($p=0.215$, \textit{n.s.}).
Writing style had a significant effect ($p=0.018$), with the control receiving higher overall quality evaluations compared to the AI-inducing style when controlling for AI suspicion.

We did not see racial differences in overall quality evaluations, nor did we see evidence of racial differences in the other quality measures, namely content and structure. 
Similar to Experiment~1, we analyze our measures with a linear mixed model that includes a three-way interaction between race, writing style, and AI suspicion. As expected, in both content and structure evaluations, AI suspicion had a significant main effect ($p<0.001$).  
Writing style did not have a significant main effect on the content index measure ($p=0.535$, \textit{n.s.}).
However, writing style had a significant main effect on the structure index ($p=0.015$), with the control receiving higher structure scores than the AI-inducing style (controlling for suspicion).

\begin{figure}[h]
\includegraphics[height=4.8cm]{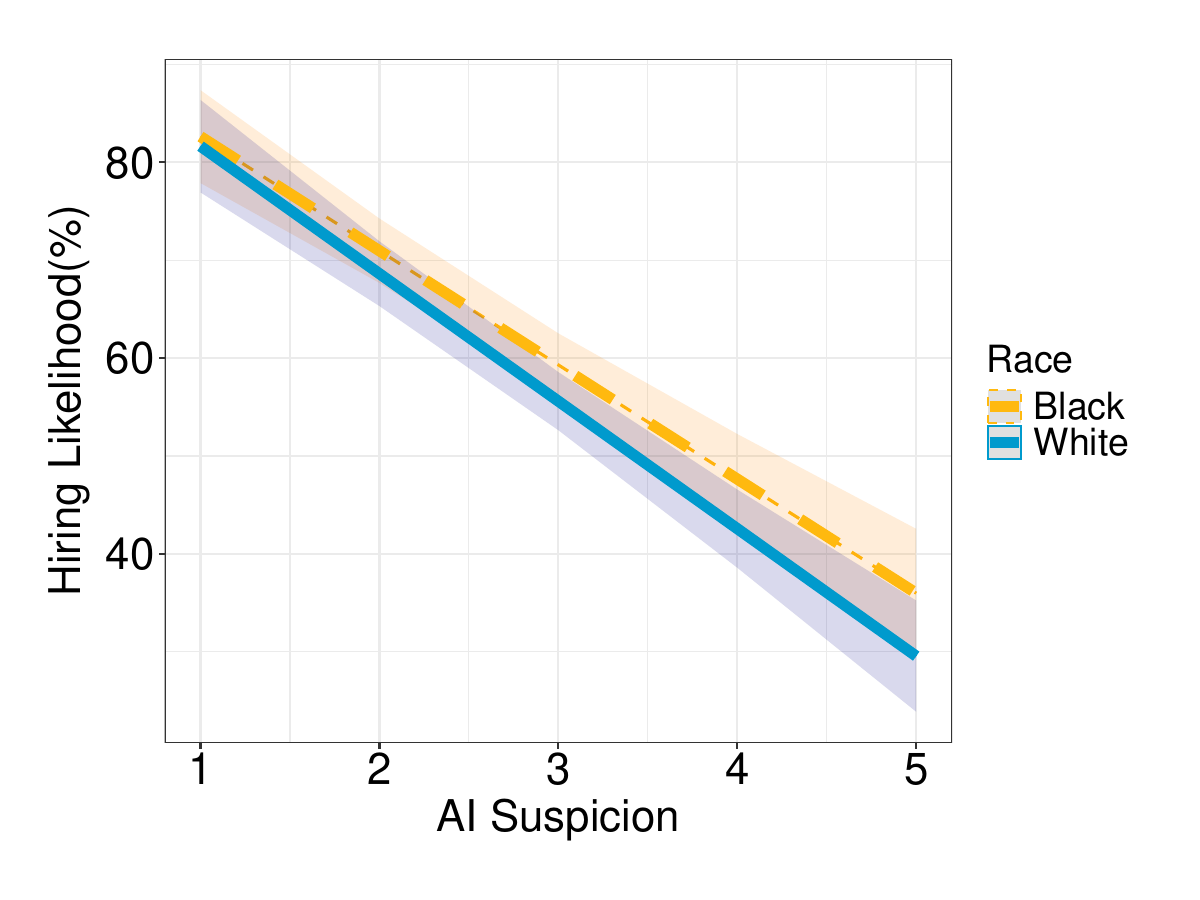}
\caption{The negative effect of AI Suspicion on Hiring Likelihood, by race.}
\label{fig:race_hiring}
\end{figure}

\minititle{Hiring Likelihood.} 
We look at the potential impact on participants' willingness to hire these freelancers. Figure~\ref{fig:race_hiring} shows that participants are less likely to want to hire a freelancer they suspect of using AI. This trend is consistent across both racial categories.
We confirmed the trend with a linear mixed model like we did in Experiment~1. We found that AI suspicion ($p<0.001$) and writing style ($p=0.024$) had a significant main effect on hiring. There were no significant racial differences ($p=0.330$, \textit{n.s.}), nor were there significant interaction effects.

\subsection{Experiment~3: Nationality}
Our third and final experiment examines the potential for perceptual harms against individuals of foreign nationalities in the U.S. 
In particular, we compare freelance profiles suggestive of East Asian identity with freelancers presented as White Americans (or of similar origin). 
We find that freelancers presented as East Asian are suspected of using AI more than those presented as White Americans. 
However, we did not observe differences in quality evaluations or job outcomes.

\begin{figure}[h]
    \centering
    \begin{subfigure}[h]{0.5\textwidth}
        \centering
        \includegraphics[width=\textwidth,height=4.8cm,keepaspectratio]
        {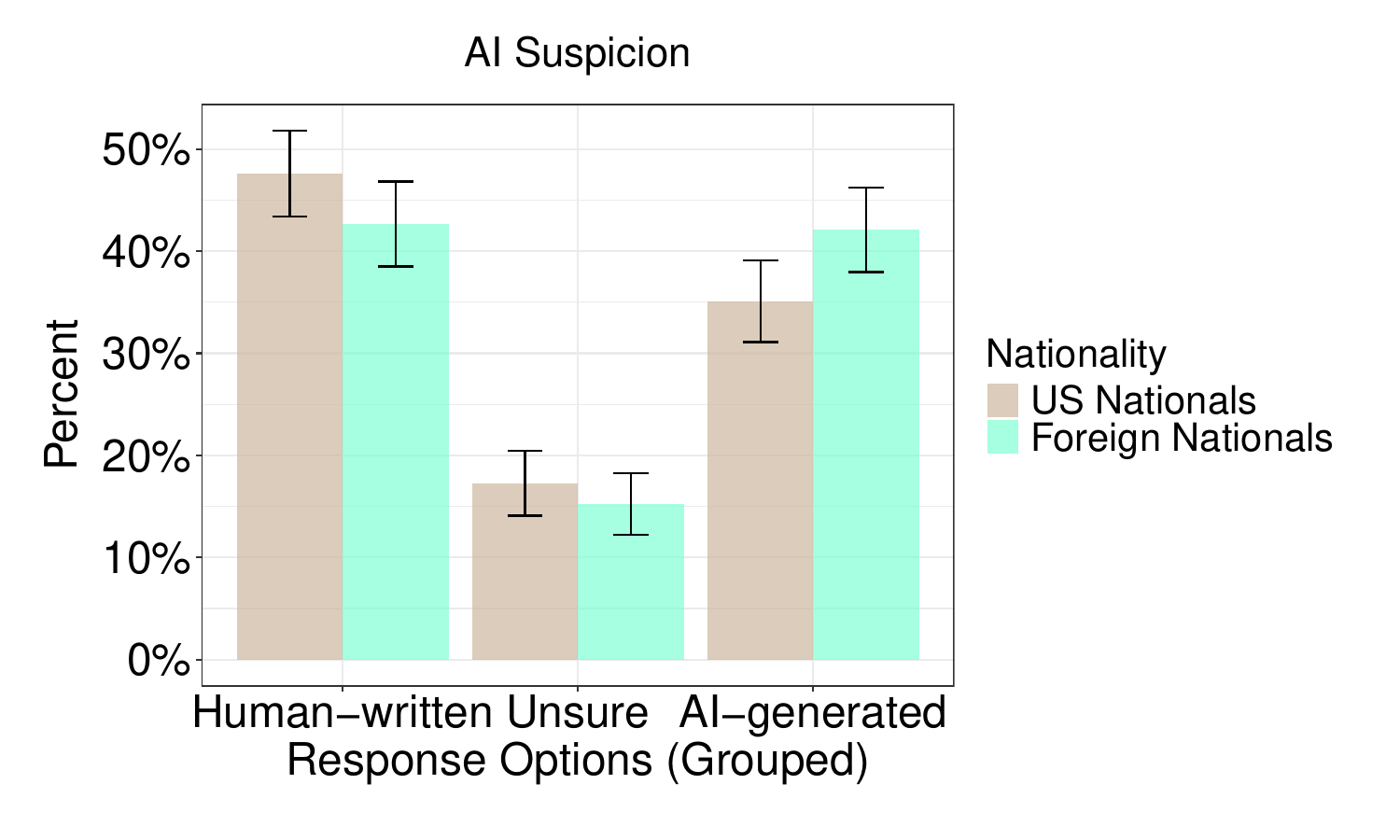}
        \caption{AI Suspicion: overall frequency of responses by nationality}
        \label{fig:esl_ai}
    \end{subfigure} 
    \begin{subfigure}[h]{0.49\textwidth}
        \centering
        \includegraphics[width=\textwidth,height=4.8cm,keepaspectratio]
        {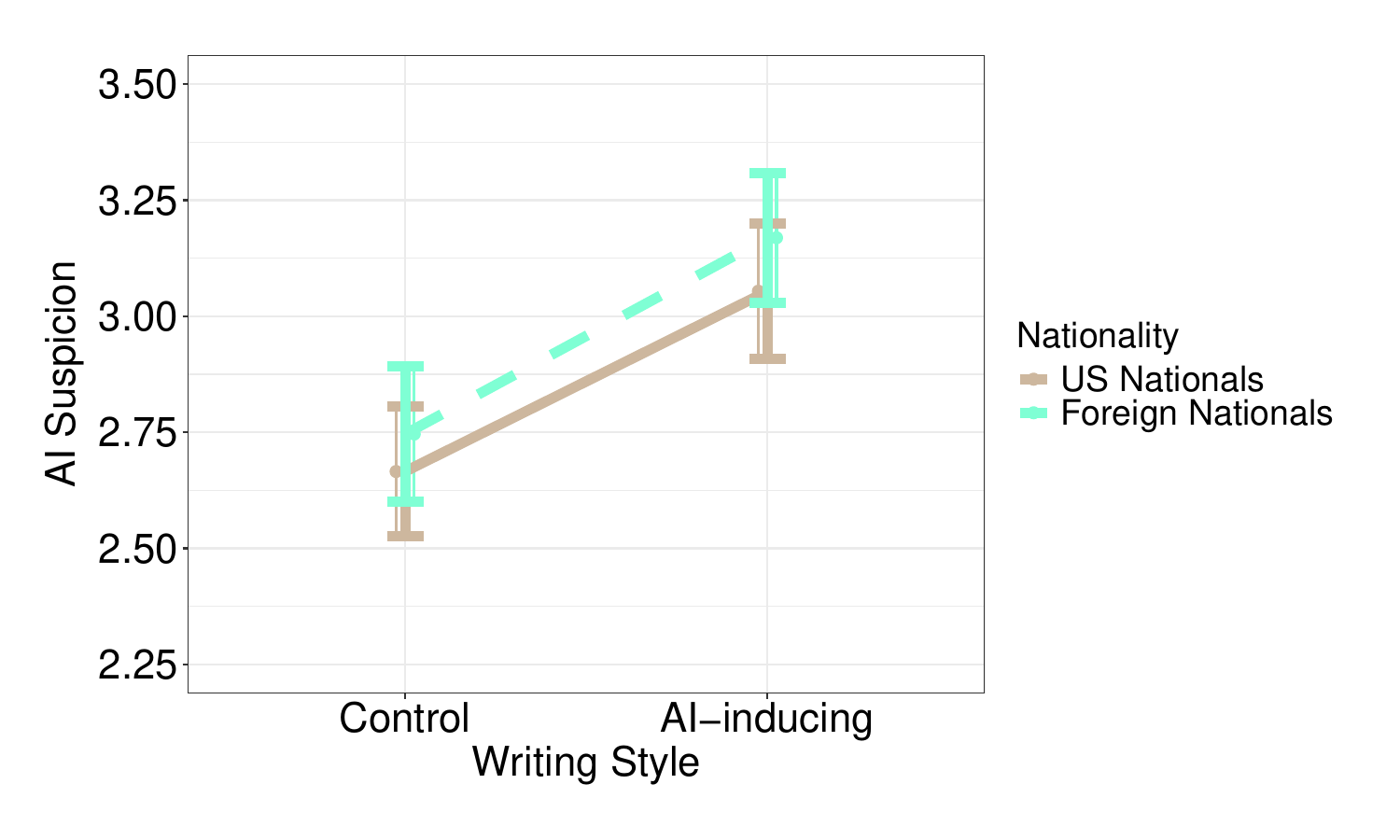}
        \caption{AI Suspicion: interaction between writing style and nationality}
        \label{fig:esl_type_ai}
    \end{subfigure}
    \caption{Participants’ AI Suspicion evaluation by the presented nationality of the evaluated freelancer.}
    \label{fig:esl_ai_overall}
\end{figure}

\minititle{AI Suspicion.}
East Asian freelancer profiles were somewhat more likely to be suspected of using AI, as shown in Figure~\ref{fig:esl_ai}. The figure shows that 42.0\% of foreign nationals (rightmost, turquoise column})  were suspected of using AI compared to 35.1\% for U.S. nationals (rightmost, tan column).
We confirmed this difference as significant ($p<0.029$) with a linear mixed model (with nationality as a fixed effect and participants as a random effect).
However, we note that the differences in AI suspicion due to nationality are no longer significant when accounting for the writing style treatment.
Figure~\ref{fig:esl_type_ai} shows that writing in an AI-inducing writing style increases AI suspicion, with both U.S. and foreign nationals' profiles being affected similarly. 
While the figure still shows that foreign nationals (dashed, turquoise line) tend to be more readily suspected, a fuller linear mixed model did not confirm this effect as significant.
Specifically, the linear mixed model (fixed effects: nationality, writing style, and their interaction; random effect: participants) confirmed that writing style has a significant main effect ($p<0.001$) on AI suspicion. Neither nationality ($p=0.261$, \textit{n.s.}) nor the interaction between nationality and writing style ($p=0.807$, \textit{n.s.}) had an effect.

\begin{figure*}[t]
    \centering
    \begin{subfigure}[h]{0.28\textwidth}
        \centering
        \includegraphics[width=\textwidth,height=5cm,keepaspectratio]{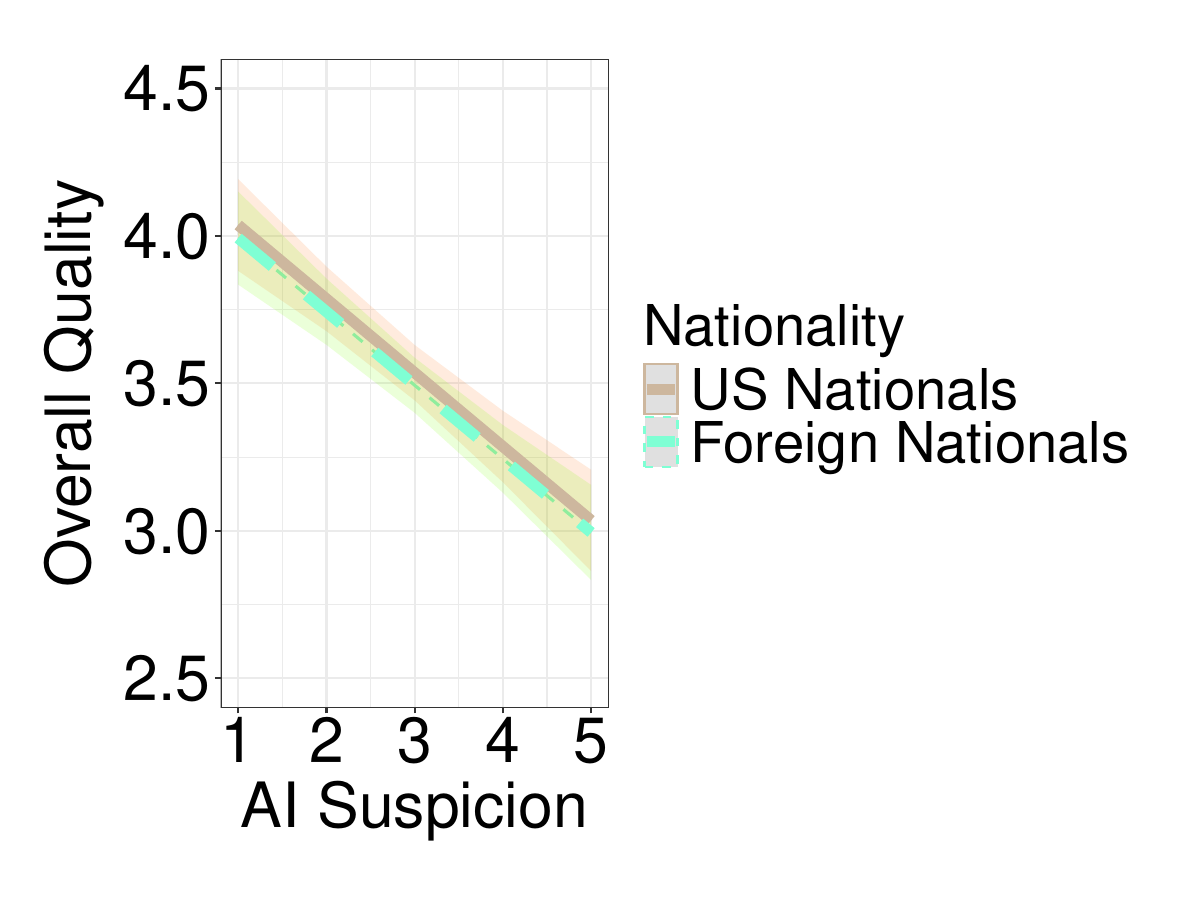}
        \caption{All Writing Styles} 
        \label{fig:esl_all_quality}
    \end{subfigure} 
    \begin{subfigure}[h]{0.28\textwidth}
        \centering
        \includegraphics[width=\textwidth,height=5cm,keepaspectratio]{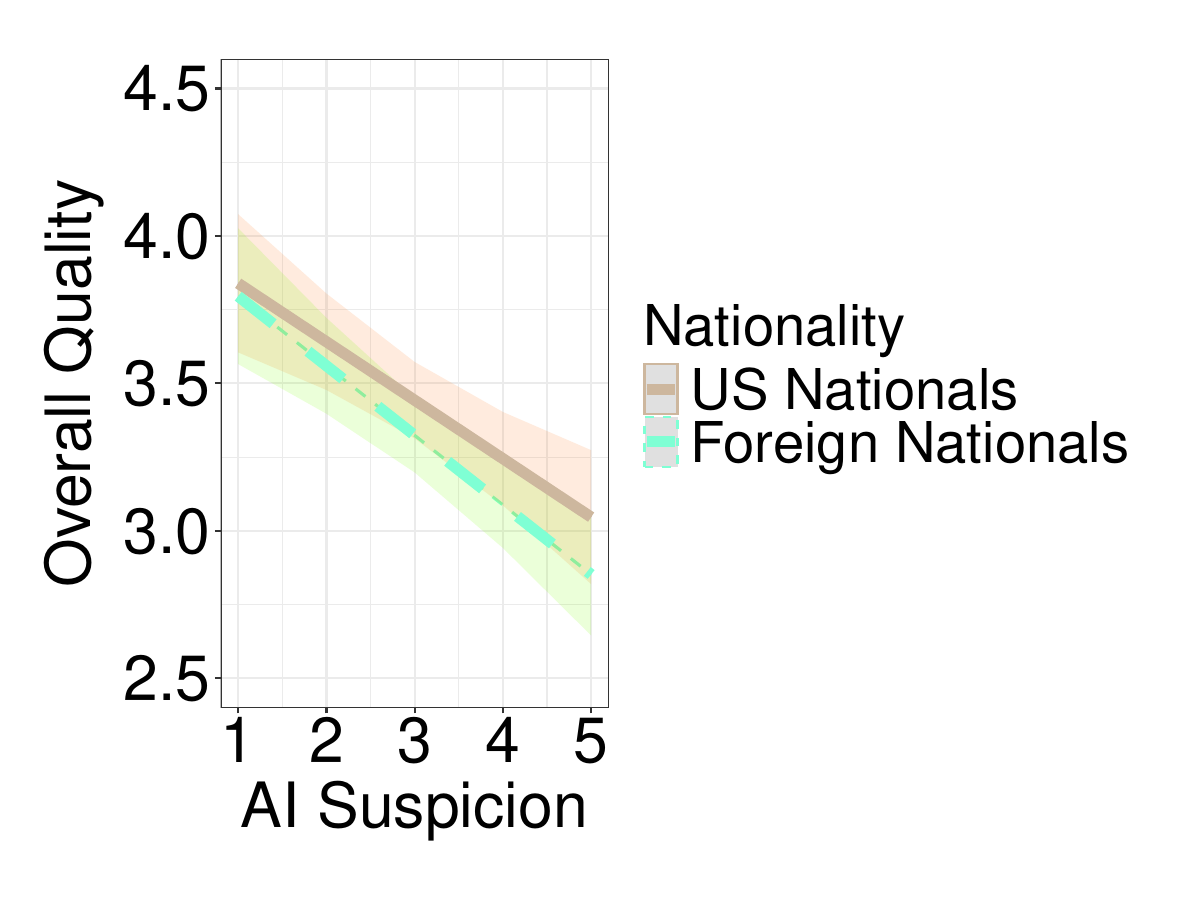}
        \caption{AI-inducing}
        \label{fig:esl_ai_quality}
    \end{subfigure}
    \begin{subfigure}[h]{0.27\textwidth}
        \centering
    \includegraphics[width=\textwidth,height=5cm,keepaspectratio]{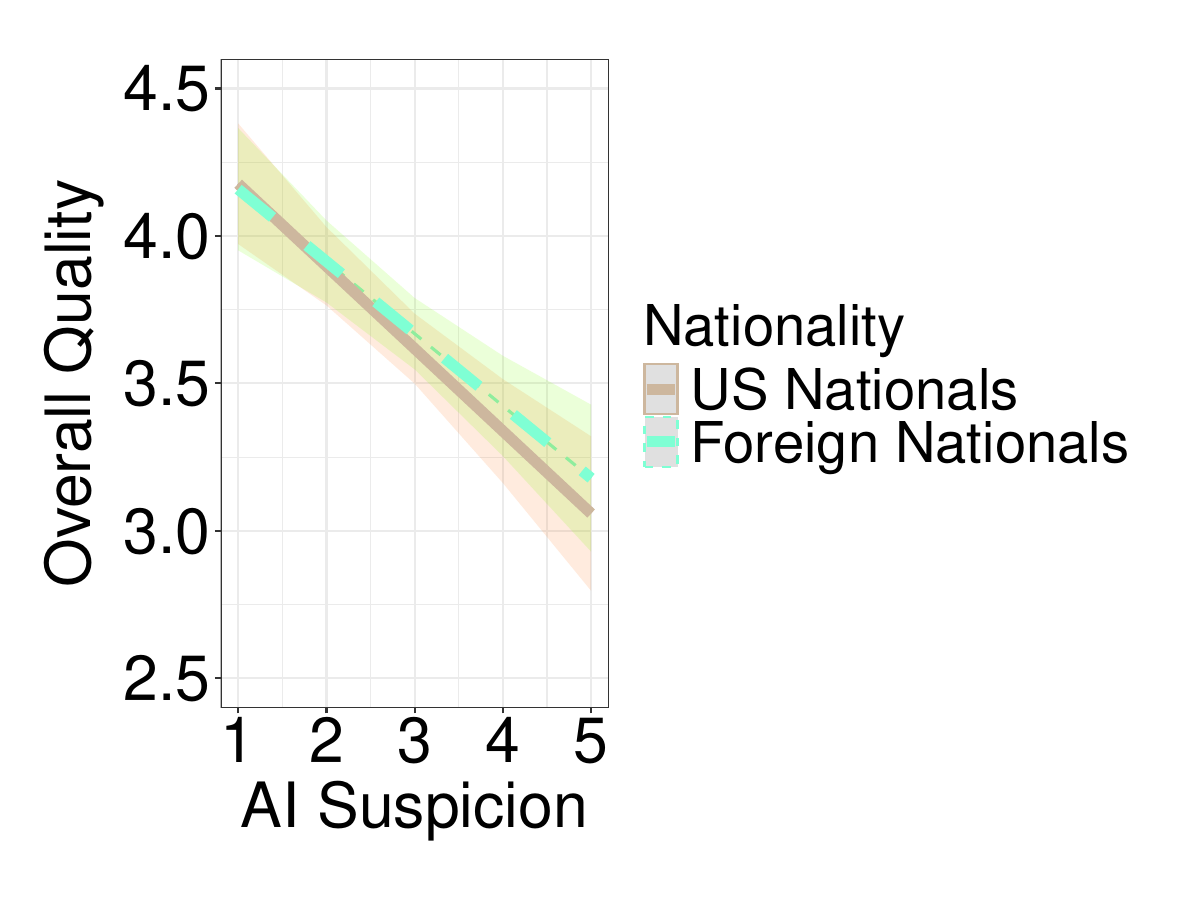}
        \caption{Control}
        \label{fig:esl_control_quality}
    \end{subfigure}
    \caption{The negative effect of AI Suspicion on Quality Evaluations, by nationality.}
    \label{fig:esl_interaction_ai_quality}
\end{figure*}

\minititle{Quality Evaluations.}  
Like in Experiments~1 and~2, we see an inverse relationship between overall quality and AI suspicion.
Figure~\ref{fig:esl_interaction_ai_quality} shows quality evaluations as a function of AI suspicion. Both sets of freelancer profiles perform similarly and show the same steep drop in quality as AI suspicion increases.
A linear mixed model predicting overall quality for all writing styles (Figure~\ref{fig:esl_interaction_ai_quality}a) that included all main effects, two-way interactions, and a three-way interaction between nationality, AI suspicion, and writing style with participants as a random effect, confirmed the statistically significant impact of AI suspicion ($p<0.001$) and writing style ($p=0.031$) on quality. There were no significant differences between nationalities ($p=0.943$, \textit{n.s.}), and there were no significant interaction effects. 

Considering the effect of AI suspicion on our other quality evaluation measures, we find similar results to our previous studies: AI suspicion negatively impacts the content and structure evaluations. This difference is statistically significant for content ($p<0.001$) and structure ($p=0.004$) according to linear mixed models that take into account the three-way interaction between nationality, AI suspicion, and writing style. For both content and structure evaluations, nationality and writing style do not have a main effect, and there were no significant interaction effects.

\begin{figure}[h]
\includegraphics[height=4.8cm]{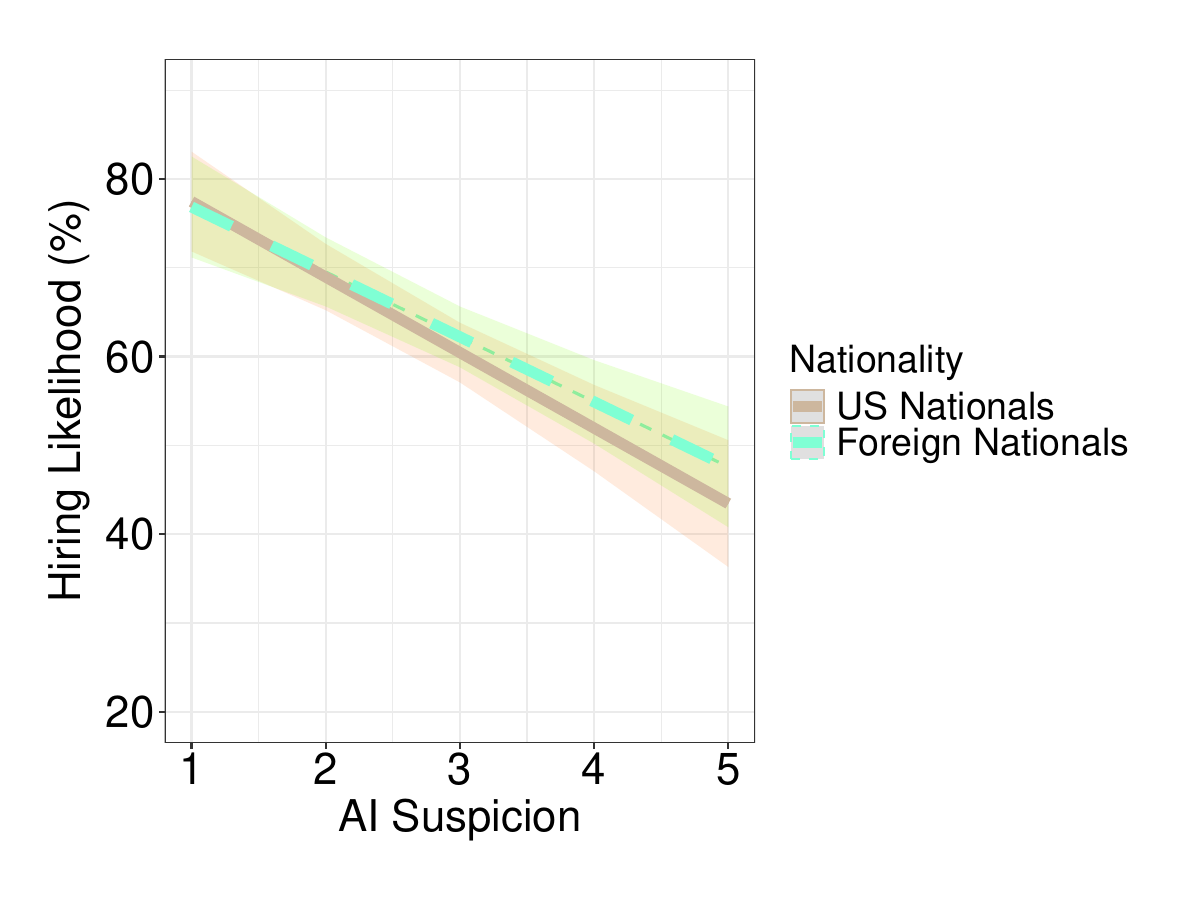}
\caption{The negative effect of AI Suspicion on Hiring Likelihood, by nationality.}
\label{fig:esl_hiring}
\end{figure}

\minititle{Hiring Likelihood.} 
Lastly, we look at participants' reported likelihood to hire.
The results are again similar to the two previous studies.
As AI suspicion increases, hiring likelihood decreases, as seen in Figure~\ref{fig:esl_hiring}. 
The figure shows that between the lowest and highest levels of AI suspicion, hiring ratings drop from around 75 out of 100 to about 40 for both East Asian and White American freelance profiles.
We confirm these results with a final linear mixed model (AI suspicion, nationality, writing style, all two-way interactions, and the three-way interaction as fixed effects; participants as a random effect). The model shows that AI suspicion had a statistically significant effect on hiring ratings ($p<0.001$).
The effect of writing style was close to significant ($p=0.054$, \textit{n.s.}).
There were no effects of nationality ($p=0.982$, \textit{n.s.}), and there were no significant interaction effects.

\section{Discussion}
Our study is the first to characterize an emergent potential harm in generative AI systems---perceptual harms.
By introducing perceptual harms, our work extends the previous research in computing and AI harms, which has focused on harms as a result of AI's development and deployment~\cite{shelby2023,weidinger-etal-2022-risks}. 
Perceptual harms, by our definition, can negatively impact individuals and groups independently of whether they are actually using AI. 
Rather, it is the \textit{appearance or perception} of AI use that results in a negative judgment against the perceived AI user.
In this study, we examine perceptual harms with textual content; however, perceptual harms could also occur in other mediums such as images, videos, and audio. Similar to writing tasks, creators may use text-to-image or text-to-audio~\cite{gozalo2023chatgpt} models to enhance their creative productivity or produce novel content~\cite{zhou2024artcreativity}. For example, although research has demonstrated that knowing whether a piece of art was created by a human or AI-generated influences how people perceive the same piece of art~\cite{hong2018biasart}, it remains unclear how merely suspecting that AI was involved impacts people's judgments about both the artwork and the creator, and how this evaluation may be different for people of different social groups.

Although most work on harms indicates burdens against historically marginalized groups, our work demonstrates that, in this case, perceptual harms can also affect socially dominant groups. 
The findings across our three experiments are somewhat mixed, but taken together, they provide some context about the conditions under which perceptual harms may occur. 
In this discussion, we try to address the question of why there were differential impacts in AI suspicion in our experiment and propose potential solutions.

One possible explanation for the disparate impact of perceptual harms we observed may stem from real or stereotypical expected differences in terms of technology use. 
In Experiment~1, where participants evaluated profiles of different genders, men were suspected of using AI more than women. This result is consistent with gender attitudes toward technology. Previous studies have shown that men have a positive view toward computer-related technologies whereas women generally have a more critical view~\cite{kim2023one,jakesch2022different}; it is possible (and should be confirmed in future work) that men were more suspected of using AI due to real gendered differences in technology uptake.
The effect may also be driven by stereotypes, with participants viewing male freelancers as more technologically savvy and, therefore, more likely to be using this relatively new technology.
The possible impact of technology-related stereotypes may also be consistent with the results of Experiment~3.
Experiment~3 showed that freelancers with East Asian names were suspected of using AI more than White Americans. Asians and Asian Americans are often seen as so-called ``model minorities'' and are overrepresented in  STEM~\cite{mcgee2017burden,chen2018models,VUE20233138}. 
In fact, early work has shown that Asians adopted generative AI more than White folks~\cite{microsoft2023work}.
However, in addition to technology-related stereotypes, it is possible that the East Asian names were suspected of using AI due to language assumptions.
Participants may have assumed the East Asian profiles were non-native English speakers who had AI assistance due to limited language proficiency. 
Again, this is a question that could be addressed in future follow-up work.

While AI suspicion was different between most of the groups in our experiments (supporting Hypothesis~1), the hypotheses about the impact of the group on quality evaluations (Hypothesis~2) and outcomes (Hypothesis~3) were not confirmed.
Taken together, the results of the three experiments show that the perceived use of AI is the main factor impacting the writing evaluation and hiring outcomes. 
Regardless of the group, when people thought the freelancer had AI assistance, the quality, content, and structure evaluations were negatively impacted. 
This effect could be due to the negative associations of AI-writing~\cite{jakesch2023human,kobis2021artificial}. For example, it is possible that people who use AI may be seen as having undesirable traits in an employee (e.g., ``lazy'' or ``incapable'').

While the current results suggest that stereotypes may cause perceptual harms, the evolving societal norms surrounding AI and the context of its use will likely influence perceptual harms in the future.
Since the inception of artificial intelligence, there have been many ideas of what AI is and what it can or cannot do~\cite{fast2017long}. In the 1990s, many people associated AI with chess supercomputers and IBM’s Deep Blue, and by the 2010s, people thought of AI as autonomous vehicles~\cite{fast2017long}. 
While public sentiment toward AI was and still is largely positive~\cite{fast2017long},
there are growing concerns about issues such as loss of control, ethical challenges, and the impact of AI on employment~\cite{fast2017long,kelley2021exciting,neri2020role}.
The release of ChatGPT in 2022 marked a technological turning point, making generative AI mainstream as many other companies developed similar models.
Shortly after ChatGPT’s release, many Americans believed that AI would reduce job opportunities, with just 10\% of the US adults believing AI does more good than harm~\cite{Ray2024AIviews}. 
While fewer Americans now perceive AI as harmful compared to a year ago, 
many remain cautious about its applications, especially in workplace settings~\cite{Ray2024AIviews}.
Notably, people who are more knowledgeable about AI are less likely to express concerns about its effects~\cite{Ray2024AIviews}. As AI becomes increasingly normalized and people are more informed of its capabilities, the negative consequences of perceptual harms, in terms of content evaluation and loss of opportunities, may diminish or shift.

The impact of perceptual harms will depend not only on the normalization of AI but also on the specific contexts in which generative AI is suspected. 
For example, in artist communities, AI suspicion could lead to worse content evaluation and loss of income and opportunities~\cite{jiang2023aiart}. 
Whereas, in interpersonal work communication tasks---like confirming times with high-profile professionals or expressing sympathy in a distressing scenario---AI suspicion could lead to the loss of trust between the sender and the recipient~\cite{jakesch2019ai,liu2022console,hohenstein2023artificial}.
The context in which generative AI is suspected can also affect the magnitude of potential outcomes.
For example, in mass communication like journalism, AI suspicion could further erode the public's trust of not only individual journalists but journalistic institutions as a whole~\cite{longoni2022newsai,fink2019biggest}.
In contrast, in interpersonal work communication, AI suspicion might result in a more localized loss of trust. 
On the other hand, if the use of AI becomes more normalized in the future, the outcomes in these three scenarios may shift. In art and journalism, for example, generative AI tools \textit{could} come to be seen as helpful aids that enhance artists and journalists rather than undermine their credibility. 
In workplace settings, using AI assistance may come to be expected---like the use of spellchecking today---and avoiding its use may reflect badly on the contributor. 

Beyond suspicion, we may see shifts in other social dynamics between individuals as people start being aware of the potential suspicion.
Increasingly, people creating content may react in various ways, preemptively trying to avoid being perceived as using AI. 
Similarly, content creators may seek new ways to signal ownership and authenticity. 
These kinds of responses will likely be also subject to direct or indirect biases. 
On the other side, content evaluators may turn to a new set of tools to validate the legitimacy of work.
For instance, in education, where concerns about generative AI and academic integrity are prevalent~\cite{sullivan2023chatgpt,perkins2023academic,xie2023aicheating}, teachers may adopt varying approaches to addressing AI-related suspicion~\cite{liang2023gptdetectorsbiasednonnative}, or ask students to submit a history tracking all work on their assignments. 
In turn, to signal effort and authenticity, students may emphasize the time they spent completing an assignment or deliberately alter their writing style to differ from AI-generated text.

Based on our findings, we encourage HCI researchers and designers to explore potential strategies to help mitigate perceptual harms. 
We note that perceptual harms are fundamentally a social problem caused by stereotypes and societal views of AI; therefore, previous approaches (e.g., collecting more training data from various dialects or diverse speakers to mitigate representational harms) to addressing AI's
sociotechnical harms~\cite{weidinger-etal-2022-risks,weidinger2023sociotechnical,shen2024bidirectionalhumanaialignmentsystematic} may not fully address the problem. 
Perceptual harms are deeply rooted in social, cultural, and systemic dynamics that often transcend the scope of technological solutions.
We recognize that while design recommendations can often be helpful, they risk oversimplifying the nuanced and multifaceted nature of social harms. 
To address this challenge, we need to critically examine the trade-offs involved in potential mitigation strategies, aiming to balance practicality with a deeper understanding of these complexities.
One potential solution to mitigate AI suspicion could be to introduce AI disclosure labels to AI-generated content.
A disclosure statement could eliminate differences in AI suspicion; however, there is evidence to suggest disclosure would not reduce the negative perceptions around AI use~\cite{bellaiche2023humans,lim2024disclosure,hong2018biasart, jakesch2019ai} and
may not reduce the negative effects people could receive for using AI. 
A potential design intervention could go beyond an AI-disclosure label (i.e., stating that AI was involved) to signal the \textit{extent} to which AI was involved in the creative process. For example, there could be a future system similar to InkSync~\cite{laban2024inksync} to demonstrate AI's involvement by providing a log of AI suggestions and tracked AI-generated content in the document.
However, such detailed reports may still result in perceptual harms, and may also be rejected by users whose agency may be challenged and may feel monitored~\cite{levy2022data}. 

Increasing tech awareness and literacy of the capabilities of AI tools could reduce perceptual harms in terms of content evaluation and outcomes as people have a better understanding of what AI is, how it works, and what it can do~\cite{Ray2024AIviews,bhat2021xai,long2020ailiteracry}.
Many people are still unsure of what AI can or cannot do~\cite{kelley2021exciting,fast2017long,long2020ailiteracry}.
Additional data we collected from our experiments was consistent with previous work showing people's inconsistent heuristics for detecting AI text~\cite{jakesch2023human}. 
When reviewing the justification for the AI suspicion responses in our studies, we found that one person's rationale for why the content was AI-generated was often another person's rationale for why the content was human-written. 
For example, like previous work~\cite{jakesch2023human}, some of our participants associated punctuation errors or grammar mistakes with AI, while others associated these with human writing. 
If people had a more accurate understanding of AI's abilities, it might shift how they perceive others' use of AI tools~\cite{Ray2024AIviews}. 
Rather than falling into extremes of either algorithmic aversion---dismissing AI's potential benefits---or algorithmic over-appreciation, a nuanced perspective of AI's abilities would allow people to more accurately assess human capabilities without unfairly attributing success or failure solely to AI~\cite{rae2024content,long2020ailiteracry}.

Finally, we note that this work has some limitations that should be considered. 
First, while we recruited a US-representative sample for our experiment, it still relied on crowd workers who are Western, Industrialized, Educated, Rich, and Democratic (WEIRD) and whose views may differ from the US general public. 
This US-centric study's results may not directly extend to other cultural contexts; however, we have no reason to believe that \textit{some} of our findings would not extend more broadly as research from the Global South demonstrates perceptions of generative AI that influence how it is used and could shape how individuals view others' use of the technology~\cite{agarwal2024aisuggestionshomogenizewriting}.
Second, since our experiment highlighted AI use, it may have primed the participants to think about \textit{and} consider it negatively. 
We attempted to phrase the question as neutrally as possible when describing the task to participants. 
In addition, our work's generalizability is limited by the fact it is based on online experiments performed in one context (freelance marketplaces). 
The results may not generalize to other contexts, mediums (e.g., AI-generated images or audio), or demographic groups. 
Finally, while we attempted to assess potential outcomes (e.g., hiring or not), the hiring measure remained hypothetical and not a behavioral measure. 
Of course, there are several factors not examined in this study that influence hiring decisions. 
A future audit study (e.g.,~\cite{bertrand2004emily}) can help provide more robust data on the impact of perceptual harms of AI. 

\section{Conclusion}
This paper extends the responsible computing literature by proposing the concept of perceptual harms: when the appearance (or perception) of AI use, regardless of whether it was used, results in differential treatment between people of various social groups.
We propose that perceptual harms occur along three axes: suspected AI use, which can lead to differences in quality evaluations and outcomes.
Through a series of online experiments, we show that dominant social groups are sometimes more likely to be suspected of using AI. 
Consistent with prior work, we see that perceptions of AI use negatively impacted quality evaluations and hiring outcomes, but these metrics were not different between groups after controlling for AI suspicion.
We encourage the research community to further explore perceptual harms and how they may change as technology's perception changes.
As AI technologies become mainstream and all types of people are seen as equally likely to use them, who might perceptual harms impact? 
Conversely, if AI technologies are seen as virtuous, will there still be harms, or will the norms of AI use and quality evaluations change? These are important questions for future work.

\begin{acks}
This research was supported by a gift to the LinkedIn-Cornell Bowers CIS Strategic Partnership. The material is also based upon work supported by the National Science Foundation under Grant No. CHS 1901151/1901329.
Any opinions, findings, and conclusions or recommendations expressed in this material are those of the authors and do not necessarily reflect the views of the National Science Foundation or LinkedIn. 
Specifically, this research was performed in 2024, when the NSF was allowed to consider race and gender as legitimate objects of study. 
\end{acks}







\bibliographystyle{ACM-Reference-Format}
\bibliography{cscw}


\begin{thebibliography}{95}


\ifx \showCODEN    \undefined \def \showCODEN     #1{\unskip}     \fi
\ifx \showDOI      \undefined \def \showDOI       #1{#1}\fi
\ifx \showISBNx    \undefined \def \showISBNx     #1{\unskip}     \fi
\ifx \showISBNxiii \undefined \def \showISBNxiii  #1{\unskip}     \fi
\ifx \showISSN     \undefined \def \showISSN      #1{\unskip}     \fi
\ifx \showLCCN     \undefined \def \showLCCN      #1{\unskip}     \fi
\ifx \shownote     \undefined \def \shownote      #1{#1}          \fi
\ifx \showarticletitle \undefined \def \showarticletitle #1{#1}   \fi
\ifx \showURL      \undefined \def \showURL       {\relax}        \fi
\providecommand\bibfield[2]{#2}
\providecommand\bibinfo[2]{#2}
\providecommand\natexlab[1]{#1}
\providecommand\showeprint[2][]{arXiv:#2}

\bibitem[Abid et~al\mbox{.}(2021)]%
        {abid2021muslimbias}
\bibfield{author}{\bibinfo{person}{Abubakar Abid}, \bibinfo{person}{Maheen Farooqi}, {and} \bibinfo{person}{James Zou}.} \bibinfo{year}{2021}\natexlab{}.
\newblock \showarticletitle{Persistent Anti-Muslim Bias in Large Language Models}. In \bibinfo{booktitle}{\emph{Proceedings of the 2021 AAAI/ACM Conference on AI, Ethics, and Society}} (Virtual Event, USA) \emph{(\bibinfo{series}{AIES '21})}. \bibinfo{publisher}{Association for Computing Machinery}, \bibinfo{address}{New York, NY, USA}, \bibinfo{pages}{298–306}.
\newblock
\showISBNx{9781450384735}
\urldef\tempurl%
\url{https://doi.org/10.1145/3461702.3462624}
\showDOI{\tempurl}


\bibitem[Agarwal et~al\mbox{.}(2024)]%
        {agarwal2024aisuggestionshomogenizewriting}
\bibfield{author}{\bibinfo{person}{Dhruv Agarwal}, \bibinfo{person}{Mor Naaman}, {and} \bibinfo{person}{Aditya Vashistha}.} \bibinfo{year}{2024}\natexlab{}.
\newblock \bibinfo{title}{AI Suggestions Homogenize Writing Toward Western Styles and Diminish Cultural Nuances}.
\newblock
\newblock
\showeprint[arxiv]{2409.11360}~[cs.HC]
\urldef\tempurl%
\url{https://arxiv.org/abs/2409.11360}
\showURL{%
\tempurl}


\bibitem[Baidoo-Anu and Ansah(2023)]%
        {baidoo2023education}
\bibfield{author}{\bibinfo{person}{David Baidoo-Anu} {and} \bibinfo{person}{Leticia~Owusu Ansah}.} \bibinfo{year}{2023}\natexlab{}.
\newblock \showarticletitle{Education in the era of generative artificial intelligence (AI): Understanding the potential benefits of ChatGPT in promoting teaching and learning}.
\newblock \bibinfo{journal}{\emph{Journal of AI}} \bibinfo{volume}{7}, \bibinfo{number}{1} (\bibinfo{year}{2023}), \bibinfo{pages}{52--62}.
\newblock


\bibitem[Bellaiche et~al\mbox{.}(2023)]%
        {bellaiche2023humans}
\bibfield{author}{\bibinfo{person}{Lucas Bellaiche}, \bibinfo{person}{Rohin Shahi}, \bibinfo{person}{Martin~Harry Turpin}, \bibinfo{person}{Anya Ragnhildstveit}, \bibinfo{person}{Shawn Sprockett}, \bibinfo{person}{Nathaniel Barr}, \bibinfo{person}{Alexander Christensen}, {and} \bibinfo{person}{Paul Seli}.} \bibinfo{year}{2023}\natexlab{}.
\newblock \showarticletitle{Humans versus AI: whether and why we prefer human-created compared to AI-created artwork}.
\newblock \bibinfo{journal}{\emph{Cognitive Research: Principles and Implications}} \bibinfo{volume}{8}, \bibinfo{number}{1} (\bibinfo{year}{2023}), \bibinfo{pages}{42}.
\newblock


\bibitem[Bertrand and Mullainathan(2004)]%
        {bertrand2004emily}
\bibfield{author}{\bibinfo{person}{Marianne Bertrand} {and} \bibinfo{person}{Sendhil Mullainathan}.} \bibinfo{year}{2004}\natexlab{}.
\newblock \showarticletitle{Are Emily and Greg more employable than Lakisha and Jamal? A field experiment on labor market discrimination}.
\newblock \bibinfo{journal}{\emph{American economic review}} \bibinfo{volume}{94}, \bibinfo{number}{4} (\bibinfo{year}{2004}), \bibinfo{pages}{991--1013}.
\newblock


\bibitem[Bhat et~al\mbox{.}(2023)]%
        {bhat2023cogproc}
\bibfield{author}{\bibinfo{person}{Advait Bhat}, \bibinfo{person}{Saaket Agashe}, \bibinfo{person}{Parth Oberoi}, \bibinfo{person}{Niharika Mohile}, \bibinfo{person}{Ravi Jangir}, {and} \bibinfo{person}{Anirudha Joshi}.} \bibinfo{year}{2023}\natexlab{}.
\newblock \showarticletitle{Interacting with Next-Phrase Suggestions: How Suggestion Systems Aid and Influence the Cognitive Processes of Writing}. In \bibinfo{booktitle}{\emph{Proceedings of the 28th International Conference on Intelligent User Interfaces}} (Sydney, NSW, Australia) \emph{(\bibinfo{series}{IUI '23})}. \bibinfo{publisher}{Association for Computing Machinery}, \bibinfo{address}{New York, NY, USA}, \bibinfo{pages}{436–452}.
\newblock
\showISBNx{9798400701061}
\urldef\tempurl%
\url{https://doi.org/10.1145/3581641.3584060}
\showDOI{\tempurl}


\bibitem[Bhat and Long(2024)]%
        {bhat2021xai}
\bibfield{author}{\bibinfo{person}{Maalvika Bhat} {and} \bibinfo{person}{Duri Long}.} \bibinfo{year}{2024}\natexlab{}.
\newblock \showarticletitle{Designing Interactive Explainable AI Tools for Algorithmic Literacy and Transparency}. In \bibinfo{booktitle}{\emph{Proceedings of the 2024 ACM Designing Interactive Systems Conference}} (Copenhagen, Denmark) \emph{(\bibinfo{series}{DIS '24})}. \bibinfo{publisher}{Association for Computing Machinery}, \bibinfo{address}{New York, NY, USA}, \bibinfo{pages}{939–957}.
\newblock
\showISBNx{9798400705830}
\urldef\tempurl%
\url{https://doi.org/10.1145/3643834.3660722}
\showDOI{\tempurl}


\bibitem[Buschek et~al\mbox{.}(2021)]%
        {buschek2021nonnative}
\bibfield{author}{\bibinfo{person}{Daniel Buschek}, \bibinfo{person}{Martin Z\"{u}rn}, {and} \bibinfo{person}{Malin Eiband}.} \bibinfo{year}{2021}\natexlab{}.
\newblock \showarticletitle{The Impact of Multiple Parallel Phrase Suggestions on Email Input and Composition Behaviour of Native and Non-Native English Writers}. In \bibinfo{booktitle}{\emph{Proceedings of the 2021 CHI Conference on Human Factors in Computing Systems}} (Yokohama, Japan) \emph{(\bibinfo{series}{CHI '21})}. \bibinfo{publisher}{Association for Computing Machinery}, \bibinfo{address}{New York, NY, USA}, Article \bibinfo{articleno}{732}, \bibinfo{numpages}{13}~pages.
\newblock
\showISBNx{9781450380966}
\urldef\tempurl%
\url{https://doi.org/10.1145/3411764.3445372}
\showDOI{\tempurl}


\bibitem[Butler et~al\mbox{.}(2024)]%
        {microsoft2023work}
\bibfield{author}{\bibinfo{person}{Jenna Butler}, \bibinfo{person}{Sonia Jaffe}, \bibinfo{person}{Nancy Baym}, \bibinfo{person}{Mary Czerwinski}, \bibinfo{person}{Shamsi Iqbal}, \bibinfo{person}{Kate Nowak}, \bibinfo{person}{Sean Rintel}, \bibinfo{person}{Abigail Sellen}, \bibinfo{person}{Mihaela Vorvoreanu}, \bibinfo{person}{Najeeb~G. Abdulhamid}, {and} \bibinfo{person}{et al.}} \bibinfo{year}{2024}\natexlab{}.
\newblock \bibinfo{title}{Microsoft New Future of Work Report 2023}.
\newblock
\newblock
\urldef\tempurl%
\url{https://www.microsoft.com/en-us/research/publication/microsoft-new-future-of-work-report-2023/}
\showURL{%
\tempurl}


\bibitem[Catenaccio(2008)]%
        {catenaccio2008press}
\bibfield{author}{\bibinfo{person}{Paola Catenaccio}.} \bibinfo{year}{2008}\natexlab{}.
\newblock \showarticletitle{Press releases as a hybrid genre: Addressing the informative/promotional conundrum}.
\newblock \bibinfo{journal}{\emph{Pragmatics. Quarterly Publication of the International Pragmatics Association (IPrA)}} \bibinfo{volume}{18}, \bibinfo{number}{1} (\bibinfo{year}{2008}), \bibinfo{pages}{9--31}.
\newblock


\bibitem[Chan and Hu(2023)]%
        {chan2023students}
\bibfield{author}{\bibinfo{person}{Cecilia Ka~Yuk Chan} {and} \bibinfo{person}{Wenjie Hu}.} \bibinfo{year}{2023}\natexlab{}.
\newblock \showarticletitle{Students’ voices on generative AI: Perceptions, benefits, and challenges in higher education}.
\newblock \bibinfo{journal}{\emph{International Journal of Educational Technology in Higher Education}} \bibinfo{volume}{20}, \bibinfo{number}{1} (\bibinfo{year}{2023}), \bibinfo{pages}{43}.
\newblock


\bibitem[Chen and Buell(2018)]%
        {chen2018models}
\bibfield{author}{\bibinfo{person}{Grace~A Chen} {and} \bibinfo{person}{Jason~Y Buell}.} \bibinfo{year}{2018}\natexlab{}.
\newblock \showarticletitle{Of models and myths: Asian (Americans) in STEM and the neoliberal racial project}.
\newblock \bibinfo{journal}{\emph{Race Ethnicity and Education}} \bibinfo{volume}{21}, \bibinfo{number}{5} (\bibinfo{year}{2018}), \bibinfo{pages}{607--625}.
\newblock


\bibitem[Chu et~al\mbox{.}(2022)]%
        {chu2022empirical}
\bibfield{author}{\bibinfo{person}{Hyeshin Chu}, \bibinfo{person}{Joohee Kim}, \bibinfo{person}{Seongouk Kim}, \bibinfo{person}{Hongkyu Lim}, \bibinfo{person}{Hyunwook Lee}, \bibinfo{person}{Seungmin Jin}, \bibinfo{person}{Jongeun Lee}, \bibinfo{person}{Taehwan Kim}, {and} \bibinfo{person}{Sungahn Ko}.} \bibinfo{year}{2022}\natexlab{}.
\newblock \showarticletitle{An empirical study on how people perceive AI-generated music}. In \bibinfo{booktitle}{\emph{Proceedings of the 31st ACM International Conference on Information \& Knowledge Management}}. \bibinfo{pages}{304--314}.
\newblock


\bibitem[Clark et~al\mbox{.}(2021)]%
        {clark-etal-2021-thats}
\bibfield{author}{\bibinfo{person}{Elizabeth Clark}, \bibinfo{person}{Tal August}, \bibinfo{person}{Sofia Serrano}, \bibinfo{person}{Nikita Haduong}, \bibinfo{person}{Suchin Gururangan}, {and} \bibinfo{person}{Noah~A. Smith}.} \bibinfo{year}{2021}\natexlab{}.
\newblock \showarticletitle{All That{'}s {`}Human{'} Is Not Gold: Evaluating Human Evaluation of Generated Text}. In \bibinfo{booktitle}{\emph{Proceedings of the 59th Annual Meeting of the Association for Computational Linguistics and the 11th International Joint Conference on Natural Language Processing (Volume 1: Long Papers)}}, \bibfield{editor}{\bibinfo{person}{Chengqing Zong}, \bibinfo{person}{Fei Xia}, \bibinfo{person}{Wenjie Li}, {and} \bibinfo{person}{Roberto Navigli}} (Eds.). \bibinfo{publisher}{Association for Computational Linguistics}, \bibinfo{address}{Online}, \bibinfo{pages}{7282--7296}.
\newblock
\urldef\tempurl%
\url{https://doi.org/10.18653/v1/2021.acl-long.565}
\showDOI{\tempurl}


\bibitem[Correll et~al\mbox{.}(2007)]%
        {correll2007mothers}
\bibfield{author}{\bibinfo{person}{Shelley J. Correll}, \bibinfo{person}{Stephen Benard}, {and} \bibinfo{person}{In Paik}.} \bibinfo{year}{2007}\natexlab{}.
\newblock \showarticletitle{Getting a Job: Is There a Motherhood Penalty?}
\newblock \bibinfo{journal}{\emph{Amer. J. Sociology}} \bibinfo{volume}{112}, \bibinfo{number}{5} (\bibinfo{year}{2007}), \bibinfo{pages}{1297--1338}.
\newblock
\showISSN{00029602, 15375390}
\urldef\tempurl%
\url{http://www.jstor.org/stable/10.1086/511799}
\showURL{%
\tempurl}


\bibitem[Crawford(2016)]%
        {crawford2016artificial}
\bibfield{author}{\bibinfo{person}{Kate Crawford}.} \bibinfo{year}{2016}\natexlab{}.
\newblock \showarticletitle{Artificial Intelligence's White Guy Problem}.
\newblock \bibinfo{journal}{\emph{New York Times}} (\bibinfo{date}{June} \bibinfo{year}{2016}).
\newblock


\bibitem[Cui et~al\mbox{.}(2020)]%
        {cui2022correct}
\bibfield{author}{\bibinfo{person}{Wenzhe Cui}, \bibinfo{person}{Suwen Zhu}, \bibinfo{person}{Mingrui~Ray Zhang}, \bibinfo{person}{H.~Andrew Schwartz}, \bibinfo{person}{Jacob~O. Wobbrock}, {and} \bibinfo{person}{Xiaojun Bi}.} \bibinfo{year}{2020}\natexlab{}.
\newblock \showarticletitle{JustCorrect: Intelligent Post Hoc Text Correction Techniques on Smartphones}. In \bibinfo{booktitle}{\emph{Proceedings of the 33rd Annual ACM Symposium on User Interface Software and Technology}} (Virtual Event, USA) \emph{(\bibinfo{series}{UIST '20})}. \bibinfo{publisher}{Association for Computing Machinery}, \bibinfo{address}{New York, NY, USA}, \bibinfo{pages}{487–499}.
\newblock
\showISBNx{9781450375146}
\urldef\tempurl%
\url{https://doi.org/10.1145/3379337.3415857}
\showDOI{\tempurl}


\bibitem[Dervakos et~al\mbox{.}(2021)]%
        {dervakos2021heuristics}
\bibfield{author}{\bibinfo{person}{Edmund Dervakos}, \bibinfo{person}{Giorgos Filandrianos}, {and} \bibinfo{person}{Giorgos Stamou}.} \bibinfo{year}{2021}\natexlab{}.
\newblock \showarticletitle{Heuristics for evaluation of AI generated music}. In \bibinfo{booktitle}{\emph{2020 25th International Conference on Pattern Recognition (ICPR)}}. IEEE, \bibinfo{pages}{9164--9171}.
\newblock


\bibitem[Dou et~al\mbox{.}(2022)]%
        {dou-etal-2022-scarecrow}
\bibfield{author}{\bibinfo{person}{Yao Dou}, \bibinfo{person}{Maxwell Forbes}, \bibinfo{person}{Rik Koncel-Kedziorski}, \bibinfo{person}{Noah~A. Smith}, {and} \bibinfo{person}{Yejin Choi}.} \bibinfo{year}{2022}\natexlab{}.
\newblock \showarticletitle{Is {GPT}-3 Text Indistinguishable from Human Text? Scarecrow: A Framework for Scrutinizing Machine Text}. In \bibinfo{booktitle}{\emph{Proceedings of the 60th Annual Meeting of the Association for Computational Linguistics (Volume 1: Long Papers)}}, \bibfield{editor}{\bibinfo{person}{Smaranda Muresan}, \bibinfo{person}{Preslav Nakov}, {and} \bibinfo{person}{Aline Villavicencio}} (Eds.). \bibinfo{publisher}{Association for Computational Linguistics}, \bibinfo{address}{Dublin, Ireland}, \bibinfo{pages}{7250--7274}.
\newblock
\urldef\tempurl%
\url{https://doi.org/10.18653/v1/2022.acl-long.501}
\showDOI{\tempurl}


\bibitem[Draxler et~al\mbox{.}(2024)]%
        {draxler2024ghostwriter}
\bibfield{author}{\bibinfo{person}{Fiona Draxler}, \bibinfo{person}{Anna Werner}, \bibinfo{person}{Florian Lehmann}, \bibinfo{person}{Matthias Hoppe}, \bibinfo{person}{Albrecht Schmidt}, \bibinfo{person}{Daniel Buschek}, {and} \bibinfo{person}{Robin Welsch}.} \bibinfo{year}{2024}\natexlab{}.
\newblock \showarticletitle{The AI Ghostwriter Effect: When Users do not Perceive Ownership of AI-Generated Text but Self-Declare as Authors}.
\newblock \bibinfo{journal}{\emph{ACM Trans. Comput.-Hum. Interact.}} \bibinfo{volume}{31}, \bibinfo{number}{2}, Article \bibinfo{articleno}{25} (\bibinfo{date}{feb} \bibinfo{year}{2024}), \bibinfo{numpages}{40}~pages.
\newblock
\showISSN{1073-0516}
\urldef\tempurl%
\url{https://doi.org/10.1145/3637875}
\showDOI{\tempurl}


\bibitem[Epstein et~al\mbox{.}(2023)]%
        {epstein2023art}
\bibfield{author}{\bibinfo{person}{Ziv Epstein}, \bibinfo{person}{Aaron Hertzmann}, \bibinfo{person}{Investigators of Human~Creativity}, \bibinfo{person}{Memo Akten}, \bibinfo{person}{Hany Farid}, \bibinfo{person}{Jessica Fjeld}, \bibinfo{person}{Morgan~R Frank}, \bibinfo{person}{Matthew Groh}, \bibinfo{person}{Laura Herman}, \bibinfo{person}{Neil Leach}, {et~al\mbox{.}}} \bibinfo{year}{2023}\natexlab{}.
\newblock \showarticletitle{Art and the science of generative AI}.
\newblock \bibinfo{journal}{\emph{Science}} \bibinfo{volume}{380}, \bibinfo{number}{6650} (\bibinfo{year}{2023}), \bibinfo{pages}{1110--1111}.
\newblock


\bibitem[Fast and Horvitz(2017)]%
        {fast2017long}
\bibfield{author}{\bibinfo{person}{Ethan Fast} {and} \bibinfo{person}{Eric Horvitz}.} \bibinfo{year}{2017}\natexlab{}.
\newblock \showarticletitle{Long-term trends in the public perception of artificial intelligence}. In \bibinfo{booktitle}{\emph{Proceedings of the AAAI conference on artificial intelligence}}, Vol.~\bibinfo{volume}{31}.
\newblock


\bibitem[Fink(2019)]%
        {fink2019biggest}
\bibfield{author}{\bibinfo{person}{Katherine Fink}.} \bibinfo{year}{2019}\natexlab{}.
\newblock \showarticletitle{The biggest challenge facing journalism: A lack of trust}.
\newblock \bibinfo{journal}{\emph{Journalism}} \bibinfo{volume}{20}, \bibinfo{number}{1} (\bibinfo{year}{2019}), \bibinfo{pages}{40--43}.
\newblock


\bibitem[Frid et~al\mbox{.}(2020)]%
        {frid2020music}
\bibfield{author}{\bibinfo{person}{Emma Frid}, \bibinfo{person}{Celso Gomes}, {and} \bibinfo{person}{Zeyu Jin}.} \bibinfo{year}{2020}\natexlab{}.
\newblock \showarticletitle{Music creation by example}. In \bibinfo{booktitle}{\emph{Proceedings of the 2020 CHI conference on human factors in computing systems}}. \bibinfo{pages}{1--13}.
\newblock


\bibitem[Fu et~al\mbox{.}(2024)]%
        {fu2024aimc}
\bibfield{author}{\bibinfo{person}{Yue Fu}, \bibinfo{person}{Sami Foell}, \bibinfo{person}{Xuhai Xu}, {and} \bibinfo{person}{Alexis Hiniker}.} \bibinfo{year}{2024}\natexlab{}.
\newblock \showarticletitle{From Text to Self: Users’ Perception of AIMC Tools on Interpersonal Communication and Self}. In \bibinfo{booktitle}{\emph{Proceedings of the CHI Conference on Human Factors in Computing Systems}} (Honolulu, HI, USA) \emph{(\bibinfo{series}{CHI '24})}. \bibinfo{publisher}{Association for Computing Machinery}, \bibinfo{address}{New York, NY, USA}, Article \bibinfo{articleno}{977}, \bibinfo{numpages}{17}~pages.
\newblock
\showISBNx{9798400703300}
\urldef\tempurl%
\url{https://doi.org/10.1145/3613904.3641955}
\showDOI{\tempurl}


\bibitem[Gaddis(2017)]%
        {gaddis2017black}
\bibfield{author}{\bibinfo{person}{S~Michael Gaddis}.} \bibinfo{year}{2017}\natexlab{}.
\newblock \showarticletitle{How black are Lakisha and Jamal? Racial perceptions from names used in correspondence audit studies}.
\newblock \bibinfo{journal}{\emph{Sociological Science}}  \bibinfo{volume}{4} (\bibinfo{year}{2017}), \bibinfo{pages}{469--489}.
\newblock


\bibitem[Gautam et~al\mbox{.}(2024)]%
        {gautam2024melting}
\bibfield{author}{\bibinfo{person}{Sanjana Gautam}, \bibinfo{person}{Pranav~Narayanan Venkit}, {and} \bibinfo{person}{Sourojit Ghosh}.} \bibinfo{year}{2024}\natexlab{}.
\newblock \showarticletitle{From melting pots to misrepresentations: Exploring harms in generative ai}.
\newblock \bibinfo{journal}{\emph{arXiv preprint arXiv:2403.10776}} (\bibinfo{year}{2024}).
\newblock


\bibitem[Gero et~al\mbox{.}(2022)]%
        {gero2022sparks}
\bibfield{author}{\bibinfo{person}{Katy~Ilonka Gero}, \bibinfo{person}{Vivian Liu}, {and} \bibinfo{person}{Lydia Chilton}.} \bibinfo{year}{2022}\natexlab{}.
\newblock \showarticletitle{Sparks: Inspiration for Science Writing using Language Models}. In \bibinfo{booktitle}{\emph{Proceedings of the 2022 ACM Designing Interactive Systems Conference}} (Virtual Event, Australia) \emph{(\bibinfo{series}{DIS '22})}. \bibinfo{publisher}{Association for Computing Machinery}, \bibinfo{address}{New York, NY, USA}, \bibinfo{pages}{1002–1019}.
\newblock
\showISBNx{9781450393584}
\urldef\tempurl%
\url{https://doi.org/10.1145/3532106.3533533}
\showDOI{\tempurl}


\bibitem[Ghosh and Caliskan(2023)]%
        {ghosh2023person}
\bibfield{author}{\bibinfo{person}{Sourojit Ghosh} {and} \bibinfo{person}{Aylin Caliskan}.} \bibinfo{year}{2023}\natexlab{}.
\newblock \showarticletitle{'Person'== Light-skinned, Western Man, and Sexualization of Women of Color: Stereotypes in Stable Diffusion}.
\newblock \bibinfo{journal}{\emph{arXiv preprint arXiv:2310.19981}} (\bibinfo{year}{2023}).
\newblock


\bibitem[Giglio and Costa(2023)]%
        {giglio2023use}
\bibfield{author}{\bibinfo{person}{Auro~Del Giglio} {and} \bibinfo{person}{Mateus Uerlei Pereira~da Costa}.} \bibinfo{year}{2023}\natexlab{}.
\newblock \showarticletitle{The use of artificial intelligence to improve the scientific writing of non-native english speakers}.
\newblock \bibinfo{journal}{\emph{Revista da Associa{\c{c}}{\~a}o M{\'e}dica Brasileira}} \bibinfo{volume}{69}, \bibinfo{number}{9} (\bibinfo{year}{2023}), \bibinfo{pages}{e20230560}.
\newblock


\bibitem[Government({[n.\,d.]})]%
        {usalanguage}
\bibfield{author}{\bibinfo{person}{United~States Government}.} \bibinfo{year}{[n.\,d.]}\natexlab{}.
\newblock
\newblock
\urldef\tempurl%
\url{https://www.usa.gov/official-language-of-us}
\showURL{%
\tempurl}


\bibitem[Gozalo-Brizuela and Garrido-Merchan(2023)]%
        {gozalo2023chatgpt}
\bibfield{author}{\bibinfo{person}{Roberto Gozalo-Brizuela} {and} \bibinfo{person}{Eduardo~C Garrido-Merchan}.} \bibinfo{year}{2023}\natexlab{}.
\newblock \showarticletitle{ChatGPT is not all you need. A State of the Art Review of large Generative AI models}.
\newblock \bibinfo{journal}{\emph{arXiv preprint arXiv:2301.04655}} (\bibinfo{year}{2023}).
\newblock


\bibitem[Hancock et~al\mbox{.}(2020)]%
        {hancock2020ai}
\bibfield{author}{\bibinfo{person}{Jeffrey~T Hancock}, \bibinfo{person}{Mor Naaman}, {and} \bibinfo{person}{Karen Levy}.} \bibinfo{year}{2020}\natexlab{}.
\newblock \showarticletitle{AI-mediated communication: Definition, research agenda, and ethical considerations}.
\newblock \bibinfo{journal}{\emph{Journal of Computer-Mediated Communication}} \bibinfo{volume}{25}, \bibinfo{number}{1} (\bibinfo{year}{2020}), \bibinfo{pages}{89--100}.
\newblock


\bibitem[Hann\'{a}k et~al\mbox{.}(2017)]%
        {hannak2017marketplaces}
\bibfield{author}{\bibinfo{person}{Anik\'{o} Hann\'{a}k}, \bibinfo{person}{Claudia Wagner}, \bibinfo{person}{David Garcia}, \bibinfo{person}{Alan Mislove}, \bibinfo{person}{Markus Strohmaier}, {and} \bibinfo{person}{Christo Wilson}.} \bibinfo{year}{2017}\natexlab{}.
\newblock \showarticletitle{Bias in Online Freelance Marketplaces: Evidence from TaskRabbit and Fiverr}. In \bibinfo{booktitle}{\emph{Proceedings of the 2017 ACM Conference on Computer Supported Cooperative Work and Social Computing}} (Portland, Oregon, USA) \emph{(\bibinfo{series}{CSCW '17})}. \bibinfo{publisher}{Association for Computing Machinery}, \bibinfo{address}{New York, NY, USA}, \bibinfo{pages}{1914–1933}.
\newblock
\showISBNx{9781450343350}
\urldef\tempurl%
\url{https://doi.org/10.1145/2998181.2998327}
\showDOI{\tempurl}


\bibitem[Haslberger et~al\mbox{.}(2023)]%
        {haslberger2023no}
\bibfield{author}{\bibinfo{person}{Matthias Haslberger}, \bibinfo{person}{Jane Gingrich}, {and} \bibinfo{person}{Jasmine Bhatia}.} \bibinfo{year}{2023}\natexlab{}.
\newblock \showarticletitle{No great equalizer: experimental evidence on AI in the UK labor market}.
\newblock \bibinfo{journal}{\emph{Available at SSRN}} (\bibinfo{year}{2023}).
\newblock


\bibitem[Hogan and Berry(2011)]%
        {hogan2011racial}
\bibfield{author}{\bibinfo{person}{Bernie Hogan} {and} \bibinfo{person}{Brent Berry}.} \bibinfo{year}{2011}\natexlab{}.
\newblock \showarticletitle{Racial and ethnic biases in rental housing: An audit study of online apartment listings}.
\newblock \bibinfo{journal}{\emph{City \& community}} \bibinfo{volume}{10}, \bibinfo{number}{4} (\bibinfo{year}{2011}), \bibinfo{pages}{351--372}.
\newblock


\bibitem[Hohenstein et~al\mbox{.}(2023)]%
        {hohenstein2023artificial}
\bibfield{author}{\bibinfo{person}{Jess Hohenstein}, \bibinfo{person}{Rene~F Kizilcec}, \bibinfo{person}{Dominic DiFranzo}, \bibinfo{person}{Zhila Aghajari}, \bibinfo{person}{Hannah Mieczkowski}, \bibinfo{person}{Karen Levy}, \bibinfo{person}{Mor Naaman}, \bibinfo{person}{Jeffrey Hancock}, {and} \bibinfo{person}{Malte~F Jung}.} \bibinfo{year}{2023}\natexlab{}.
\newblock \showarticletitle{Artificial intelligence in communication impacts language and social relationships}.
\newblock \bibinfo{journal}{\emph{Scientific Reports}} \bibinfo{volume}{13}, \bibinfo{number}{1} (\bibinfo{year}{2023}), \bibinfo{pages}{5487}.
\newblock


\bibitem[Hong(2018)]%
        {hong2018biasart}
\bibfield{author}{\bibinfo{person}{Joo-Wha Hong}.} \bibinfo{year}{2018}\natexlab{}.
\newblock \showarticletitle{Bias in Perception of Art Produced by Artificial Intelligence}. In \bibinfo{booktitle}{\emph{Human-Computer Interaction. Interaction in Context: 20th International Conference, HCI International 2018, Las Vegas, NV, USA, July 15–20, 2018, Proceedings, Part II}} (Las Vegas, NV, USA). \bibinfo{publisher}{Springer-Verlag}, \bibinfo{address}{Berlin, Heidelberg}, \bibinfo{pages}{290–303}.
\newblock
\showISBNx{978-3-319-91243-1}
\urldef\tempurl%
\url{https://doi.org/10.1007/978-3-319-91244-8_24}
\showDOI{\tempurl}


\bibitem[House(2016)]%
        {house2016preparing}
\bibfield{author}{\bibinfo{person}{White House}.} \bibinfo{year}{2016}\natexlab{}.
\newblock \bibinfo{title}{Preparing for the future of artificial intelligence. Executive Office of the President National Science and Technology Council. Committee on Technology}.
\newblock
\newblock


\bibitem[Hui et~al\mbox{.}(2023)]%
        {hui2023short}
\bibfield{author}{\bibinfo{person}{Xiang Hui}, \bibinfo{person}{Oren Reshef}, {and} \bibinfo{person}{Luofeng Zhou}.} \bibinfo{year}{2023}\natexlab{}.
\newblock \showarticletitle{The short-term effects of generative artificial intelligence on employment: Evidence from an online labor market}.
\newblock \bibinfo{journal}{\emph{Available at SSRN 4527336}} (\bibinfo{year}{2023}).
\newblock


\bibitem[Hwang et~al\mbox{.}(2023)]%
        {hwang2023chatgpt}
\bibfield{author}{\bibinfo{person}{Sung~Il Hwang}, \bibinfo{person}{Joon~Seo Lim}, \bibinfo{person}{Ro~Woon Lee}, \bibinfo{person}{Yusuke Matsui}, \bibinfo{person}{Toshihiro Iguchi}, \bibinfo{person}{Takao Hiraki}, {and} \bibinfo{person}{Hyungwoo Ahn}.} \bibinfo{year}{2023}\natexlab{}.
\newblock \showarticletitle{Is ChatGPT a “fire of prometheus” for non-native English-speaking researchers in academic writing?}
\newblock \bibinfo{journal}{\emph{Korean Journal of Radiology}} \bibinfo{volume}{24}, \bibinfo{number}{10} (\bibinfo{year}{2023}), \bibinfo{pages}{952}.
\newblock


\bibitem[Ippolito et~al\mbox{.}(2019)]%
        {ippolito2019automatic}
\bibfield{author}{\bibinfo{person}{Daphne Ippolito}, \bibinfo{person}{Daniel Duckworth}, \bibinfo{person}{Chris Callison-Burch}, {and} \bibinfo{person}{Douglas Eck}.} \bibinfo{year}{2019}\natexlab{}.
\newblock \showarticletitle{Automatic detection of generated text is easiest when humans are fooled}.
\newblock \bibinfo{journal}{\emph{arXiv preprint arXiv:1911.00650}} (\bibinfo{year}{2019}).
\newblock


\bibitem[Jakesch et~al\mbox{.}(2022)]%
        {jakesch2022different}
\bibfield{author}{\bibinfo{person}{Maurice Jakesch}, \bibinfo{person}{Zana Bu{\c{c}}inca}, \bibinfo{person}{Saleema Amershi}, {and} \bibinfo{person}{Alexandra Olteanu}.} \bibinfo{year}{2022}\natexlab{}.
\newblock \showarticletitle{How different groups prioritize ethical values for responsible AI}. In \bibinfo{booktitle}{\emph{Proceedings of the 2022 ACM Conference on Fairness, Accountability, and Transparency}}. \bibinfo{pages}{310--323}.
\newblock


\bibitem[Jakesch et~al\mbox{.}(2019)]%
        {jakesch2019ai}
\bibfield{author}{\bibinfo{person}{Maurice Jakesch}, \bibinfo{person}{Megan French}, \bibinfo{person}{Xiao Ma}, \bibinfo{person}{Jeffrey~T Hancock}, {and} \bibinfo{person}{Mor Naaman}.} \bibinfo{year}{2019}\natexlab{}.
\newblock \showarticletitle{AI-mediated communication: How the perception that profile text was written by AI affects trustworthiness}. In \bibinfo{booktitle}{\emph{Proceedings of the 2019 CHI Conference on Human Factors in Computing Systems}}. \bibinfo{pages}{1--13}.
\newblock


\bibitem[Jakesch et~al\mbox{.}(2023)]%
        {jakesch2023human}
\bibfield{author}{\bibinfo{person}{Maurice Jakesch}, \bibinfo{person}{Jeffrey~T Hancock}, {and} \bibinfo{person}{Mor Naaman}.} \bibinfo{year}{2023}\natexlab{}.
\newblock \showarticletitle{Human heuristics for AI-generated language are flawed}.
\newblock \bibinfo{journal}{\emph{Proceedings of the National Academy of Sciences}} \bibinfo{volume}{120}, \bibinfo{number}{11} (\bibinfo{year}{2023}), \bibinfo{pages}{e2208839120}.
\newblock


\bibitem[Jiang and Nachum(2020)]%
        {jiang2020identifying}
\bibfield{author}{\bibinfo{person}{Heinrich Jiang} {and} \bibinfo{person}{Ofir Nachum}.} \bibinfo{year}{2020}\natexlab{}.
\newblock \showarticletitle{Identifying and correcting label bias in machine learning}. In \bibinfo{booktitle}{\emph{International conference on artificial intelligence and statistics}}. PMLR, \bibinfo{pages}{702--712}.
\newblock


\bibitem[Jiang et~al\mbox{.}(2023)]%
        {jiang2023aiart}
\bibfield{author}{\bibinfo{person}{Harry~H. Jiang}, \bibinfo{person}{Lauren Brown}, \bibinfo{person}{Jessica Cheng}, \bibinfo{person}{Mehtab Khan}, \bibinfo{person}{Abhishek Gupta}, \bibinfo{person}{Deja Workman}, \bibinfo{person}{Alex Hanna}, \bibinfo{person}{Johnathan Flowers}, {and} \bibinfo{person}{Timnit Gebru}.} \bibinfo{year}{2023}\natexlab{}.
\newblock \showarticletitle{AI Art and its Impact on Artists}. In \bibinfo{booktitle}{\emph{Proceedings of the 2023 AAAI/ACM Conference on AI, Ethics, and Society}} (Montr\'{e}al, QC, Canada) \emph{(\bibinfo{series}{AIES '23})}. \bibinfo{publisher}{Association for Computing Machinery}, \bibinfo{address}{New York, NY, USA}, \bibinfo{pages}{363–374}.
\newblock
\showISBNx{9798400702310}
\urldef\tempurl%
\url{https://doi.org/10.1145/3600211.3604681}
\showDOI{\tempurl}


\bibitem[Kadoma et~al\mbox{.}(2024)]%
        {kadoma2024writer}
\bibfield{author}{\bibinfo{person}{Kowe Kadoma}, \bibinfo{person}{Marianne Aubin Le~Quere}, \bibinfo{person}{Xiyu~Jenny Fu}, \bibinfo{person}{Christin Munsch}, \bibinfo{person}{Dana\"{e} Metaxa}, {and} \bibinfo{person}{Mor Naaman}.} \bibinfo{year}{2024}\natexlab{}.
\newblock \showarticletitle{The Role of Inclusion, Control, and Ownership in Workplace AI-Mediated Communication}. In \bibinfo{booktitle}{\emph{Proceedings of the CHI Conference on Human Factors in Computing Systems}} (Honolulu, HI, USA) \emph{(\bibinfo{series}{CHI '24})}. \bibinfo{publisher}{Association for Computing Machinery}, \bibinfo{address}{New York, NY, USA}, Article \bibinfo{articleno}{1016}, \bibinfo{numpages}{10}~pages.
\newblock
\showISBNx{9798400703300}
\urldef\tempurl%
\url{https://doi.org/10.1145/3613904.3642650}
\showDOI{\tempurl}


\bibitem[Kelley et~al\mbox{.}(2021)]%
        {kelley2021exciting}
\bibfield{author}{\bibinfo{person}{Patrick~Gage Kelley}, \bibinfo{person}{Yongwei Yang}, \bibinfo{person}{Courtney Heldreth}, \bibinfo{person}{Christopher Moessner}, \bibinfo{person}{Aaron Sedley}, \bibinfo{person}{Andreas Kramm}, \bibinfo{person}{David~T Newman}, {and} \bibinfo{person}{Allison Woodruff}.} \bibinfo{year}{2021}\natexlab{}.
\newblock \showarticletitle{Exciting, useful, worrying, futuristic: Public perception of artificial intelligence in 8 countries}. In \bibinfo{booktitle}{\emph{Proceedings of the 2021 AAAI/ACM Conference on AI, Ethics, and Society}}. \bibinfo{pages}{627--637}.
\newblock


\bibitem[Kim et~al\mbox{.}(2023)]%
        {kim2023one}
\bibfield{author}{\bibinfo{person}{Taenyun Kim}, \bibinfo{person}{Maria~D Molina}, \bibinfo{person}{Minjin Rheu}, \bibinfo{person}{Emily~S Zhan}, {and} \bibinfo{person}{Wei Peng}.} \bibinfo{year}{2023}\natexlab{}.
\newblock \showarticletitle{One AI does not fit all: A cluster analysis of the laypeople’s perception of AI roles}. In \bibinfo{booktitle}{\emph{Proceedings of the 2023 CHI Conference on Human Factors in Computing Systems}}. \bibinfo{pages}{1--20}.
\newblock


\bibitem[K{\"o}bis and Mossink(2021)]%
        {kobis2021artificial}
\bibfield{author}{\bibinfo{person}{Nils K{\"o}bis} {and} \bibinfo{person}{Luca~D Mossink}.} \bibinfo{year}{2021}\natexlab{}.
\newblock \showarticletitle{Artificial intelligence versus Maya Angelou: Experimental evidence that people cannot differentiate AI-generated from human-written poetry}.
\newblock \bibinfo{journal}{\emph{Computers in human behavior}}  \bibinfo{volume}{114} (\bibinfo{year}{2021}), \bibinfo{pages}{106553}.
\newblock


\bibitem[Kotek et~al\mbox{.}(2023)]%
        {kotek2023genderbias}
\bibfield{author}{\bibinfo{person}{Hadas Kotek}, \bibinfo{person}{Rikker Dockum}, {and} \bibinfo{person}{David Sun}.} \bibinfo{year}{2023}\natexlab{}.
\newblock \showarticletitle{Gender bias and stereotypes in Large Language Models}. In \bibinfo{booktitle}{\emph{Proceedings of The ACM Collective Intelligence Conference}} (Delft, Netherlands) \emph{(\bibinfo{series}{CI '23})}. \bibinfo{publisher}{Association for Computing Machinery}, \bibinfo{address}{New York, NY, USA}, \bibinfo{pages}{12–24}.
\newblock
\showISBNx{9798400701139}
\urldef\tempurl%
\url{https://doi.org/10.1145/3582269.3615599}
\showDOI{\tempurl}


\bibitem[Laban et~al\mbox{.}(2024)]%
        {laban2024inksync}
\bibfield{author}{\bibinfo{person}{Philippe Laban}, \bibinfo{person}{Jesse Vig}, \bibinfo{person}{Marti Hearst}, \bibinfo{person}{Caiming Xiong}, {and} \bibinfo{person}{Chien-Sheng Wu}.} \bibinfo{year}{2024}\natexlab{}.
\newblock \showarticletitle{Beyond the Chat: Executable and Verifiable Text-Editing with LLMs}. In \bibinfo{booktitle}{\emph{Proceedings of the 37th Annual ACM Symposium on User Interface Software and Technology}} (Pittsburgh, PA, USA) \emph{(\bibinfo{series}{UIST '24})}. \bibinfo{publisher}{Association for Computing Machinery}, \bibinfo{address}{New York, NY, USA}, Article \bibinfo{articleno}{20}, \bibinfo{numpages}{23}~pages.
\newblock
\showISBNx{9798400706288}
\urldef\tempurl%
\url{https://doi.org/10.1145/3654777.3676419}
\showDOI{\tempurl}


\bibitem[Landivar(2013)]%
        {landivar2013disparities}
\bibfield{author}{\bibinfo{person}{Liana~Christin Landivar}.} \bibinfo{year}{2013}\natexlab{}.
\newblock \showarticletitle{Disparities in STEM employment by sex, race, and Hispanic origin}.
\newblock \bibinfo{journal}{\emph{Education Review}} \bibinfo{volume}{29}, \bibinfo{number}{6} (\bibinfo{year}{2013}), \bibinfo{pages}{911--922}.
\newblock


\bibitem[Levy(2022)]%
        {levy2022data}
\bibfield{author}{\bibinfo{person}{Karen Levy}.} \bibinfo{year}{2022}\natexlab{}.
\newblock \showarticletitle{Data driven: truckers, technology, and the new workplace surveillance}.
\newblock  (\bibinfo{year}{2022}).
\newblock


\bibitem[Liang et~al\mbox{.}(2023)]%
        {liang2023gptdetectorsbiasednonnative}
\bibfield{author}{\bibinfo{person}{Weixin Liang}, \bibinfo{person}{Mert Yuksekgonul}, \bibinfo{person}{Yining Mao}, \bibinfo{person}{Eric Wu}, {and} \bibinfo{person}{James Zou}.} \bibinfo{year}{2023}\natexlab{}.
\newblock \bibinfo{title}{GPT detectors are biased against non-native English writers}.
\newblock
\newblock
\showeprint[arxiv]{2304.02819}~[cs.CL]
\urldef\tempurl%
\url{https://arxiv.org/abs/2304.02819}
\showURL{%
\tempurl}


\bibitem[Lim and Schmälzle(2024)]%
        {lim2024disclosure}
\bibfield{author}{\bibinfo{person}{Sue Lim} {and} \bibinfo{person}{Ralf Schmälzle}.} \bibinfo{year}{2024}\natexlab{}.
\newblock \showarticletitle{The effect of source disclosure on evaluation of AI-generated messages}.
\newblock \bibinfo{journal}{\emph{Computers in Human Behavior: Artificial Humans}} \bibinfo{volume}{2}, \bibinfo{number}{1} (\bibinfo{year}{2024}), \bibinfo{pages}{100058}.
\newblock
\showISSN{2949-8821}
\urldef\tempurl%
\url{https://doi.org/10.1016/j.chbah.2024.100058}
\showDOI{\tempurl}


\bibitem[Liu et~al\mbox{.}(2023)]%
        {liu2023generate}
\bibfield{author}{\bibinfo{person}{Jin Liu}, \bibinfo{person}{Xingchen Xu}, \bibinfo{person}{Yongjun Li}, {and} \bibinfo{person}{Yong Tan}.} \bibinfo{year}{2023}\natexlab{}.
\newblock \showarticletitle{" Generate" the Future of Work through AI: Empirical Evidence from Online Labor Markets}.
\newblock \bibinfo{journal}{\emph{arXiv preprint arXiv:2308.05201}} (\bibinfo{year}{2023}).
\newblock


\bibitem[Liu et~al\mbox{.}(2022)]%
        {liu2022console}
\bibfield{author}{\bibinfo{person}{Yihe Liu}, \bibinfo{person}{Anushk Mittal}, \bibinfo{person}{Diyi Yang}, {and} \bibinfo{person}{Amy Bruckman}.} \bibinfo{year}{2022}\natexlab{}.
\newblock \showarticletitle{Will AI Console Me when I Lose my Pet? Understanding Perceptions of AI-Mediated Email Writing}. In \bibinfo{booktitle}{\emph{Proceedings of the 2022 CHI Conference on Human Factors in Computing Systems}} (New Orleans, LA, USA) \emph{(\bibinfo{series}{CHI '22})}. \bibinfo{publisher}{Association for Computing Machinery}, \bibinfo{address}{New York, NY, USA}, Article \bibinfo{articleno}{474}, \bibinfo{numpages}{13}~pages.
\newblock
\showISBNx{9781450391573}
\urldef\tempurl%
\url{https://doi.org/10.1145/3491102.3517731}
\showDOI{\tempurl}


\bibitem[Long and Magerko(2020)]%
        {long2020ailiteracry}
\bibfield{author}{\bibinfo{person}{Duri Long} {and} \bibinfo{person}{Brian Magerko}.} \bibinfo{year}{2020}\natexlab{}.
\newblock \showarticletitle{What is AI Literacy? Competencies and Design Considerations}. In \bibinfo{booktitle}{\emph{Proceedings of the 2020 CHI Conference on Human Factors in Computing Systems}} (Honolulu, HI, USA) \emph{(\bibinfo{series}{CHI '20})}. \bibinfo{publisher}{Association for Computing Machinery}, \bibinfo{address}{New York, NY, USA}, \bibinfo{pages}{1–16}.
\newblock
\showISBNx{9781450367080}
\urldef\tempurl%
\url{https://doi.org/10.1145/3313831.3376727}
\showDOI{\tempurl}


\bibitem[Longoni et~al\mbox{.}(2022)]%
        {longoni2022newsai}
\bibfield{author}{\bibinfo{person}{Chiara Longoni}, \bibinfo{person}{Andrey Fradkin}, \bibinfo{person}{Luca Cian}, {and} \bibinfo{person}{Gordon Pennycook}.} \bibinfo{year}{2022}\natexlab{}.
\newblock \showarticletitle{News from Generative Artificial Intelligence Is Believed Less}. In \bibinfo{booktitle}{\emph{Proceedings of the 2022 ACM Conference on Fairness, Accountability, and Transparency}} (Seoul, Republic of Korea) \emph{(\bibinfo{series}{FAccT '22})}. \bibinfo{publisher}{Association for Computing Machinery}, \bibinfo{address}{New York, NY, USA}, \bibinfo{pages}{97–106}.
\newblock
\showISBNx{9781450393522}
\urldef\tempurl%
\url{https://doi.org/10.1145/3531146.3533077}
\showDOI{\tempurl}


\bibitem[Lucy and Bamman(2021)]%
        {lucy-bamman-2021-gender}
\bibfield{author}{\bibinfo{person}{Li Lucy} {and} \bibinfo{person}{David Bamman}.} \bibinfo{year}{2021}\natexlab{}.
\newblock \showarticletitle{Gender and Representation Bias in {GPT}-3 Generated Stories}. In \bibinfo{booktitle}{\emph{Proceedings of the Third Workshop on Narrative Understanding}}, \bibfield{editor}{\bibinfo{person}{Nader Akoury}, \bibinfo{person}{Faeze Brahman}, \bibinfo{person}{Snigdha Chaturvedi}, \bibinfo{person}{Elizabeth Clark}, \bibinfo{person}{Mohit Iyyer}, {and} \bibinfo{person}{Lara~J. Martin}} (Eds.). \bibinfo{publisher}{Association for Computational Linguistics}, \bibinfo{address}{Virtual}, \bibinfo{pages}{48--55}.
\newblock
\urldef\tempurl%
\url{https://doi.org/10.18653/v1/2021.nuse-1.5}
\showDOI{\tempurl}


\bibitem[McGee et~al\mbox{.}(2017)]%
        {mcgee2017burden}
\bibfield{author}{\bibinfo{person}{Ebony~O McGee}, \bibinfo{person}{Bhoomi~K Thakore}, {and} \bibinfo{person}{Sandra~S LaBlance}.} \bibinfo{year}{2017}\natexlab{}.
\newblock \showarticletitle{The burden of being “model”: Racialized experiences of Asian STEM college students.}
\newblock \bibinfo{journal}{\emph{Journal of Diversity in Higher Education}} \bibinfo{volume}{10}, \bibinfo{number}{3} (\bibinfo{year}{2017}), \bibinfo{pages}{253}.
\newblock


\bibitem[Mehrabi et~al\mbox{.}(2021)]%
        {mehrabi2021survey}
\bibfield{author}{\bibinfo{person}{Ninareh Mehrabi}, \bibinfo{person}{Fred Morstatter}, \bibinfo{person}{Nripsuta Saxena}, \bibinfo{person}{Kristina Lerman}, {and} \bibinfo{person}{Aram Galstyan}.} \bibinfo{year}{2021}\natexlab{}.
\newblock \showarticletitle{A survey on bias and fairness in machine learning}.
\newblock \bibinfo{journal}{\emph{ACM computing surveys (CSUR)}} \bibinfo{volume}{54}, \bibinfo{number}{6} (\bibinfo{year}{2021}), \bibinfo{pages}{1--35}.
\newblock


\bibitem[Mengesha et~al\mbox{.}(2021)]%
        {mengesha2021inclusive}
\bibfield{author}{\bibinfo{person}{Zion Mengesha}, \bibinfo{person}{Courtney Heldreth}, \bibinfo{person}{Michal Lahav}, \bibinfo{person}{Juliana Sublewski}, {and} \bibinfo{person}{Elyse Tuennerman}.} \bibinfo{year}{2021}\natexlab{}.
\newblock \showarticletitle{"I don't think these devices are very culturally sensitive." - The impact of errors on African Americans in Automated Speech Recognition}.
\newblock \bibinfo{journal}{\emph{Frontiers in Artificial Intelligence}}  \bibinfo{volume}{26} (\bibinfo{year}{2021}).
\newblock
\urldef\tempurl%
\url{https://www.frontiersin.org/journals/artificial-intelligence/articles/10.3389/frai.2021.725911}
\showURL{%
\tempurl}


\bibitem[Metaxa-Kakavouli et~al\mbox{.}(2018)]%
        {metaxa2018gender}
\bibfield{author}{\bibinfo{person}{Dana{\"e} Metaxa-Kakavouli}, \bibinfo{person}{Kelly Wang}, \bibinfo{person}{James~A Landay}, {and} \bibinfo{person}{Jeff Hancock}.} \bibinfo{year}{2018}\natexlab{}.
\newblock \showarticletitle{Gender-inclusive design: Sense of belonging and bias in web interfaces}. In \bibinfo{booktitle}{\emph{Proceedings of the 2018 CHI Conference on human factors in computing systems}}. \bibinfo{pages}{1--6}.
\newblock


\bibitem[Mieczkowski et~al\mbox{.}(2021)]%
        {mieczkowski2021ai}
\bibfield{author}{\bibinfo{person}{Hannah Mieczkowski}, \bibinfo{person}{Jeffrey~T Hancock}, \bibinfo{person}{Mor Naaman}, \bibinfo{person}{Malte Jung}, {and} \bibinfo{person}{Jess Hohenstein}.} \bibinfo{year}{2021}\natexlab{}.
\newblock \showarticletitle{AI-mediated communication: Language use and interpersonal effects in a referential communication task}.
\newblock \bibinfo{journal}{\emph{Proceedings of the ACM on Human-Computer Interaction}} \bibinfo{volume}{5}, \bibinfo{number}{CSCW1} (\bibinfo{year}{2021}), \bibinfo{pages}{1--14}.
\newblock


\bibitem[Morgan et~al\mbox{.}(2016)]%
        {morgan2016student}
\bibfield{author}{\bibinfo{person}{Helen~K Morgan}, \bibinfo{person}{Joel~A Purkiss}, \bibinfo{person}{Annie~C Porter}, \bibinfo{person}{Monica~L Lypson}, \bibinfo{person}{Sally~A Santen}, \bibinfo{person}{Jennifer~G Christner}, \bibinfo{person}{Cyril~M Grum}, {and} \bibinfo{person}{Maya~M Hammoud}.} \bibinfo{year}{2016}\natexlab{}.
\newblock \showarticletitle{Student evaluation of faculty physicians: gender differences in teaching evaluations}.
\newblock \bibinfo{journal}{\emph{Journal of Women's Health}} \bibinfo{volume}{25}, \bibinfo{number}{5} (\bibinfo{year}{2016}), \bibinfo{pages}{453--456}.
\newblock


\bibitem[Nadeem et~al\mbox{.}(2021)]%
        {nadeem-etal-2021-stereoset}
\bibfield{author}{\bibinfo{person}{Moin Nadeem}, \bibinfo{person}{Anna Bethke}, {and} \bibinfo{person}{Siva Reddy}.} \bibinfo{year}{2021}\natexlab{}.
\newblock \showarticletitle{{S}tereo{S}et: Measuring stereotypical bias in pretrained language models}. In \bibinfo{booktitle}{\emph{Proceedings of the 59th Annual Meeting of the Association for Computational Linguistics and the 11th International Joint Conference on Natural Language Processing (Volume 1: Long Papers)}}, \bibfield{editor}{\bibinfo{person}{Chengqing Zong}, \bibinfo{person}{Fei Xia}, \bibinfo{person}{Wenjie Li}, {and} \bibinfo{person}{Roberto Navigli}} (Eds.). \bibinfo{publisher}{Association for Computational Linguistics}, \bibinfo{address}{Online}, \bibinfo{pages}{5356--5371}.
\newblock
\urldef\tempurl%
\url{https://doi.org/10.18653/v1/2021.acl-long.416}
\showDOI{\tempurl}


\bibitem[Neri and Cozman(2020)]%
        {neri2020role}
\bibfield{author}{\bibinfo{person}{Hugo Neri} {and} \bibinfo{person}{Fabio Cozman}.} \bibinfo{year}{2020}\natexlab{}.
\newblock \showarticletitle{The role of experts in the public perception of risk of artificial intelligence}.
\newblock \bibinfo{journal}{\emph{AI \& society}} \bibinfo{volume}{35}, \bibinfo{number}{3} (\bibinfo{year}{2020}), \bibinfo{pages}{663--673}.
\newblock


\bibitem[Noy and Zhang(2023)]%
        {Zhang-2023-productivity}
\bibfield{author}{\bibinfo{person}{Shakked Noy} {and} \bibinfo{person}{Whitney Zhang}.} \bibinfo{year}{2023}\natexlab{}.
\newblock \showarticletitle{Experimental evidence on the productivity effects of generative artificial intelligence}.
\newblock \bibinfo{journal}{\emph{Science}} \bibinfo{volume}{381}, \bibinfo{number}{6654} (\bibinfo{year}{2023}), \bibinfo{pages}{187--192}.
\newblock
\urldef\tempurl%
\url{https://doi.org/10.1126/science.adh2586}
\showDOI{\tempurl}


\bibitem[Perkins(2023)]%
        {perkins2023academic}
\bibfield{author}{\bibinfo{person}{Mike Perkins}.} \bibinfo{year}{2023}\natexlab{}.
\newblock \showarticletitle{Academic Integrity considerations of AI Large Language Models in the post-pandemic era: ChatGPT and beyond}.
\newblock \bibinfo{journal}{\emph{Journal of University Teaching and Learning Practice}} \bibinfo{volume}{20}, \bibinfo{number}{2} (\bibinfo{year}{2023}).
\newblock


\bibitem[Rae(2024)]%
        {rae2024content}
\bibfield{author}{\bibinfo{person}{Irene Rae}.} \bibinfo{year}{2024}\natexlab{}.
\newblock \showarticletitle{The Effects of Perceived AI Use On Content Perceptions}. In \bibinfo{booktitle}{\emph{Proceedings of the 2024 CHI Conference on Human Factors in Computing Systems}} (Honolulu, HI, USA) \emph{(\bibinfo{series}{CHI '24})}. \bibinfo{publisher}{Association for Computing Machinery}, \bibinfo{address}{New York, NY, USA}, Article \bibinfo{articleno}{978}, \bibinfo{numpages}{14}~pages.
\newblock
\showISBNx{9798400703300}
\urldef\tempurl%
\url{https://doi.org/10.1145/3613904.3642076}
\showDOI{\tempurl}


\bibitem[Ray(2024)]%
        {Ray2024AIviews}
\bibfield{author}{\bibinfo{person}{Julie Ray}.} \bibinfo{year}{2024}\natexlab{}.
\newblock \bibinfo{title}{Americans express real concerns about Artificial Intelligence}.
\newblock
\newblock
\urldef\tempurl%
\url{https://news.gallup.com/poll/648953/americans-express-real-concerns-artificial-intelligence.aspx}
\showURL{%
\tempurl}


\bibitem[Rothschild and Klingenberg(1990)]%
        {rothschild1990self}
\bibfield{author}{\bibinfo{person}{Dennie Rothschild} {and} \bibinfo{person}{Felicia Klingenberg}.} \bibinfo{year}{1990}\natexlab{}.
\newblock \showarticletitle{Self and peer evaluation of writing in the interactive ESL classroom: An exploratory study}.
\newblock \bibinfo{journal}{\emph{TESL Canada Journal}} (\bibinfo{year}{1990}), \bibinfo{pages}{52--65}.
\newblock


\bibitem[Sager(1973)]%
        {sager1973sager}
\bibfield{author}{\bibinfo{person}{Carol Sager}.} \bibinfo{year}{1973}\natexlab{}.
\newblock \showarticletitle{Sager Writing Scale.}
\newblock  (\bibinfo{year}{1973}).
\newblock


\bibitem[Sashaborm(2021)]%
        {Sashaborm_2021}
\bibfield{author}{\bibinfo{person}{Sashaborm}.} \bibinfo{year}{2021}\natexlab{}.
\newblock \bibinfo{title}{Thispersondoesnotexist - random AI generated photos of fake persons}.
\newblock
\newblock
\urldef\tempurl%
\url{https://this-person-does-not-exist.com/en}
\showURL{%
\tempurl}


\bibitem[Segal(2023)]%
        {Segal_2023}
\bibfield{author}{\bibinfo{person}{Edward Segal}.} \bibinfo{year}{2023}\natexlab{}.
\newblock \bibinfo{title}{New report provides reality check about freelancers in the workforce}.
\newblock
\newblock
\urldef\tempurl%
\url{https://www.forbes.com/sites/edwardsegal/2023/12/12/new-report-provides-reality-check-about-freelancers-in-the-workforce/}
\showURL{%
\tempurl}


\bibitem[Selbst et~al\mbox{.}(2019)]%
        {selbst2019fairness}
\bibfield{author}{\bibinfo{person}{Andrew~D Selbst}, \bibinfo{person}{Danah Boyd}, \bibinfo{person}{Sorelle~A Friedler}, \bibinfo{person}{Suresh Venkatasubramanian}, {and} \bibinfo{person}{Janet Vertesi}.} \bibinfo{year}{2019}\natexlab{}.
\newblock \showarticletitle{Fairness and abstraction in sociotechnical systems}. In \bibinfo{booktitle}{\emph{Proceedings of the conference on fairness, accountability, and transparency}}. \bibinfo{pages}{59--68}.
\newblock


\bibitem[Shelby et~al\mbox{.}(2023)]%
        {shelby2023}
\bibfield{author}{\bibinfo{person}{Renee Shelby}, \bibinfo{person}{Shalaleh Rismani}, \bibinfo{person}{Kathryn Henne}, \bibinfo{person}{AJung Moon}, \bibinfo{person}{Negar Rostamzadeh}, \bibinfo{person}{Paul Nicholas}, \bibinfo{person}{N'Mah Yilla-Akbari}, \bibinfo{person}{Jess Gallegos}, \bibinfo{person}{Andrew Smart}, \bibinfo{person}{Emilio Garcia}, {and} \bibinfo{person}{Gurleen Virk}.} \bibinfo{year}{2023}\natexlab{}.
\newblock \showarticletitle{Sociotechnical Harms of Algorithmic Systems: Scoping a Taxonomy for Harm Reduction}. In \bibinfo{booktitle}{\emph{Proceedings of the 2023 AAAI/ACM Conference on AI, Ethics, and Society}} (Montr\'{e}al, QC, Canada) \emph{(\bibinfo{series}{AIES '23})}. \bibinfo{publisher}{Association for Computing Machinery}, \bibinfo{address}{New York, NY, USA}, \bibinfo{pages}{723–741}.
\newblock
\showISBNx{9798400702310}
\urldef\tempurl%
\url{https://doi.org/10.1145/3600211.3604673}
\showDOI{\tempurl}


\bibitem[Shen et~al\mbox{.}(2024)]%
        {shen2024bidirectionalhumanaialignmentsystematic}
\bibfield{author}{\bibinfo{person}{Hua Shen}, \bibinfo{person}{Tiffany Knearem}, \bibinfo{person}{Reshmi Ghosh}, \bibinfo{person}{Kenan Alkiek}, \bibinfo{person}{Kundan Krishna}, \bibinfo{person}{Yachuan Liu}, \bibinfo{person}{Ziqiao Ma}, \bibinfo{person}{Savvas Petridis}, \bibinfo{person}{Yi-Hao Peng}, \bibinfo{person}{Li Qiwei}, \bibinfo{person}{Sushrita Rakshit}, \bibinfo{person}{Chenglei Si}, \bibinfo{person}{Yutong Xie}, \bibinfo{person}{Jeffrey~P. Bigham}, \bibinfo{person}{Frank Bentley}, \bibinfo{person}{Joyce Chai}, \bibinfo{person}{Zachary Lipton}, \bibinfo{person}{Qiaozhu Mei}, \bibinfo{person}{Rada Mihalcea}, \bibinfo{person}{Michael Terry}, \bibinfo{person}{Diyi Yang}, \bibinfo{person}{Meredith~Ringel Morris}, \bibinfo{person}{Paul Resnick}, {and} \bibinfo{person}{David Jurgens}.} \bibinfo{year}{2024}\natexlab{}.
\newblock \bibinfo{title}{Towards Bidirectional Human-AI Alignment: A Systematic Review for Clarifications, Framework, and Future Directions}.
\newblock
\newblock
\showeprint[arxiv]{2406.09264}~[cs.HC]
\urldef\tempurl%
\url{https://arxiv.org/abs/2406.09264}
\showURL{%
\tempurl}


\bibitem[Singh et~al\mbox{.}(2023)]%
        {singh2023elephant}
\bibfield{author}{\bibinfo{person}{Nikhil Singh}, \bibinfo{person}{Guillermo Bernal}, \bibinfo{person}{Daria Savchenko}, {and} \bibinfo{person}{Elena~L. Glassman}.} \bibinfo{year}{2023}\natexlab{}.
\newblock \showarticletitle{Where to Hide a Stolen Elephant: Leaps in Creative Writing with Multimodal Machine Intelligence}.
\newblock \bibinfo{journal}{\emph{ACM Trans. Comput.-Hum. Interact.}} \bibinfo{volume}{30}, \bibinfo{number}{5}, Article \bibinfo{articleno}{68} (\bibinfo{date}{sep} \bibinfo{year}{2023}), \bibinfo{numpages}{57}~pages.
\newblock
\showISSN{1073-0516}
\urldef\tempurl%
\url{https://doi.org/10.1145/3511599}
\showDOI{\tempurl}


\bibitem[Sullivan et~al\mbox{.}(2023)]%
        {sullivan2023chatgpt}
\bibfield{author}{\bibinfo{person}{Miriam Sullivan}, \bibinfo{person}{Andrew Kelly}, {and} \bibinfo{person}{Paul McLaughlan}.} \bibinfo{year}{2023}\natexlab{}.
\newblock \showarticletitle{ChatGPT in higher education: Considerations for academic integrity and student learning}.
\newblock  (\bibinfo{year}{2023}).
\newblock


\bibitem[Sun et~al\mbox{.}(2024)]%
        {sun2024smile}
\bibfield{author}{\bibinfo{person}{Luhang Sun}, \bibinfo{person}{Mian Wei}, \bibinfo{person}{Yibing Sun}, \bibinfo{person}{Yoo~Ji Suh}, \bibinfo{person}{Liwei Shen}, {and} \bibinfo{person}{Sijia Yang}.} \bibinfo{year}{2024}\natexlab{}.
\newblock \showarticletitle{{Smiling women pitching down: auditing representational and presentational gender biases in image-generative AI}}.
\newblock \bibinfo{journal}{\emph{Journal of Computer-Mediated Communication}} \bibinfo{volume}{29}, \bibinfo{number}{1} (\bibinfo{date}{02} \bibinfo{year}{2024}), \bibinfo{pages}{zmad045}.
\newblock
\showISSN{1083-6101}
\urldef\tempurl%
\url{https://doi.org/10.1093/jcmc/zmad045}
\showDOI{\tempurl}


\bibitem[Team(2024)]%
        {upworkdata}
\bibfield{author}{\bibinfo{person}{The~Upwork Team}.} \bibinfo{year}{2024}\natexlab{}.
\newblock
\newblock
\urldef\tempurl%
\url{https://www.upwork.com/resources/highest-paying-freelance-jobs}
\showURL{%
\tempurl}


\bibitem[Vue et~al\mbox{.}(2023)]%
        {VUE20233138}
\bibfield{author}{\bibinfo{person}{Zer Vue}, \bibinfo{person}{Chia Vang}, \bibinfo{person}{Neng Vue}, \bibinfo{person}{Vijayvardhan Kamalumpundi}, \bibinfo{person}{Taylor Barongan}, \bibinfo{person}{Bryanna Shao}, \bibinfo{person}{Sunny Huang}, \bibinfo{person}{Larry Vang}, \bibinfo{person}{Mein Vue}, \bibinfo{person}{Nancy Vang}, \bibinfo{person}{Jianqiang Shao}, \bibinfo{person}{CoohleenAnn Coombes}, \bibinfo{person}{Prasanna Katti}, \bibinfo{person}{Kaihua Liu}, \bibinfo{person}{Kailee Yoshimura}, \bibinfo{person}{Michelle Biete}, \bibinfo{person}{Dao-Fu Dai}, \bibinfo{person}{Mark~A. Phillips}, {and} \bibinfo{person}{Richard~R. Behringer}.} \bibinfo{year}{2023}\natexlab{}.
\newblock \showarticletitle{Asian Americans in STEM are not a monolith}.
\newblock \bibinfo{journal}{\emph{Cell}} \bibinfo{volume}{186}, \bibinfo{number}{15} (\bibinfo{year}{2023}), \bibinfo{pages}{3138--3142}.
\newblock
\showISSN{0092-8674}
\urldef\tempurl%
\url{https://doi.org/10.1016/j.cell.2023.06.017}
\showDOI{\tempurl}


\bibitem[Waldman and Avolio(1991)]%
        {waldman1991race}
\bibfield{author}{\bibinfo{person}{David~A Waldman} {and} \bibinfo{person}{Bruce~J Avolio}.} \bibinfo{year}{1991}\natexlab{}.
\newblock \showarticletitle{Race effects in performance evaluations: Controlling for ability, education, and experience.}
\newblock \bibinfo{journal}{\emph{Journal of applied psychology}} \bibinfo{volume}{76}, \bibinfo{number}{6} (\bibinfo{year}{1991}), \bibinfo{pages}{897}.
\newblock


\bibitem[Weidinger et~al\mbox{.}(2023)]%
        {weidinger2023sociotechnical}
\bibfield{author}{\bibinfo{person}{Laura Weidinger}, \bibinfo{person}{Maribeth Rauh}, \bibinfo{person}{Nahema Marchal}, \bibinfo{person}{Arianna Manzini}, \bibinfo{person}{Lisa~Anne Hendricks}, \bibinfo{person}{Juan Mateos-Garcia}, \bibinfo{person}{Stevie Bergman}, \bibinfo{person}{Jackie Kay}, \bibinfo{person}{Conor Griffin}, \bibinfo{person}{Ben Bariach}, {et~al\mbox{.}}} \bibinfo{year}{2023}\natexlab{}.
\newblock \showarticletitle{Sociotechnical safety evaluation of generative ai systems}.
\newblock \bibinfo{journal}{\emph{arXiv preprint arXiv:2310.11986}} (\bibinfo{year}{2023}).
\newblock


\bibitem[Weidinger et~al\mbox{.}(2022)]%
        {weidinger-etal-2022-risks}
\bibfield{author}{\bibinfo{person}{Laura Weidinger}, \bibinfo{person}{Jonathan Uesato}, \bibinfo{person}{Maribeth Rauh}, \bibinfo{person}{Conor Griffin}, \bibinfo{person}{Po-Sen Huang}, \bibinfo{person}{John Mellor}, \bibinfo{person}{Amelia Glaese}, \bibinfo{person}{Myra Cheng}, \bibinfo{person}{Borja Balle}, \bibinfo{person}{Atoosa Kasirzadeh}, \bibinfo{person}{Courtney Biles}, \bibinfo{person}{Sasha Brown}, \bibinfo{person}{Zac Kenton}, \bibinfo{person}{Will Hawkins}, \bibinfo{person}{Tom Stepleton}, \bibinfo{person}{Abeba Birhane}, \bibinfo{person}{Lisa~Anne Hendricks}, \bibinfo{person}{Laura Rimell}, \bibinfo{person}{William Isaac}, \bibinfo{person}{Julia Haas}, \bibinfo{person}{Sean Legassick}, \bibinfo{person}{Geoffrey Irving}, {and} \bibinfo{person}{Iason Gabriel}.} \bibinfo{year}{2022}\natexlab{}.
\newblock \showarticletitle{Taxonomy of Risks Posed by Language Models}. In \bibinfo{booktitle}{\emph{Proceedings of the 2022 ACM Conference on Fairness, Accountability, and Transparency}} (Seoul, Republic of Korea) \emph{(\bibinfo{series}{FAccT '22})}. \bibinfo{publisher}{Association for Computing Machinery}, \bibinfo{address}{New York, NY, USA}, \bibinfo{pages}{214–229}.
\newblock
\showISBNx{9781450393522}
\urldef\tempurl%
\url{https://doi.org/10.1145/3531146.3533088}
\showDOI{\tempurl}


\bibitem[Wenzel et~al\mbox{.}(2023)]%
        {wenzel2023voice}
\bibfield{author}{\bibinfo{person}{Kimi Wenzel}, \bibinfo{person}{Nitya Devireddy}, \bibinfo{person}{Cam Davison}, {and} \bibinfo{person}{Geoff Kaufman}.} \bibinfo{year}{2023}\natexlab{}.
\newblock \showarticletitle{Can Voice Assistants Be Microaggressors? Cross-Race Psychological Responses to Failures of Automatic Speech Recognition}. In \bibinfo{booktitle}{\emph{Proceedings of the 2023 CHI Conference on Human Factors in Computing Systems}} (Hamburg, Germany) \emph{(\bibinfo{series}{CHI '23})}. \bibinfo{publisher}{Association for Computing Machinery}, \bibinfo{address}{New York, NY, USA}, Article \bibinfo{articleno}{109}, \bibinfo{numpages}{14}~pages.
\newblock
\showISBNx{9781450394215}
\urldef\tempurl%
\url{https://doi.org/10.1145/3544548.3581357}
\showDOI{\tempurl}


\bibitem[Wu and Kelly(2021)]%
        {wu2020dating}
\bibfield{author}{\bibinfo{person}{Yihan Wu} {and} \bibinfo{person}{Ryan~M. Kelly}.} \bibinfo{year}{2021}\natexlab{}.
\newblock \showarticletitle{Online Dating Meets Artificial Intelligence: How the Perception of Algorithmically Generated Profile Text Impacts Attractiveness and Trust}. In \bibinfo{booktitle}{\emph{Proceedings of the 32nd Australian Conference on Human-Computer Interaction}} (Sydney, NSW, Australia) \emph{(\bibinfo{series}{OzCHI '20})}. \bibinfo{publisher}{Association for Computing Machinery}, \bibinfo{address}{New York, NY, USA}, \bibinfo{pages}{444–453}.
\newblock
\showISBNx{9781450389754}
\urldef\tempurl%
\url{https://doi.org/10.1145/3441000.3441074}
\showDOI{\tempurl}


\bibitem[Xie et~al\mbox{.}(2023)]%
        {xie2023aicheating}
\bibfield{author}{\bibinfo{person}{Ying Xie}, \bibinfo{person}{Shaoen Wu}, {and} \bibinfo{person}{Sumit Chakravarty}.} \bibinfo{year}{2023}\natexlab{}.
\newblock \showarticletitle{AI meets AI: Artificial Intelligence and Academic Integrity - A Survey on Mitigating AI-Assisted Cheating in Computing Education}. In \bibinfo{booktitle}{\emph{Proceedings of the 24th Annual Conference on Information Technology Education}} (Marietta, GA, USA) \emph{(\bibinfo{series}{SIGITE '23})}. \bibinfo{publisher}{Association for Computing Machinery}, \bibinfo{address}{New York, NY, USA}, \bibinfo{pages}{79–83}.
\newblock
\showISBNx{9798400701306}
\urldef\tempurl%
\url{https://doi.org/10.1145/3585059.3611449}
\showDOI{\tempurl}


\bibitem[Xueliang and Qingxia(1988)]%
        {xueliang1988language}
\bibfield{author}{\bibinfo{person}{Ma Xueliang} {and} \bibinfo{person}{Dai Qingxia}.} \bibinfo{year}{1988}\natexlab{}.
\newblock \showarticletitle{Language and nationality}.
\newblock \bibinfo{journal}{\emph{Chinese Sociology \& Anthropology}} \bibinfo{volume}{21}, \bibinfo{number}{1} (\bibinfo{year}{1988}), \bibinfo{pages}{81--104}.
\newblock


\bibitem[Yuan et~al\mbox{.}(2022)]%
        {yuan2022wordcraft}
\bibfield{author}{\bibinfo{person}{Ann Yuan}, \bibinfo{person}{Andy Coenen}, \bibinfo{person}{Emily Reif}, {and} \bibinfo{person}{Daphne Ippolito}.} \bibinfo{year}{2022}\natexlab{}.
\newblock \showarticletitle{Wordcraft: Story Writing With Large Language Models}. In \bibinfo{booktitle}{\emph{Proceedings of the 27th International Conference on Intelligent User Interfaces}} (Helsinki, Finland) \emph{(\bibinfo{series}{IUI '22})}. \bibinfo{publisher}{Association for Computing Machinery}, \bibinfo{address}{New York, NY, USA}, \bibinfo{pages}{841–852}.
\newblock
\showISBNx{9781450391443}
\urldef\tempurl%
\url{https://doi.org/10.1145/3490099.3511105}
\showDOI{\tempurl}


\bibitem[Zhou and Lee(2024)]%
        {zhou2024artcreativity}
\bibfield{author}{\bibinfo{person}{Eric Zhou} {and} \bibinfo{person}{Dokyun Lee}.} \bibinfo{year}{2024}\natexlab{}.
\newblock \showarticletitle{{Generative artificial intelligence, human creativity, and art}}.
\newblock \bibinfo{journal}{\emph{PNAS Nexus}} \bibinfo{volume}{3}, \bibinfo{number}{3} (\bibinfo{date}{03} \bibinfo{year}{2024}), \bibinfo{pages}{pgae052}.
\newblock
\showISSN{2752-6542}
\urldef\tempurl%
\url{https://doi.org/10.1093/pnasnexus/pgae052}
\showDOI{\tempurl}
\showeprint{https://academic.oup.com/pnasnexus/article-pdf/3/3/pgae052/57464715/pgae052.pdf}


\end{thebibliography}

\appendix

\section{Appendix}

\subsection{Deviation from Preregistration}
\label{sec:dev}
For full transparency, we expand here where we deviated from our preregistered submissions in terms of hypotheses. 
First, we only preregistered Experiments~1 and~3.
Second, H2 was different in the Experiment~3 preregistration, where we hypothesized that controlling for suspicion, people presented as foreign nationals will receive \textit{higher} quality evaluations. 
In hindsight, we are not sure this direction of the hypothesis was well-justified. For simplicity, we aligned the hypothesis with the other H2 hypotheses above (H2 was not confirmed either way). 
Finally, for Experiment~1, we only preregistered Hypotheses~1 and~2 and did not include Hypothesis~3 regarding outcomes. We developed the outcomes after the preregistration but before the Experiment~1 launch. 

\subsection{Treatment Validation}
\label{sec:manip_check}
\begin{figure*}[h]
    \centering
    \begin{subfigure}[h]{0.3\textwidth}
        \centering
        \includegraphics[width=\textwidth]
        {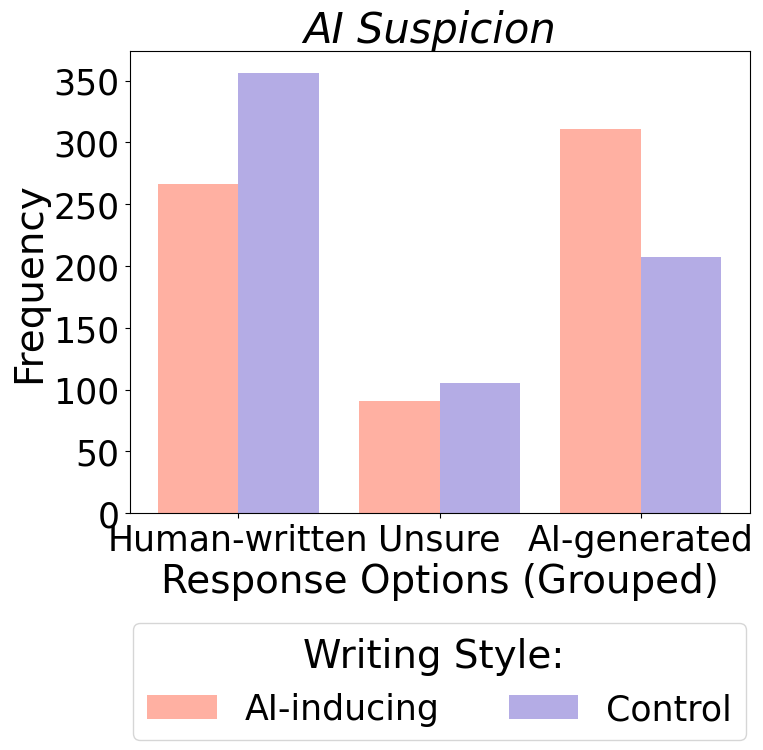}
        \caption{Experiment~1--Gender}
        \label{fig:gender_validation}
    \end{subfigure}
    \begin{subfigure}[h]{0.3\textwidth}
        \centering
        \includegraphics[width=\textwidth]
        {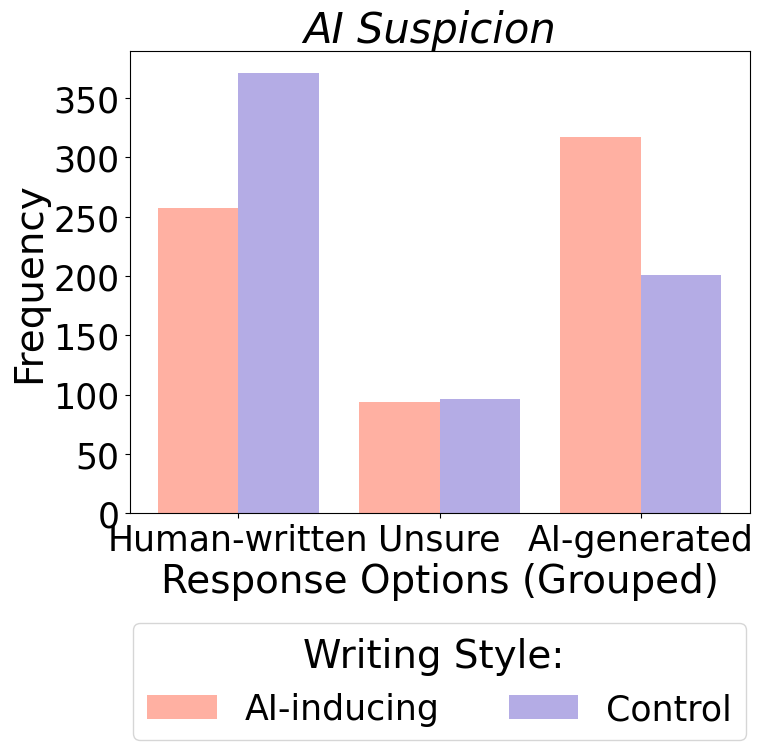}
        \caption{Experiment~2--Race}
        \label{fig:race_validation}
    \end{subfigure}
    \begin{subfigure}[h]{0.3\textwidth}
        \centering
        \includegraphics[width=\textwidth]
        {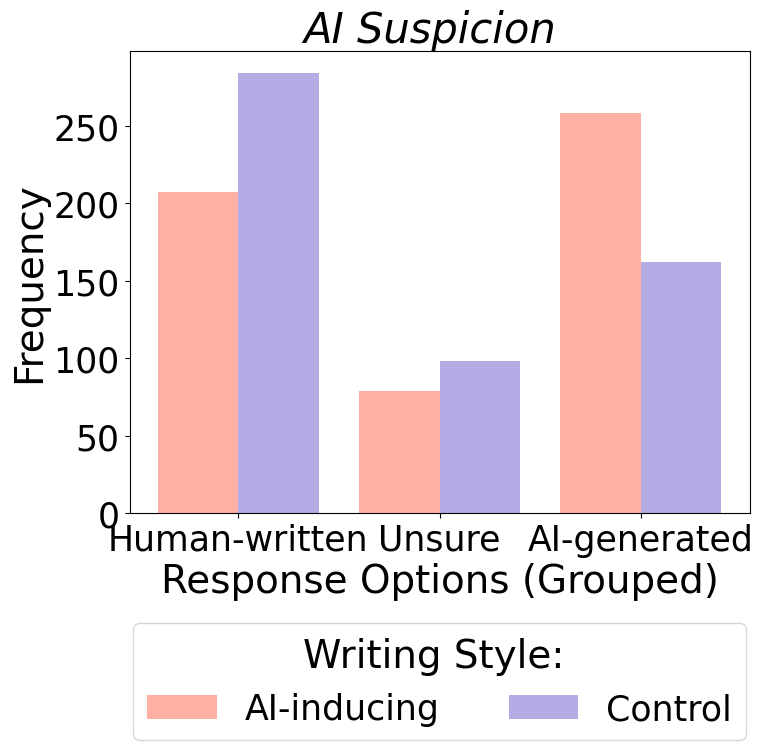}
        \caption{Experiment~3--Nationality}
        \label{fig:esl_validation}
    \end{subfigure}
    \caption{\textbf{Participants Assessment of Writing Styles.} The AI-inducing writing style was more suspected of being AI-generated compared to the control.}
    \label{fig:validation}
\end{figure*}

We manually manipulated half of the press releases in an AI-inducing style to raise participants' suspicion of AI use.
Table~\ref{table:treatment} presents an example of AI-inducing sentences, which were created by replacing words with less common synonyms and employing verbose language. We validated the effect of writing style post hoc, as shown in Figure~\ref{fig:validation}. 
In all three studies, the AI-inducing writing style was more likely to be seen as AI-generated compared to the control. 
A linear mixed model to predict AI suspicion with writing style as a fixed effect and participant as a random effect confirmed the visual finding. The result is statistically significant across all three studies ($p<0.001$).

\subsection{Participant Demographics}
\label{sec:detailed_demographics}

\begin{table*}[h]
\caption{\textbf{AI-inducing sentences}. We manipulated half of the press releases to be AI-inducing by modifying a three sentences in the sample. We provide a sample of AI-Inducing sentences below. Noticeable differences in the sentences are highlighted in blue.}
\centering
\begin{tabular}{|p{8cm}|p{8cm}|} 
\hline
\textbf{Original Sentence} & \textbf{Modified Sentence} \\ \hline
Klarna, a leading global retail bank, payments, and shopping service is excited to announce its new collaboration with OpenAI, which will level up the shopping experience. \newline \newline

Trade the bustling city for the peaceful countryside and a world of mystery with the release of InnoGames’ new exploration and farming simulation game Sunrise Village.\newline \newline

RainFocus is the next-generation event marketing platform built to capture and analyze unprecedented amounts of first-party data for exceptional events and optimized engagement throughout the customer journey.  
&  
ReBank, a \textcolor{blue}{preeminent entity} in the global retail banking, payments, and shopping services sector, is excited to announce its novel integration of generative AI, which will enhance the shopping paradigm. \newline 

Inugamis is elated to unveil Sunset Village, an enchanting mobile game that invites players to \textcolor{blue}{exchange the frenetic city life} for the serene countryside and a realm brimming with mystery. \newline 

Focus is the vanguard of next-generation event marketing platforms, \textcolor{blue}{meticulously engineered} to capture and analyze \textcolor{blue}{unparalleled amounts} of first-party data, thereby facilitating exceptional events and optimized engagement throughout the customer journey.  
\\ \hline
\end{tabular}
\label{table:treatment}
\end{table*}

We recruited a US representative sample from Prolific and present the demographic data of the participants in all three experiments in Table~\ref{table:demographics}. 
We also conduct an exploratory analysis to determine if age had a significant effect on AI suspicion and hiring. 
Since our largest age group was participants aged 55-64 years old, we removed this group and re-ran our AI suspicion analysis and hiring analysis for each experiment in our study.
The exploratory analysis shows the same trends for all three experiments.

In Experiment~1, we conducted a linear mixed model analysis to predict AI suspicion based on gender and a separate model incorporating both gender and writing style as predictors.
In both models, we excluded data from participants aged 55-64 years old. Our findings indicate a trend where men are more likely to be suspected of using generative AI; however, this trend is not statistically significant ($p=0.116$, \textit{n.s.}), likely due to reduced statistical power. The model including both gender and writing style did show the same pattern, with statistically significant results: men are more often suspected ($p<0.001$), with suspicion being notably higher when the writing style is perceived as AI-generated ($p<0.001$). 
Lastly, we conducted a linear mixed model with this subset of participants to predict hiring likelihood based on gender, AI suspicion, and writing style. AI suspicion negatively impacted hiring likelihood ($p<0.001$).

We performed the same analysis, again excluding the data from participants aged 55-64 years old in Experiment~2 (race). Like the findings in our main results section, while White freelancers were suspected of using AI more than Black freelancers, we do not see any significant racial differences ($p=0.155$, \textit{n.s.}). When we include race and writing style in the AI suspicion model, we still do not see any significant racial differences ($p=0.594$, \textit{n.s.}). We do, however, see the effect of writing style ($p<0.001$), where the AI-inducing style is much more likely to be suspected of being AI-generated. As expected, in our hiring prediction model, we see that AI suspicion negatively impacts hiring likelihood ($p<0.001$).

The re-analysis of the Experiment~3 data, excluding the data from participants aged 55-64 years old, again showed similar results. Foreign Nationals are more likely to be suspected of using generative AI ($p=0.017$). When we include writing style in the AI suspicion model, we still observe the same trend ($p=0.011$). Furthermore, we see that the AI-inducing style is more likely to be suspected of being AI-generated ($p<0.01$). Similar to the previous experiments, we see that AI suspicion reduces hiring likelihood ($p<0.001$).

\begin{table*}[h]
\caption{\textbf{Participant Demographics}. An overview of the participants' demographic backgrounds in all three experiments.}
\begin{tabular}{|lrrr|}
\hline
\multicolumn{1}{|c|}{} & \multicolumn{1}{c|}{Experiment~1} & \multicolumn{1}{c|}{Experiment~2} & \multicolumn{1}{l|}{Experiment~3} \\ \hline
\multicolumn{4}{|l|}{\textbf{Age}} \\ \hline
\multicolumn{1}{|l|}{18-24 years old} & \multicolumn{1}{r|}{10.18\%} & \multicolumn{1}{r|}{12.28\%} & 12.87\% \\ 
\multicolumn{1}{|l|}{25-34 years old} & \multicolumn{1}{r|}{19.76\%} & \multicolumn{1}{r|}{17.96\%} & 18.38\% \\ 
\multicolumn{1}{|l|}{35-44 years old} & \multicolumn{1}{r|}{18.56\%} & \multicolumn{1}{r|}{16.77\%} & 15.81\% \\ 
\multicolumn{1}{|l|}{45-54 years old} & \multicolumn{1}{r|}{15.57\%} & \multicolumn{1}{r|}{16.47\%} & 17.28\% \\ 
\multicolumn{1}{|l|}{55-64 years old} & \multicolumn{1}{r|}{25.75\%} & \multicolumn{1}{r|}{24.25\%} & 24.26\% \\ 
\multicolumn{1}{|l|}{65+ years old} & \multicolumn{1}{r|}{10.18\%} & \multicolumn{1}{r|}{12.28\%} & 11.40\% \\ \hline
\multicolumn{4}{|l|}{\textbf{Sex}} \\ \hline
\multicolumn{1}{|l|}{Female} & \multicolumn{1}{r|}{49.40\%} & \multicolumn{1}{r|}{49.70\%} & 48.90\% \\ 
\multicolumn{1}{|l|}{Male} & \multicolumn{1}{r|}{48.50\%} & \multicolumn{1}{r|}{47.60\%} & 48.53\% \\ 
\multicolumn{1}{|l|}{Non-binary / third gender} & \multicolumn{1}{r|}{1.80\%} & \multicolumn{1}{r|}{2.10\%} & 2.21\% \\ 
\multicolumn{1}{|l|}{Prefer to self-describe} & \multicolumn{1}{r|}{0.30\%} & \multicolumn{1}{r|}{0.60\%} & 0.368\% \\ \hline
\multicolumn{4}{|l|}{\textbf{Race}} \\ \hline
\multicolumn{1}{|l|}{Asian} & \multicolumn{1}{r|}{6.29\%} & \multicolumn{1}{r|}{6.59\%} & 6.62\% \\ 
\multicolumn{1}{|l|}{Black/African American} & \multicolumn{1}{r|}{13.17\%} & \multicolumn{1}{r|}{16.17\%} & 12.5\% \\ 
\multicolumn{1}{|l|}{Hispanic} & \multicolumn{1}{r|}{9.28\%} & \multicolumn{1}{r|}{10.48\%} & 8.09\% \\ 
\multicolumn{1}{|l|}{Native American} & \multicolumn{1}{r|}{3.29\%} & \multicolumn{1}{r|}{1.50\%} & 0.74\% \\ 
\multicolumn{1}{|l|}{Pacific Islander} & \multicolumn{1}{r|}{0\%} & \multicolumn{1}{r|}{0.30\%} & 0\% \\ 
\multicolumn{1}{|l|}{Prefer to self-describe} & \multicolumn{1}{r|}{4.49\%} & \multicolumn{1}{r|}{3.29\%} & 5.15\% \\ 
\multicolumn{1}{|l|}{White/Caucasian} & \multicolumn{1}{r|}{63.47\%} & \multicolumn{1}{r|}{61.68\%} & 66.91\% \\ \hline
\multicolumn{4}{|l|}{\textbf{Sexuality}} \\ \hline
\multicolumn{1}{|l|}{Heterosexual or straight} & \multicolumn{1}{r|}{80.54\%} & \multicolumn{1}{r|}{83.23\%} & 79.04\% \\ 
\multicolumn{1}{|l|}{Bisexual} & \multicolumn{1}{r|}{12.87\%} & \multicolumn{1}{r|}{9.28\%} & 13.98\% \\ 
\multicolumn{1}{|l|}{Gay or Lesbian} & \multicolumn{1}{r|}{4.50\%} & \multicolumn{1}{r|}{4.50\%} & 5.15\% \\ 
\multicolumn{1}{|l|}{Prefer to self-describe} & \multicolumn{1}{r|}{2.10\%} & \multicolumn{1}{r|}{3.00\%} & 1.84\% \\ \hline
\multicolumn{4}{|l|}{\textbf{Education}} \\ \hline
\multicolumn{1}{|l|}{Less than high school} & \multicolumn{1}{r|}{0\%} & \multicolumn{1}{r|}{0.30\%} & 0\% \\ 
\multicolumn{1}{|l|}{Some college but no degree} & \multicolumn{1}{r|}{22.46\%} & \multicolumn{1}{r|}{20.36\%} & 18.01\% \\ 
\multicolumn{1}{|l|}{Associate degree in college (2-year)} & \multicolumn{1}{r|}{14.07\%} & \multicolumn{1}{r|}{11.38\%} & 9.19\% \\ 
\multicolumn{1}{|l|}{Bachelor's degree in college (4-year)} & \multicolumn{1}{r|}{35.03\%} & \multicolumn{1}{r|}{35.33\%} & 42.65\% \\ 
\multicolumn{1}{|l|}{Master's degree} & \multicolumn{1}{r|}{17.07\%} & \multicolumn{1}{r|}{18.86\%} & 15.44\% \\ 
\multicolumn{1}{|l|}{Professional degree (JD, MD)} & \multicolumn{1}{r|}{2.40\%} & \multicolumn{1}{r|}{1.80\%} & 1.100\% \\ 
\multicolumn{1}{|l|}{Doctoral degree} & \multicolumn{1}{r|}{0.60\%} & \multicolumn{1}{r|}{2.40\%} & 1.10\% \\ \hline
\end{tabular}
\label{table:demographics}
\end{table*}

\end{document}